\documentclass[11pt, a4paper]{article}
\usepackage{slashed,paperstyle,multirow,relsize,soul} \usepackage[normalem]{ulem}
\usepackage{color} \usepackage{subcaption} \usepackage{siunitx}
\usepackage[capitalise]{cleveref}
\usepackage[normalem]{ulem}
\usepackage{graphicx}

\newcommand{\refeq}[1]{Eq.~(\ref{#1})}

\newcommand{\refeqss}[3]{Eqs.~(\ref{#1}), (\ref{#2})~and~(\ref{#3})}
\newcommand{\reffig}[1]{Fig.~\ref{#1}}

\newcommand{\refsec}[1]{Section~\ref{#1}}
\newcommand{\refapp}[1]{Appendix~\ref{#1}}
\newcommand{\reftab}[1]{Table~\ref{#1}}
\newcommand{\refref}[1]{Ref.~\cite{#1}}
\newcommand{\refrefs}[2]{Refs.~\cite{#1}~and~\cite{#2}}

\newcounter{CommentCount}
\definecolor{PB}{rgb}{0.9,0,0}
\definecolor{TC}{rgb}{0,0,0.9}

\def\eg{\emph{e.g.}}

\def\anu{\overline{\nu}}

\title{Sensitivities and synergies of DUNE and T2HK}

\author[a]{Peter Ballett,}
\author[b]{Stephen F. King,}
\author[a]{Silvia Pascoli,}
\author[b,c]{Nick W. Prouse}
\author[a]{and TseChun Wang}

\affiliation[a]{Institute for Particle Physics Phenomenology, Department of
Physics, Durham University, South Road, Durham DH1 3LE, United Kingdom.}
\affiliation[b]{School of Physics and Astronomy, University of Southampton, SO17 1BJ Southampton, United Kingdom.} 
\affiliation[c]{Particle Physics Research Centre, School of Physics and Astronomy, Queen Mary University of London, Mile End Road, London E1 4NS, United Kingdom.} 

\emailAdd{peter.ballett@durham.ac.uk}
\emailAdd{king@soton.ac.uk}
\emailAdd{silvia.pascoli@durham.ac.uk}
\emailAdd{n.prouse@soton.ac.uk}
\emailAdd{tse-chun.wang@durham.ac.uk}

\preprint{IPPP/16/90}

\abstract{
Long-baseline neutrino oscillation experiments, in particular Deep Underground
Neutrino Experiment (DUNE) and Tokai to Hyper-Kamiokande (T2HK), will lead the
effort in the precision determination of the as yet unknown parameters of the
leptonic mixing matrix. In this article, we revisit the potential of DUNE, T2HK
and their combination in light of the most recent experimental information. As
well as addressing more conventional questions, we pay particular attention to
the attainable precision on $\delta$, which is playing an increasingly
important role in the physics case of the long-baseline programme.
We analyse the complementarity of the two designs, identify the benefit of a
programme comprising distinct experiments and consider how best to optimise the
global oscillation programme.
This latter question is particularly pertinent in light of a number of
alternative design options which have recently been mooted: a Korean second
detector for T2HK and different beams options at DUNE.
We study the impact of these options and quantify the synergies between
alternative proposals, identifying the best means of furthering our knowledge
of the fundamental physics of neutrino oscillation. 
}

\begin{document} 

\maketitle

\section{Introduction}

Our knowledge of the neutrino sector has undergone a sea-change over the last
decade. The oscillation mechanism has been well established as the explanation
of the anomalous solar and atmospheric neutrino flavour ratios, and the
paradigm has been subjected to scrutiny from long-baseline accelerator and
reactor experiments resulting in a measurement of the final mixing angle
$\theta_{13}$ \cite{Abe:2011sj, An:2012eh, Ahn:2012nd, An:2013uza}.  Although
some short-baseline anomalies still remain unexplained
\cite{AguilarArevalo:2008rc, Aguilar-Arevalo:2013pmq, Aguilar:2001ty}, the
oscillation mechanism has leapt many hurdles to become a part of the new
Standard Model (SM). However, some significant unknowns remain: the ordering of
neutrino masses parameterized by the sign of $\Delta m^2_{31}$, the existence
and extent of CP violation (CPV) or maximal CP violation in leptonic mixing,
and the precise value, including crucially the octant, of $\theta_{23}$. In
addition, the current precision on the oscillation parameters is insufficient
to rule out many theoretical models, for example those discussed recently in
Refs.~\cite{Ballett:2016yod, Ma:2016nkf, Lu:2016jit, Berryman:2016mfg}. These
models can offer predictions for $\delta$ --- potentially explaining maximally
CP violating or CP conserving values --- as well as the octant, and the mass
ordering.

With the intention of building on the progress of the oscillation programme,
the international community has conceived a range of future facilities with
the potential to explore the final unknowns in the conventional oscillation
paradigm, and to hunt for tensions in the data which might indicate that a
richer extension of the SM is required.  There are three major strands in the
future experimental neutrino oscillation programme: short-baseline experiments
such as those comprising the SBN programme \cite{Antonello:2015lea},
intermediate baseline reactor facilities, RENO-50 and JUNO \cite{An:2015jdp,
Djurcic:2015vqa, Kim:2014rfa}, and long-baseline experiments (LBL) such as LBNF-DUNE
and T2HK \cite{Adamson:2016tbq, Adamson:2016xxw, Abe:2015awa, Abe:2014oxa,
Adams:2013qkq, Acciarri:2015uup}. In this article we focus on these latter two
proposals for novel long-baseline facilities: Long-Baseline Neutrino
Facility-Deep Underground Neutrino Experiment (LBNF-DUNE, referred to
subsequently as DUNE) and Tokai to Hyper-Kamiokande (T2HK). 
DUNE is the flag-ship long baseline experiment of the Fermilab neutrino
programme \cite{Adams:2013qkq,Acciarri:2015uup}. It consists of a new beam
sourced at Fermilab and a detector complex at Sanford Underground Research
Facility (SURF) in South Dakota separated by a distance of 1300 km.  Over this
distance, neutrinos produced in the decays of secondary particles from proton
collisions at Fermilab will propagate, undergoing oscillations and scattering
processes in the matter of the Earth. The appreciable matter effects will
modify the probability of detecting a given flavour of neutrino, in a way that
will ultimately make the facility highly sensitive to the mass ordering while the broad spectrum of
events arising from its on-axis flux also allows for significant sensitivity to
the unknown CPV phase $\delta$. 
The detector will use Liquid Argon Time Projection Chamber (LAr-TPC)
technology, allowing for strong event reconstruction. As a result, a high
signal to background ratio is expected.
T2HK \cite{Abe:2014oxa} in contrast was conceived with a smaller baseline of
295 km and a different detector technology. Building on the successes of
Kamiokande and Super-Kamiokande \cite{Fukuda:2002uc}, Hyper-Kamiokande will
employ Water \v{C}erenkov technology at a significantly larger scale, with
fiducial volumes on the order of hundreds of kilotonnes. Matter effects for
this facility will be smaller due to the shorter baseline (although
non-negligible), and the significantly enhanced event rate will allow for a
high-statistics comparison between neutrino and anti-neutrino modes, searching
for fundamental asymmetries due to the CP violating phase $\delta$. 

Much work has been done over the years assessing the physics reach of T2HK
\cite{Abe:2014oxa, Soumya:2014ika, Coloma:2012ji} and DUNE \cite{Ghosh:2014rna,
Acciarri:2015uup, DeRomeri:2016qwo, Nath:2015kjg, Barger:2014dfa} (along with
its predecessor designs LBNE \cite{Adams:2013qkq, Barger:2013rha, Bora:2014zwa,
Bishai:2012ss, Coloma:2012ji} and LBNO \cite{Agarwalla:2013kaa, Ghosh:2013pfa,
Coloma:2012ji}). In this article, we revisit the physics sensitivity of DUNE
and T2HK for key measurements relating to the mass ordering, $\delta$ phase and
the mixing angle $\theta_{23}$, focusing in particular on the combined reach of
these designs.  Recently, as the designs for T2HK and DUNE have matured, both
collaborations have considered significant alterations to the benchmark
proposals in \refrefs{Abe:2014oxa}{Ghosh:2014rna,Acciarri:2015uup}.  The nuPIL
(neutrinos from a PIon beam Line) design \cite{Lagrange:2015jyv, Liu:2016kbe,
Lagrange:2016mmg}, developed at Fermilab, is a novel beam technology building
on accelerator R\&D work done for the neutrino factory \cite{Choubey:2011zzq}. 
Although nuPIL is no longer in consideration by the DUNE collaboration, its
unique design leads to phenomenology which may be of interest to future work.
nuPIL foresees the collection and sign selection of pions from a conventional
beam, which are directed though a beam line and decay to produce neutrinos.
This selection and manipulation of the secondary beam forces unwanted parent
particles out of the beam resulting in a particularly clean flux. 
This screening process presents a particular advantage over conventional
neutrino beams, where the contamination of the flux due to mesons of the wrong
sign can limit the sensitivity of the antineutrino channel. In the latter case,
the contamination from intrinsic $\nu_\mu$ is effectively enhanced by the
cross-section differences. This increases the relative number of wrong-sign
events, and reduces the signal over background ratio.
The simulated flux is also notably narrower than the DUNE reference design
(although this could be changed through modification of the design) which will
alter the sensitivity to the oscillation probability.
In a parallel development, T2HK has reconsidered the location of its second
detector module.
The current design divides the detector into two modules installed at Kamioka
following a staged implementation \cite{HKDR}: an initial data-taking period
would use a single tank during which the second tank would be constructed and
would start taking data after $6$ years to further boost the statistical power
of the experiment. 
Instead of this plan, the suggestion has been made to locate the second tank in
South Korea at a baseline distance of between $1000$ -- $1300$ km from J-PARC
\cite{T2HKK, Hagiwara:2005pe, Hagiwara:2006vn, Kajita:2006bt,
Ishitsuka:2005qi}.  This would allow T2HK + Korea (T2HKK) to collect data from
two different baselines and with two different off-axis angles (and
consequently energy spectra), crucially altering the phenomenology of the
experiment.
 
Although the question of the combined sensitivity of DUNE and T2HK has been
studied before (most recently in \cite{Fukasawa:2016yue}), our work brings
three new elements to the discussion. Firstly, our work incorporates the
significant redesign and development work that has been performed in the last
few years on both designs.  Our simulation of T2HK is particularly noteworthy,
departing significantly from those used in previous comparable analyses
\cite{Fukasawa:2016yue} by incorporating up-to-date information about detector
performance from the collaboration's in-house simulation, and has been
carefully calibrated against previously published results. Secondly, we
thoroughly address the precision measurement of $\delta$ and its phenomenology,
often deemed a secondary question in earlier studies, but one which is
increasingly central to the aims of the long-baseline programme, and which has
significant theoretical implications. Finally, we provide a detailed discussion
of the differences between the two designs as well as their potential redesigns
(nuPIL, T2HKK) and a quantification of their complementarity in an attempt to
identify the optimal choice from a global perspective.

We start our discussion with a brief recap of the relevant phenomenology of
oscillation physics in \refsec{sec:pheno}. In \refsec{sec:experiments}, we
describe the details of DUNE and T2HK (including their alternative designs)
taken into account in our simulations. \refsec{sec:results} is devoted to the
results of our simulations assuming the standard configurations of each
experiment which look at mass ordering sensitivity, CP violation discovery, the
ability to exclude maximally CP violating values of $\delta$, the
expected precision on $\theta_{23}$ and the ability to resolve the octant.  We
present an analysis of the complementarity for precision on $\delta$ in
\refsec{sec:delta}, taking care to discuss the interplay of factors which
influence this measurement. In \refsec{sec:alternative_designs}, we reconsider
these physics goals in light of the alternative deigns for DUNE and T2HK. We
end our study with some concluding remarks in \refsec{sec:conclusions}. 

\section{\label{sec:pheno}Oscillation phenomenology at DUNE and T2HK}

The fundamental parameters which describe the oscillation phenomenon are the
angles and Dirac phase of the Pontecorvo-Maki-Nakagawa-Sakata (PMNS) mixing
matrix as well as two independent mass-squared splittings, \eg\ $\Delta
m^2_{21}$ and $\Delta m^2_{31}$. The PMNS matrix is the mapping between the
bases of mass and flavour states (denoted with Latin and Greek indices,
respectively), which can be written as 
\[ \nu_\alpha = U_{\alpha i}^*\nu_i,   \] 
where $U$ will be expressed by the conventional factorization
\cite{Beringer:1900zz}:
\begin{align*} U_{\mathrm{PMNS}} &= U_{23}U_{13}U_{12}P, \\ &=
\left(\begin{matrix} 1 & 0 & 0\\ 0 & c_{23} & s_{23} \\ 0 & -s_{23} & c_{23}
\end{matrix}\right)  \left(\begin{matrix} c_{13} & 0 &
s_{13}e^{-\mathrm{i}\delta}\\ 0 & 1 & 0 \\ -s_{13}e^{\mathrm{i}\delta} & 0 &
c_{13} \end{matrix}\right) \left(\begin{matrix} c_{12} & s_{12} & 0\\ -s_{12} &
c_{12} & 0 \\ 0 & 0 & 1
\end{matrix}\right)\left(\begin{matrix}e^{\mathrm{i}\alpha_1} & 0 & 0\\ 0 &
e^{\mathrm{i}\alpha_2} & 0 \\ 0 & 0 & 1  \end{matrix}\right),\\ &=
\left(\begin{matrix}c_{12}c_{13} & s_{12}c_{13} &  s_{13}e^{-\mathrm{i}\delta}
\\ -s_{12}c_{23} - c_{12}s_{23}s_{13}e^{\mathrm{i}\delta} & c_{12}c_{23} -
s_{12}s_{23}s_{13}e^{\mathrm{i}\delta} & s_{23}c_{13} \\ s_{12}s_{23} -
c_{12}c_{23}s_{13}e^{\mathrm{i}\delta} &  -c_{12}s_{23} -
s_{12}c_{23}s_{13}e^{\mathrm{i}\delta} & c_{23}c_{13} \end{matrix}\right)P,
\end{align*}
where $P$ is a diagonal matrix containing two Majorana phases $\alpha_1$ and
$\alpha_2$ which play no role in oscillation physics. The mixing angles
$\theta_{12}$, $\theta_{13}$ and $\theta_{23}$ are often referred to as the
solar, reactor and atmospheric mixing angles respectively; all of these angles
are now known to be non-zero \cite{Gonzalez-Garcia:2014bfa}.
The remaining parameter in $U$ is the phase $\delta$, which is currently poorly
constrained by data. This parameter dictates the size of CP violating effects
in vacuum during oscillation. All such effects will be proportional to the
Jarlskog invariant of $U_{\text{PMNS}}$, 
\[  J = \frac{1}{8}\sin\delta \sin\left(2\theta_{23}\right)
\sin\left(2\theta_{13}\right) \sin\left(2\theta_{12}\right) \cos\theta_{13}. \]
For the theory to manifest CP violating effects, $J$ must be non-zero. Given
our knowledge of the mixing angles, the exclusion of $\delta \notin\{0,\pi\}$
would be sufficient to establish fundamental leptonic CP violation.

Long-baseline experiments such as DUNE and T2HK aim to improve our knowledge of
$U$, as well as the atmospheric mass-squared splitting, by the precision
measurement of both the appearance $\nu_\mu\to\nu_e$ and disappearance
oscillation channels $\nu_\mu \to \nu_\mu$, as well as their CP conjugates.
In the following section, we will discuss the key aims of the long-baseline
program and the important design features of these experiments which lead to
their sensitivities. To facilitate this discussion, we introduce an
approximation of the appearance channel probability following
\refref{Asano:2011nj}, which is derived by performing a perturbative expansion
in the small parameter $\epsilon \equiv \Delta m^2_{21}/\Delta m^2_{31} \approx
0.03 $ under the assumption that $\sin^2\theta_{13} =
\mathcal{O}(\epsilon)$\footnote{For alternative schemes of approximation, see
\refref{Agarwalla:2013tza, Johnson:2015psa, Minakata:2015gra,
Denton:2016wmg}.}.  The expression for the oscillation probability is
decomposed into terms of increasing power of $\epsilon$, 
\begin{equation} 
P(\nu_\mu \to \nu_e; E, L) \equiv P_1 + P_\frac{3}{2} +
\mathcal{O}\left(\epsilon^2\right),
\label{eq:Asano_prob1} \end{equation}
where $E$ is the neutrino energy, $L$ the oscillation baseline, and the ordered
terms $P_n = \mathcal{O}(\epsilon^n)$ are given by
\begin{align}
P_1 &= \frac{4}{(1-r_A)^2} \sin^2\theta_{23}\sin^2\theta_{13}
\sin^2\left(\frac{(1-r_A)\Delta L}{2}\right), \label{eq:Asano_prob2a} \\
P_\frac{3}{2} &= 8 J_r \frac{\epsilon}{r_A(1-r_A)} \cos\left(\delta +
\frac{\Delta L}{2}\right)\sin\left(\frac{r_A \Delta
L}{2}\right)\sin\left(\frac{(1-r_A) \Delta L}{2}\right),\label{eq:Asano_prob2b}
\end{align}
where $J_r =
\cos\theta_{12}\sin\theta_{12}\cos\theta_{23}\sin\theta_{23}\sin\theta_{13}$,
$r_A = 2\sqrt{2}G_\text{F}N_eE/\Delta m^2_{31}$, with $N_e$ denoting the
electron density in the medium, and $\Delta = \Delta m^2_{31}/2E$.
Using the same scheme, the disappearance channel can be written at leading
order as
\begin{equation} P(\nu_\mu\to\nu_\mu; E, L) = 1 -
\sin^2(2\theta_{23})\sin^2\left(\frac{\Delta L}{2}\right) +
\mathcal{O}(\epsilon). \label{eq:Asano_dis}\end{equation}
For both channels, equivalent expressions for antineutrino probabilities can be
obtained by the mapping $r_A \to -r_A$ and $\delta \to -\delta$. 

\subsection{Mass ordering, CPV and the octant of $\theta_{23}$}

The sensitivity of long-baseline experiments to the questions of the neutrino
mass ordering, the existence of CPV and the octant of $\theta_{23}$, are by now
well studied topics (for a recent review see \eg\ \refref{Agarwalla:2014fva}).
To help us clarify the role of the designs of DUNE and T2HK, as well as their
possible modifications, we will briefly recap how experiments on these scales
derive their sensitivities using the approximate formulae expressed by
\refeqss{eq:Asano_prob2a}{eq:Asano_prob2b}{eq:Asano_dis}. 

The dependence on the sign of $\Delta m^2_{31}$, and therefore the mass
ordering, arises at long-baselines from the interplay with matter, where
forward elastic scattering can significantly enhance or suppress the
oscillation probability. This is governed by the parameter $r_A$ in
\refeq{eq:Asano_prob1} and goes to zero in the absence of matter. Changing from
Normal Ordering (NO, $\Delta m^2_{31}>0$) to Inverted Ordering (IO, $\Delta
m^2_{31}<0$) requires the replacements $\Delta \to -\Delta$ and $r_A \to -
r_A$.  However, in vacuum ($r_A = 0$) the leading-order term in
\refeq{eq:Asano_prob1} remains invariant under this mapping. This invariance is
broken once a matter term is included ($r_A\neq0$), and the oscillation
probability acquires a measurable enhancement or suppression dependent on the
sign of $\Delta m^2_{31}$. The size of this enhancement increases with baseline
length, and this effect is expected to be very relevant for appearance channels
at a long-baseline experiment $\nu_\mu\rightarrow\nu_e$ and
$\bar{\nu}_\mu\rightarrow\bar{\nu}_e$.  However, the determination of the mass
ordering is further facilitated by the contrasting behaviour of neutrinos and
antineutrinos. Due to the dependence on $r_A$, for NO larger values of the
matter density cause an enhancement and a shift in the probability for $\nu_\mu\to\nu_e$
oscillation at the first maximum, whilst suppressing the probability for
$\overline{\nu}_\mu\to\overline{\nu}_e$. This behaviour is reversed for IO,
with neutrinos seeing a suppression and antineutrinos, an enhancement.
Moreover, matter effects also affect the energies of the first oscillation
maxima for neutrinos and antineutrinos. Through precise measurements around the
first maxima, these shifts can be observed allowing long-baseline oscillation
experiments to determine the mass ordering.

To detect CPV in neutrino oscillation an experiment requires sensitivity to
$\delta$. Unfortunately, the leading order appearance probability $P_1$ is
independent of the CP phase $\delta$ in vacuum, as seen in
\refeq{eq:Asano_prob2a}.
CP asymmetries between neutrino and antineutrino channels first appear with the
subdominant term $P_\frac{3}{2}$. In the presence of a background medium, CP
violating effects are instead introduced in $P_1$ due to $r_A$ which differs by
a sign for neutrinos and antineutrinos; however, these offer no sensitivity to
the fundamental CP violating parameter $\delta$. As the sensitivity to $\delta$
is subdominant and masked by CP asymmetry arising from matter effects,
extracting the CP phase is a more challenging measurement, requiring greater
experimental sensitivity. 
Long baseline (LBL) experiments can obtain sensitivity to $\delta$ by looking
not only at the first maximum but also at the spectral differences between CP
conjugate channels. In particular, an important role is played by low-energy
events in the sensitive determination of $\delta$ \cite{Bishai:2012ss,
Huber:2010dx, Geer:2007kn, Coloma:2011pg}: around the second maximum, CP
dependent terms of the oscillation probability are more significant. Although
accessing these events can be a challenging experimental problem, and low
statistics or large backgrounds could limit their potential
\cite{Huber:2010dx}, their benefit is clear from recent experimental work
\cite{Agarwalla:2014tca}. 

The atmospheric mixing angle is known to be large and close to maximal
$\theta_{23}\approx \pi/4$, but it is not currently established if it lies in
the first octant $\theta_{23}<\pi/4$ or the second octant $\theta_{23}>\pi/4$.
We see in \refeq{eq:Asano_prob2a} that the
appearance channel is sensitive to the octant. However, we also see that
changing the octant enhances or suppresses the first maximum of the appearance
channel in much the same way as the matter enhancement. For this reason, the
sensitivity to these questions can be expected to be correlated; however, this
correlation will be reduced when data from both neutrino and antineutrino is
available as this effect is the same in both CP conjugate channels. The
determination of $\theta_{23}$ is also known to be beset by issues of
degeneracy with $\delta$ which can complicate its determination
\cite{Minakata:2013hgk, Coloma:2014kca, Agarwalla:2014fva}.  As
both of these parameters enter the second-order terms in
\refeq{eq:Asano_prob2a}, the freedom to vary $\delta$ can be used
to mask the effects of a wrong octant, making their joint determination more
challenging. Fortunately, a precise measurement of $\sin(2\theta_{23})$ is
possible through the disappearance channel, helping to break this degeneracy.
Also, spectral information is expected to mitigate this problem.

\subsection{\label{sec:precision}Precision on $\delta$}

Although the question of the \emph{existence} of leptonic CP violation often
dominates discussions about $\delta$, the precision measurement of $\delta$
could prove to be the most valuable contribution of the long-baseline
programme. To determine the existence of fundamental leptonic CP violation it
suffices to exclude the CP conserving values $\delta = 0$ and $\delta =\pi$,
those values corresponding to a vanishing Jarlskog invariant.
Therefore the discovery potential of a facility to CP violation is
fundamentally linked to the precision attainable for measurements of $\delta$
in the neighbourhood of $0$ and $\pi$.
However, the question of precision on $\delta$ goes beyond CP violation
discovery. Many models of flavour symmetries, for example, are consistent with
the known oscillation data and make predictions for $\delta$.\footnote{For
example, recent studies of mixing sum rules can be seen as predicting $\delta$
for long-baseline experiments \cite{Antusch:2007rk, Ballett:2013wya,
Girardi:2014faa, Ballett:2014dua, Girardi:2015vha}. For a review of the
predictions from such models, see \eg\ \refrefs{King:2015aea}{King:2017guk}.}
No experiment on comparable time-scales is expected to be able to compete with
precision measurements of $\delta$ from DUNE and T2HK. 

It can be shown that the precision expected on $\delta$ worsens significantly
around $\delta=\pm\frac{\pi}{2}$, and that this is because of the probability
itself \cite{Coloma:2012wq}. 
Looking at the CP sensitive term in \refeq{eq:Asano_prob2b} at energies around
the first maximum, where $\Delta L/2 \approx \pi/2$, we can approximate the
probability by  
\[  P_{\frac{3}{2}} \approx -8 J_r \frac{\epsilon}{r_A(1-r_A)} \sin\delta
\sin\left(\frac{r_A \Delta L}{2}\right)\sin\left(\frac{(1-r_A) \Delta
L}{2}\right). \]
The highest sensitivity to $\delta$ is found when this function is most
sensitive to changes in $\delta$, information naturally encoded in the
function's first derivative. Due to the sinusoidal nature of the function, when
the CP term has its largest effect ($|\sin\delta|=1$), it is at a maximum and
consequently its gradient is at a minimum.  Therefore, we expect the errors on
$\delta$ to be small around $0$ and $\pi$, when even though the absolute size
of the CP sensitive terms are small, they are most sensitive to parameter
shifts. Taking matter into account moves the location of the worst sensitivity
away from $\delta=\pm\frac{\pi}{2}$. Assuming we are close to the first
maximum, and introducing a dimensionless parameter $\xi$ to describe the
deviation from this point (where $\xi=0$ corresponds to the first maximum), the
relevant parameter governing the phase of the sinusoidal terms can be expressed
by
\begin{equation} \Delta L = \pi\frac{1+\xi}{1-r_A},
\label{eq:maximum}\end{equation}
we can find the value of $\delta$ for which we expect the worst sensitivity by
minimising the gradient of \refeq{eq:Asano_prob2b}, which occurs for the values
\begin{equation}   \delta \approx -\frac{\pi}{2} \frac{1+\xi}{1-r_A} + \pi n,
\label{eq:worstdelta}\end{equation}
for $n\in \mathbb{Z}$. From this formula it is clear that the value of $\delta$
with the worst sensitivity shifts away from $\frac{(2n+1)\pi}{2}$ in a
direction governed by the signs of $r_A$ and $\xi$.  Specifically, the
dependence on $r_A$ means that the neutrino and anti-neutrino mode
sensitivities at fixed energy have their worst sensitivity for different true
values of $\delta$. Running both CP conjugate channels in a single experiment
allows each channel to compensate for the poorer performance of the other at
certain values of $\delta$, helping to smooth out the expected precision
\cite{Coloma:2012wq}. In this way, the multichannel nature of LBL experiments
allows for a greater physics reach than a single channel experiment.
 
The argument above assumed that all events came from a fixed energy defined
implicitly by $\xi$ in \refeq{eq:maximum}. Due to the dependence on $\xi$ in
\refeq{eq:worstdelta}, having information from different energies will also be
complementary, acting analogously to the combination of neutrino and
antineutrino data by mitigating the poorest performance. 
Although all LBL experiments aim to include the first maximum, where event
rates are highest, none have a purely monochromatic beam and so-called
wide-band beams include considerable information from other energies. Therefore
such experiments can be expected to avoid the significant loss of sensitivity
predicted by the simple analytic formula. We can infer, however, that a narrow
beam focused on the first maximum in the presence of small matter effects
should have a worse sensitivity at maximal values of $\delta$ compared to CP
conserving values \cite{Coloma:2012wq}.

With reference to the traditional designs of T2HK and DUNE, from the above
discussion we can infer that T2HK can be expected to have a greater range of
expected precisions as we vary $\delta$ than DUNE. In particular, due to its
narrower beam and small matter effects, we expect markedly poorer performance
for T2HK at $\delta\in\{-\frac{\pi}{2},\frac{\pi}{2}\}$. DUNE on the other hand
will be less variable as its broad band mitigates the total loss of sensitivity
at certain energies, and its large matter effect helps to stabilise
performance, but it can be expected to see its worst sensitivity at values of
$\delta$ slightly displaced from $0$ and $\pi$, where the sensitivity at the
first maximum is worst. This suggests a degree of complementarity of the
wide-band and narrow-band beams when it comes to precision measurements of
$\delta$: a narrow-band focused on the first maximum is optimal for precision
around $0$ and $\pi$ (and by implication, for CPV discovery) while a wide-band
beam should perform better for precision measurements around $\delta=\pm
\frac{\pi}{2}$. This general behaviour will be relevant not only for the
traditional designs of DUNE and T2HK, but also their possible redesigns: nuPIL
could lead to a narrowing of the neutrino flux, and T2HKK could see a
wider-band component in its flux, or a narrow-band component focused away from
the first maximum. The interplay of these factors will be explored in more
detail in \refsec{sec:delta}. 

\section{\label{sec:experiments}Simulation details}

To better understand the sensitivities and complementarity of DUNE and T2HK
(including their potential redesigns), we have performed a simulation of the
experiments in isolation and in combination. We are using the General Long
Baseline Experiment Simulator (GLoBES) libraries \cite{Huber:2004ka,
Huber:2007ji} and in the following sections, we will describe the features of
our modelling of the two facilities and the statistical treatment. 

\subsection{\label{sec:details_DUNE}DUNE}

\begin{figure}[t]

\centering

\includegraphics[width=0.49\textwidth, clip, trim = 0 5 30 15]{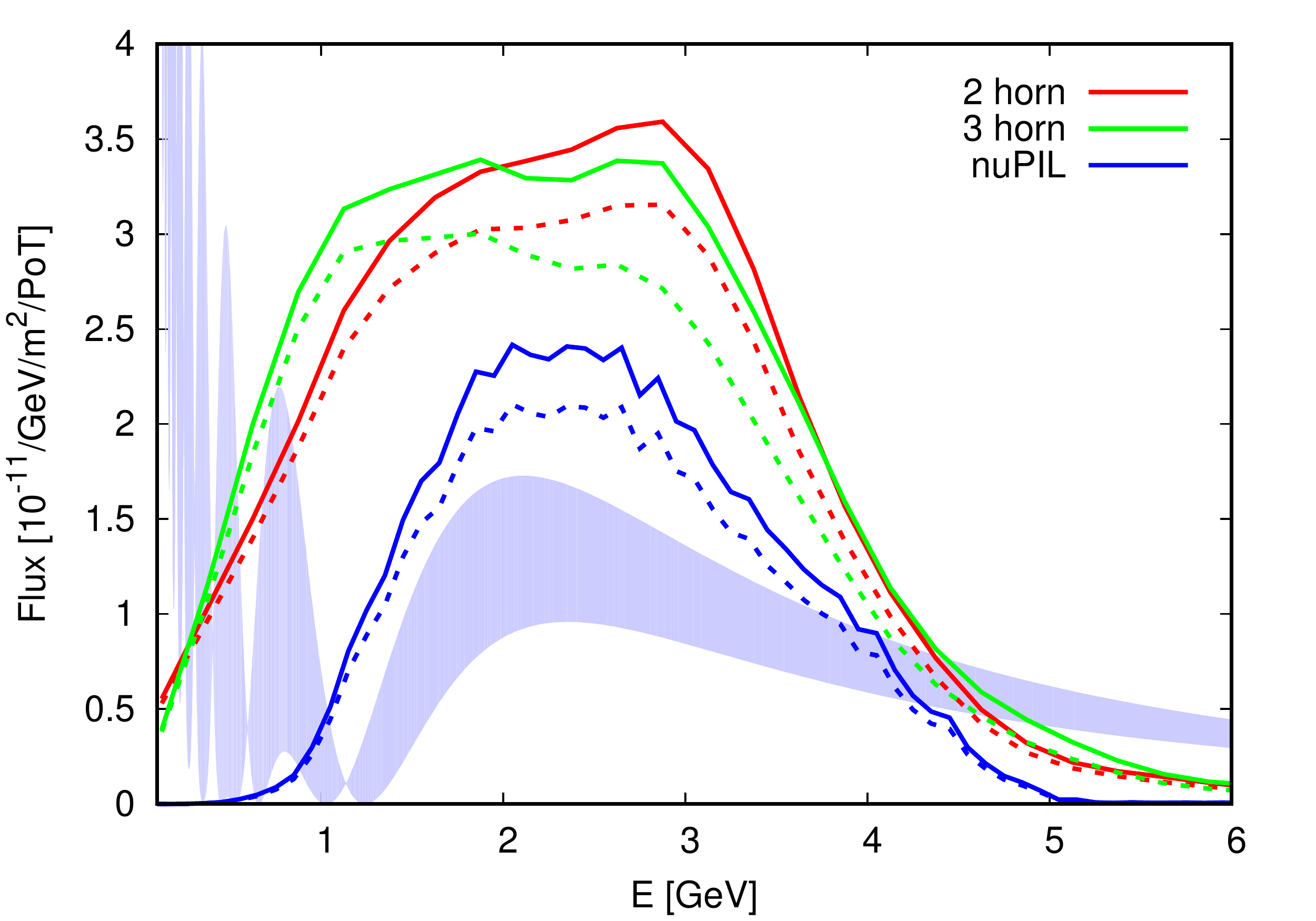}
\includegraphics[width=0.49\textwidth, clip, trim = 0 5 20 15]{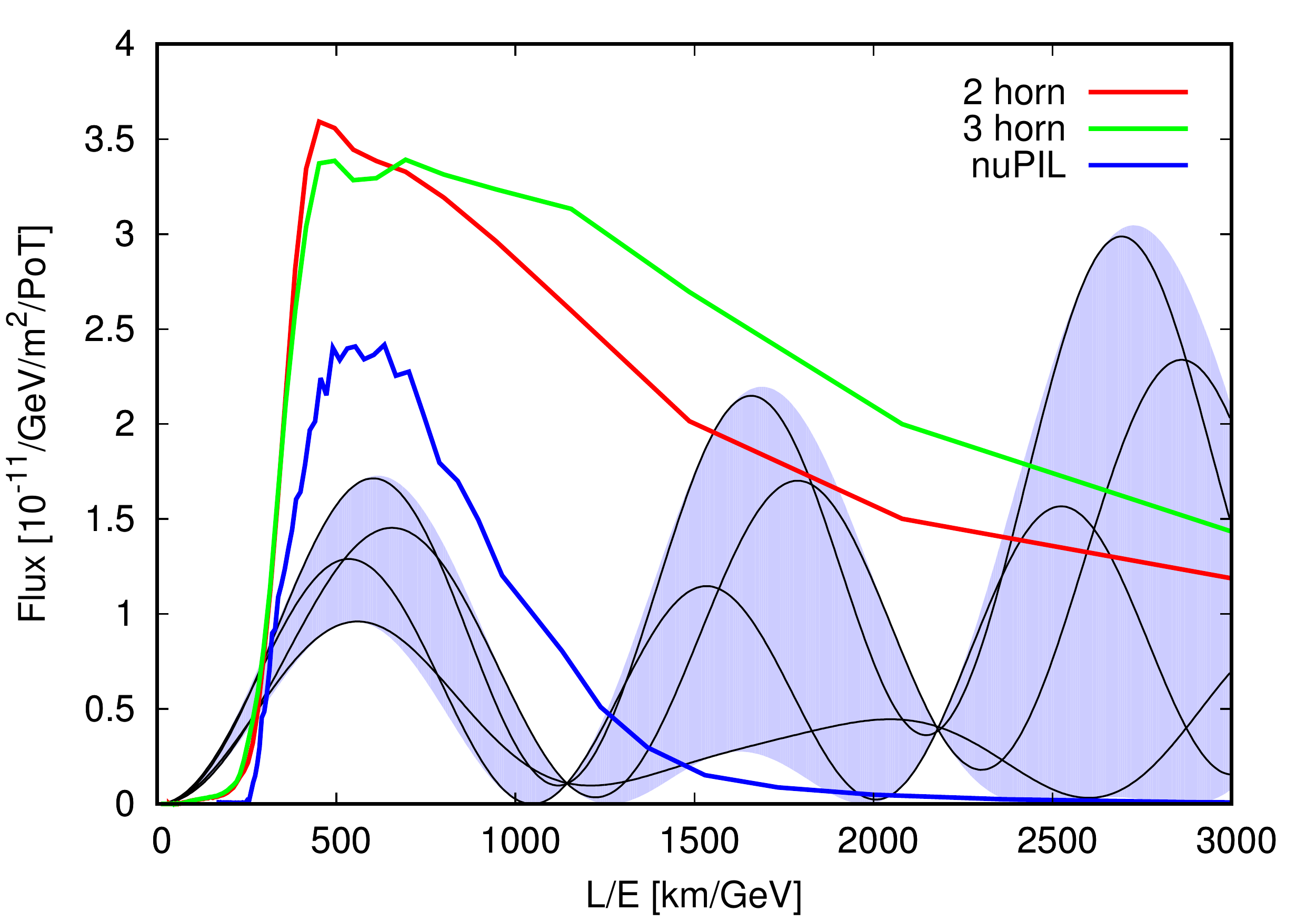}

\caption{\label{fig:dune_fluxes}Left: $\nu_\mu$ ($\overline{\nu}_\mu$) flux
component in $\nu$-mode ($\bar{\nu}$-mode) shown as solid (dashed) lines for
2-horn optimised, 3-horn optimised, and nuPIL beam designs. Right: the fluxes
for $\nu$-mode shown as a function of $L/E$. In both panels, the shaded region
shows the envelope of the oscillation probability as $\delta$ is varied over
its full range. The black lines in the right panel show the probability for
$\delta\in \{0,\frac{\pi}{2}, \pi, \frac{3\pi}{2}\}$.}

\end{figure}

The DUNE experiment consists of a new neutrino source, known as Long Baseline
Neutrino Facility (LBNF), a near detector based at Fermilab and a LArTPC
detector complex located in SURF a distance of 1300~km away.  Several variants
of the LBNF beam have been developed.  In this work, we study three neutrino
fluxes: a 2-horn optimised beam design \cite{Acciarri:2015uup, 
Alion:2016uaj}, a 3-horn optimised beam design \cite{Papadimitriou:2016apo,
3hornflux}, and the neutrinos from a PIon beam Line (nuPIL)
\cite{Lagrange:2015jyv, Liu:2016kbe, Lagrange:2016mmg, nuPILflux}. We show all
three fluxes used in our simulations in \reffig{fig:dune_fluxes}.
 
The 2-horn optimised beam has been designed to maximise the
sensitivity to CP violation \cite{Acciarri:2015uup}.
In our simulation, we take the proton energy to be $80$ GeV, and follow a
staged implementation of the beam power in line with the DUNE proposal,
which assumes the beam power will double after 6 years \cite{Acciarri:2016crz}. 
Our simulation assumes a power of $1.07$ MW and $1.47\times 10^{21}$ protons on
target (POT) per year for the first 6 years, and $2.14$ MW ($2.94 \times
10^{21}$ POT per year) afterwards.  
Thanks to constant development work by the DUNE collaboration, an additional
optimised beam has also been designed. This 3-horn design has a stronger focus
on producing lower energy events, leading to an increase in flux
between $0.5$ GeV and $1.5$ GeV. This leads to a greater number of expected
events from around the second oscillation maximum, which is well-known to be
particularly sensitive to the phase $\delta$. 
For this design, the proton energy is assumed to be $62.5$ GeV and the POT per
year is taken as $1.83\times10^{21}$, before doubling at the 6th year in line
with the expected beam upgrade.
We also consider the nuPIL design.
Although this design is no longer considered to be an option for the LBNF beam,
its novelty leads to interesting phenomenological consequences and we study it
alongside the main beam design.
nuPIL foresees the collection and sign selection of pions from proton
collisions with a target which are then directed though a beam line and
ultimately decay to produce neutrinos. 
This selection and manipulation of the secondary beam forces unwanted parent
particles out of the beam resulting in lower intrinsic contamination of the
neutrino (antineutrino) flux by antineutrinos (neutrinos).  In particular, this
would improve the signal to background ratio of the antineutrino mode compared
to a conventional neutrino beam. The proton energy for this design is assumed
to be $80$ GeV, and the corresponding POT per year is $1.47\times10^{21}$ which
again doubles after 6 years. Compared to the other two designs, nuPIL offers a
lower intrinsic contamination from other flavours and CP states while
maintaining low systematic uncertainties. We note that nuPIL also expects a
smaller total flux, although this might be avoidable through further design
effort. 
Another characteristic of the nuPIL design is its notably narrower flux. As
events from the second oscillation maximum are expected to be highly
informative about the true value of $\delta$, this may impact the sensitivity
to $\delta$. The coverage of first and second maxima is seen clearly in the
right-hand panel of \reffig{fig:dune_fluxes}, where the fluxes are shown as a
function of $L/E$.  The first maximum ($L/E \approx 600$ km/GeV) is covered
comparably well for all three flux designs, while the flux at the second
maximum ($L/E\approx 1800$ km/GeV) varies significantly. The 2-horn design is
seen to be similar to the 3-horn design: the two designs are very similar
around the first maximum, but the 2-horn design sees slightly fewer events at
higher values of $L/E$.

Although we consider alternative fluxes, we always assume the same detector
configuration of four 10-kiloton LArTPC detectors at 1300 km from the neutrino
source. We neglect the possibility of staging, assuming that all four tanks are
operational at the same time, and do not account for the expected improvement
in performance throughout the lifetime of the detectors.  
LArTPC technology has a particularly strong particle identification capability
as well as good energy resolution which are both crucial in providing high
efficiency searches and low backgrounds. 
We model the LArTPC detector response with migration matrices incorporating the
results of parameterized Monte Carlo simulations undertaken by the
collaboration \cite{Alion:2016uaj}. We use fourteen migration matrices ---
seven each for the disappearance and appearance channels --- describing the
detection and reconstruction of all three flavours of neutrino, and
antineutrino, as well as generic flavour blind NC events. 

We include both appearance and disappearance searches in our study. The
appearance channel signal is taken as the combination of $\nu_e$ and
$\bar{\nu}_e$ charged-current (CC) events. For the disappearance channel, we
study $\nu_\mu$ and $\bar{\nu}_\mu$ for neutrino and antineutrino modes,
respectively. The backgrounds to the appearance channel are taken 
to be neutral-current (NC) events, mis-identified $\nu_\mu+ \bar{\nu}_\mu$  CC 
interactions, intrinsic $\nu_e+ \bar{\nu}_e$ CC events, and $\nu_\tau+ \bar{\nu}_\tau$ 
CC  events. On the other hand, in $\nu_\mu$ and $\bar{\nu}_\mu$ disappearance 
we consider NC events, $\nu_\mu+ \bar{\nu}_\mu$ CC events, and 
$\nu_\tau+\bar{\nu}_\tau$ CC events. These assumptions follow the collaboration's 
own analysis \cite{Acciarri:2015uup}.
The rates of these backgrounds are governed by the migration matrices.

We assume the same systematic errors for all beam designs. The reduction of the
systematic errors is an ongoing task in the collaboration, and our values are
based on the conservative end of the current estimates of $1$--$2\%$
\cite{Acciarri:2015uup,Alion:2016uaj}. As such, we take an overall normalization 
error on the signal ($2\%$ for appearance and $5\%$ for disappearance) and on the 
background rates ($5\%$ for $\nu_e$, $\bar{\nu}_e$, $\nu_\mu$, and 
$\bar{\nu}_\mu$ CC events, $10\%$ for NC interactions, and $20\%$ for $\nu_\tau$ 
and $\bar{\nu}_\tau$ CC events). 
This accounts for fully correlated uncertainties on the event rates in each
bin, and we do not consider uncorrelated uncertainties. We note the nuPIL
design could lower the systematic error with respect to the conventional
design, although the extent of this is unknown, and beating $1\%$ systematics
will be challenging.
 
\subsection{\label{sec:details_T2HK}T2HK}

\begin{figure}[t]

\centering

\includegraphics[width=0.49\textwidth, clip, trim = 0 5 30 15]{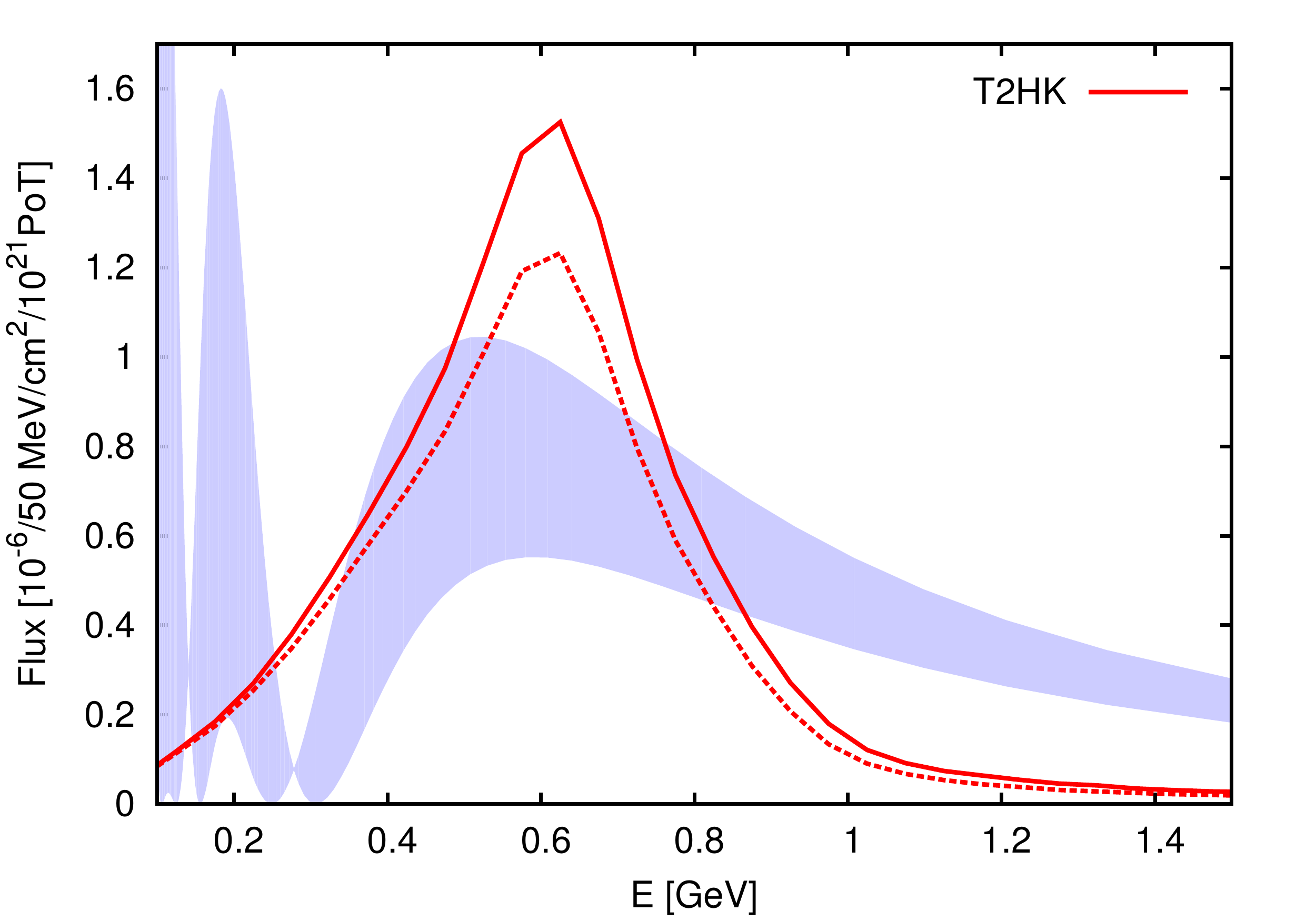}
\includegraphics[width=0.49\textwidth, clip, trim = 0 5 20 15]{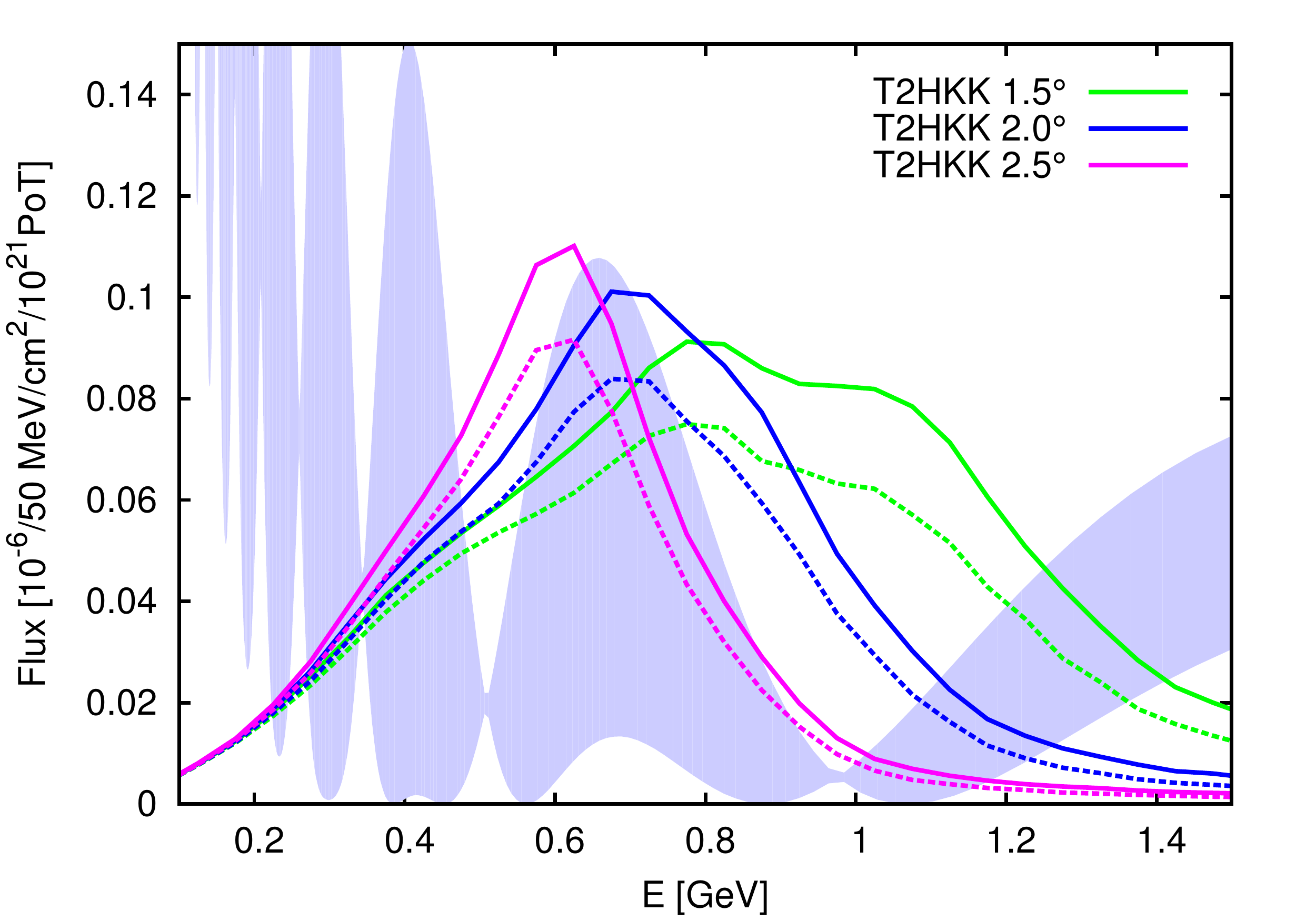}

\caption{\label{fig:t2hk_fluxes}Left: T2HK's flux plotted against neutrino
energy for $\nu$-mode (solid) and $\overline{\nu}$-mode (dashed). Right: the
T2HKK fluxes plotted against energy for $\nu$ and $\overline{\nu}$ modes. The
shaded region shows the envelope of the probability found by varying the true
value of $\delta$. Due to T2HKK's longer baseline but comparable energy range
to T2HK, the fluxes on the right sample a very different part of the
probability.}

\end{figure}

The Tokai to Hyper-Kamiokande (T2HK) experiment~\cite{HKDR} is the
proposed next-generation long-baseline experiment using
a neutrino beam produced at the synchrotron at J-PARC in Tokai directed
\ang{2.5} off-axis to Hyper-Kamiokande (Hyper-K), a new water \v{C}erenkov
detector to be built near Kamioka, \SI{295}{\km} from the beam source. The
narrow-band beam comprises mostly of $\nu_\mu$ (or $\anu_\mu$), with the
energy peaked near \SI{600}{\MeV} corresponding to the first oscillation maximum
at \SI{295}{\km}. Hyper-K is capable of detecting interactions of $\nu_\mu$,
$\anu_\mu$, $\nu_e$ and $\anu_e$, allowing measurements of the oscillation
probabilities $P(\nu_\mu\to\nu_e)$, $P(\nu_\mu\to\nu_\mu)$,
$P(\anu_\mu\to\anu_e)$, $P(\anu_\mu\to\anu_\mu)$ with the primary goal of
searching for CP violation and measuring $\delta_{CP}$.

The J-PARC neutrino beam will be upgraded from that used for the T2K experiment
to provide a beam power of \SI{1.3}{\MW}~\cite{j-parc1,j-parc2}. The beam is
produced from \SI{30}{\GeV} protons colliding with a graphite target. Charged
pions produced in these collisions are focused through magnetic horns into a
decay volume, where the majority of the neutrinos in the beam are the $\nu_\mu$
($\anu_\mu$) produced from the $\pi^+$ ($\pi^-$) decay. The polarity of the
\SI{320}{\kA} horn current can be reversed to focus pions of positive or
negative charge in order to produce a beam of neutrinos or antineutrinos
respectively. A small contamination (less than 1\% of the neutrino flux) of
$\nu_e$ or $\anu_e$ in the beam and $\anu_\mu$ ($\nu_\mu$) in the $\nu_\mu$
($\anu_\mu$) beam result from the decay of the $\mu^+$ ($\mu^-$) produced in
the pion decay, however the majority of the $\mu^\pm$ are stopped after
reaching the end of the decay volume before decaying.

The baseline design for the Hyper-Kamiokande detector consists of two water
tanks each with a total (fiducial) mass of \SI{258}{\kilo\tonne}
(\SI{187}{\kilo\tonne})~\cite{newHK}. Each tank is surrounded by approximately
40,000 inward facing \SI{50}{\cm} diameter photosensors corresponding to a 40\%
photocoverage, equivalent to that currently used at Super-Kamiokande. The tanks
would be built and commissioned in a staged process with the second tank
starting to take data six years after the first. The detectors use the water
\v{C}erenkov ring-imaging technique as used at Super-Kamiokande, capable of
detecting the charged leptons produced in neutrino interactions on nuclei in
water. At these energies, most neutrino--nucleus interactions are 
quasi-elastic, and the measurement of the outgoing charged lepton allows for an 
accurate reconstruction of the energy and flavour of the initial neutrino.

\begin{figure}[t]

\centering

\includegraphics[width=0.65\textwidth, clip, trim = 0 5 30 15]{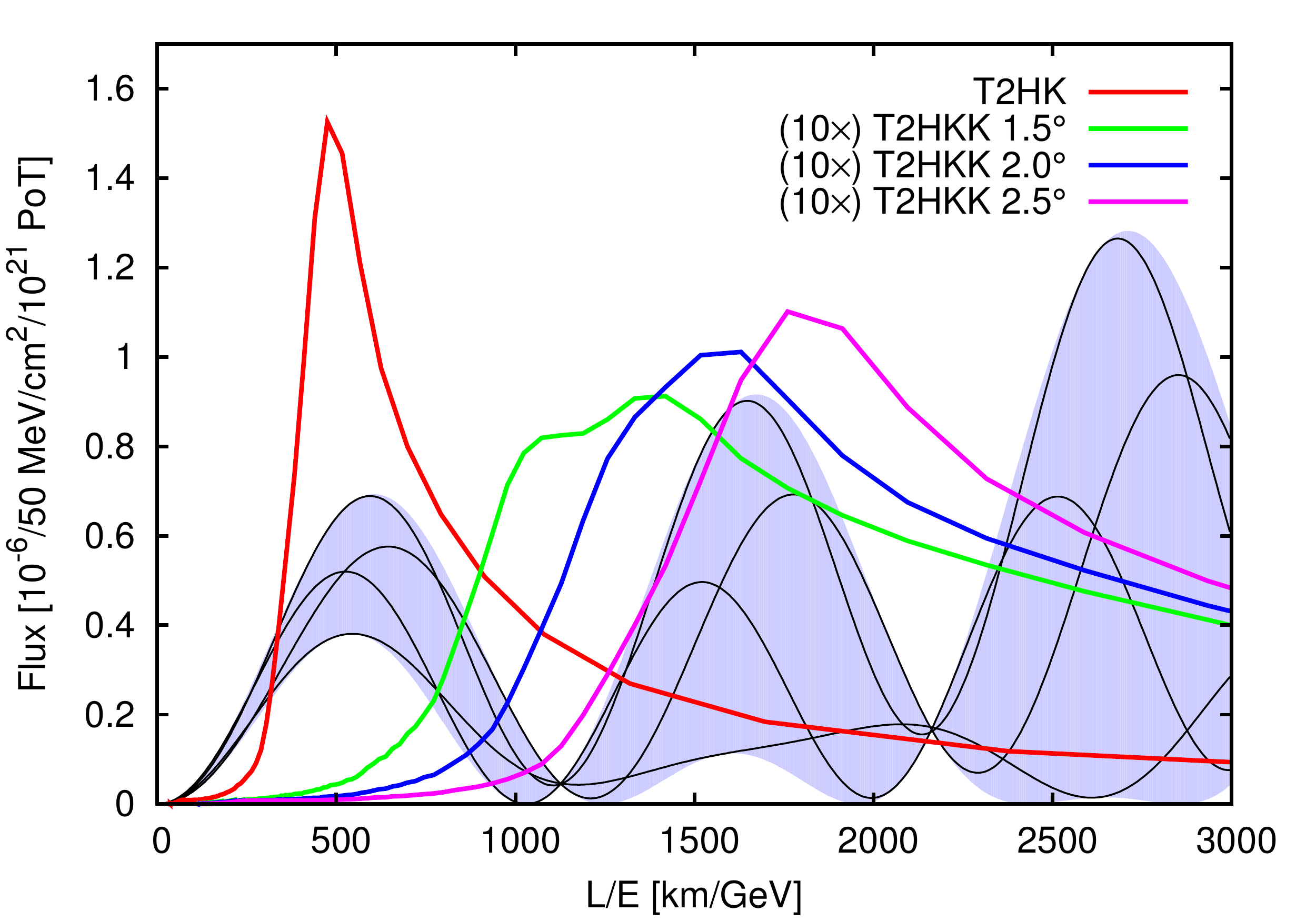}

\caption{\label{fig:t2hk_fluxes_LE}The T2HK and T2HKK fluxes shown as a
function of $L/E$. The shaded region shows the envelope of the probability for
$L=1100$ km and the black lines indicate the specific behaviour for $\delta
\in \{ 0, \frac{\pi}{2}, \pi, \frac{3\pi}{2}\}$. Note that the T2HK flux
actually samples from the probability with a smaller matter effect
corresponding to its shorter baseline $L=295$ km; however, on this scale the
location of the first maximum does not deviate much from what is shown here.}

\end{figure}

We have developed an up-to-date GLoBES implementation of T2HK, incorporating
the collaboration's latest estimates for detector performance\footnote{We thank
the Hyper-Kamiokande proto-collaboration for kindly providing us
with this information.}.
Our simulation is based on the GLoBES implementation of
T2HK~\cite{Huber:2002mx} with comprehensive modifications to match the latest
experimental design. The beam power and fiducial mass have been updated to
\SI{1.3}{\MW} and \SI{187}{\kilo\tonne} per tank. For our studies we have used
the staged design with one tank operational for $6$ years followed by two
operational tanks beyond that time. In cases where we show results against the
run time of the experiment, we have used additional simulations with just a single
tank operational throughout to highlight the discontinuous nature of this design.
The neutrino flux and channel definitions have been updated to match those of
\refref{HKDR}, with separate channels for four interaction types (charged
current quasielastic, charged current with one pion, other charged current and
neutral current), for the $\nu_\mu\to\nu_e$ and $\anu_\mu\to\anu_e$ signals, and
unoscillated $\nu_e$, $\anu_e$, $\nu_\mu$ and $\anu_\mu$ backgrounds. New tables
of pre-smearing efficiencies and migration matrices have been created for each
channel based on the full detector simulations used in \refref{HKDR}. New
cross-sections for interactions on water for the four interaction types have
been generated using the GENIE Monte-Carlo neutrino interaction event
generator~\cite{Andreopoulos:2009rq}.

The simulation determines the event rates for signal and background components
for each of $\nu_\mu/\anu_\mu\to\nu_e/\anu_e$ appearance and
$\nu_\mu/\anu_\mu\to\nu_\mu/\anu_\mu$ disappearance measurements in neutrino
mode and antineutrino mode. The rates are determined for 12 energy bins, given
in \refapp{app:t2hk_details}. For the appearance measurements, the energy range is
restricted to \SIrange{0}{1.25}{\GeV}, so only bins \numrange{1}{8} are
included. All bins are included in the disappearance measurements. Separate
uncorrelated systematic errors are assumed on the total signal and background
rates for each of the four measurements, where the size of the errors assumed,
summarised in \cref{tab:hk-systematics}, are the same as in the official
Hyper-K studies after an adjustment to account for correlations between
systematics not included in our simulations.

The design of T2HKK~\cite{T2HKK} and the location of the second detector module
are still under development. As such, physics studies are being performed for a
number of simulated fluxes with varying off-axis angles, generally ranging from
on-axis to $2.5^\circ$ off-axis, which is aligned with the first detector in
Kamioka.
The novelty of this design is not only the longer baseline
distance, which will enhance the role of matter effects, but also the fact that
the energy profile of the flux remains similar to that at the detector at 295
km, meaning that the oscillation probability is sampled at very different
values of $L/E$. This results in the second detector having access to increased
spectral information, which can help to break degeneracies and enhance overall
sensitivity \cite{Ishitsuka:2005qi}.
This is clearly seen in \reffig{fig:t2hk_fluxes}, where the left panel shows
how the flux aligns with the first maximum of the probability at Kamioka while
the right panel shows that the fluxes align around the second maximum for the
Korean detector. When plotted against $L/E$, as in \reffig{fig:t2hk_fluxes_LE},
we see that the T2HK flux has only minor coverage of the second maximum in
contrast to T2HKK. The fluxes used in our simulation were provided by the
Hyper-Kamiokande proto-collaboration and were produced in the same way as the
fluxes used in~\cite{HKDR} but with a baseline of $1100$~km and off-axis angles
of $1.5^\circ$, $2.0^\circ$ and $2.5^\circ$.

\subsection{\label{sec:runtime}Experimental run times and $\nu$ : $\overline{\nu}$ ratios}

The previous sections have discussed our models of the experimental details of
DUNE and T2HK. However, in the present study, we will consider a number of
different exposures for these experiments and their combination. This section
is intended to clarify our terminology and explain our choices of run time,
neutrino--antineutrino sharing, and staging adopted in the following analyses.

First, we comment that although the ratio of the run time between $\nu$ and
$\bar{\nu}$ beam modes is also known to affect the sensitivities of
long-baseline experiments, we stick to the ratios defined by each experiment's
official designs throughout our work. For DUNE and T2HK, the ratio of $\nu$ to
$\bar{\nu}$ are 1:1 and 1:3, respectively. We have investigated the impact of
changing these ratios, but they do not significantly impact the
results, and for both experiments the optimal ratio was close to those assumed
here. In the study for alternative designs, we stick with the same ratios as
the standard configurations of DUNE and T2HK.

\begin{table}[t]
\centering
\begin{tabular}{ |c|c|c|c| } 
\hline
~ & Label & $\nu$ : $\overline{\nu}$ at DUNE  & $\nu$ : $\overline{\nu}$ at T2HK  \\ \hline\hline
\multirow{3}{*}{Fixed run time} & DUNE & $5$ : $5$ & 0 : 0          \\ \cline{2-4}
 & T2HK & 0 : 0 & $2.5$ : $7.5$          \\\cline{2-4}
 & DUNE + T2HK & $5$ : $5$ & $2.5$ : $7.5$  \\\hline\hline
\multirow{3}{*}{Variable run time} & DUNE & $T/2$ : $T/2$ & 0 : 0        \\\cline{2-4}
~ & T2HK & 0 : 0 & $T/4$ : $3T/4$      \\ \cline{2-4}
~ & DUNE/2 + T2HK/2 & $T/4$ : $T/4$ & $T/8$ : $3T/8$     \\ \hline
\end{tabular}

\caption{\label{tab:standard_run_times}The run times in years for each
component of DUNE, T2HK, and their combination (DUNE + T2HK) for both the
standard full data taking period (top 3 rows) and when considered with variable
run times (bottom 3 rows). Plots with cumulative run time $T$ on the $x$-axis
are for the ``variable run time'' configurations, whilst all other plots are
for the ``fixed run time'' configurations. We specify the details for
configurations without staged power or mass increases when relevant in the
text. We note here that the fixed run-time configuration of DUNE (T2HK)
corresponds to $600$ ($3400$) kiloton$\times$MW$\times$ years of exposure.}

\end{table}

Most of our plots deal with three configurations labelled as DUNE, T2HK and
DUNE + T2HK, and the sensitivities shown assume the full data taking periods
for these experiments have ended. These are our standard configurations, and
are defined in terms of run times and neutrino--antineutrino sharing in the
rows labelled ``fixed run time'' in \reftab{tab:standard_run_times}. 
We point out that as we are interested in comparing experimental performance,
we take our standard configuration of DUNE to have 10 years runtime, equal to
the baseline configuration of T2HK \cite{HKDR}. This does, however, differ from
the 7 years considered in \refref{Acciarri:2015uup}, and our sensitivities are
correspondingly better.

However, we will also plot quantities against run time, and for these figures
we define the sharing of run time between components in terms of a quantity we
call the cumulative run time $T$; these are shown in the rows labelled
``variable run time'' in \reftab{tab:standard_run_times}. The cumulative run
time for the combination of DUNE and T2HK is defined to be the sum of the
individual experiments' run times, \emph{i.e.} if the two experiments were run
back to back, with no overlapping period of operation, then our definition of
cumulative run time is identical to the calendar time taken for the full data
set to be collected\footnote{In the interests of clarity, let us point out that
we use the term \emph{calendar time} to denote the actual time passed on the
calendar. This is highly dependent on staging and the relative placements of
individual experiment schedules, and is only used later in the text as an
informal means of comparison for certain staging options.}. Of course, if the
experiments run in parallel, with identical start and end dates, our definition
of cumulative run time would be double the calendar time required to collect
the data.
To remind readers of our definitions, we label this variable run time
configuration as DUNE/2 + T2HK/2, as half of the cumulative run time goes to
each experiment. 
Note also that, as per the official studies of each experiment, we assume
$10^7$~seconds per year of active beam time for T2HK ($2.7\times10^{21}$
POT/year at 1.3~MW with 30~GeV protons) and combined accelerator uptime and
efficiency of 56\% ($1.47\times10^{21}$  POT/year at 1.07~MW with 80~GeV
protons up to the 6th year, doubling the POT thereafter) for DUNE.

The possible staging options for the two modules of T2HK and the power of LBNF
cause some added complication when plotting sensitivities against run time. In
this study, we assume that our standard configurations of T2HK and DUNE follow
the staging scenarios suggested by the collaborations: $6$ years
of 1-tank ($187$~kt of total volume) running followed by $4$ with an additional
tank for T2HK ($374$ kiloton of total volume), and $6$ years of $1.07$~MW
($1.47\times10^{21}$ POT/year) followed by $4$ of $2.14$~MW
($2.54\times10^{21}$ POT/year) for DUNE with 2-horn 80-GeV-proton design.  In
practice, we implement an effective mass for T2HK which depends on the run time
$t$ assigned to T2HK defined by
\[   M(t) = M_0\left[1 + \Theta(t-6)\frac{t-6}{t}\right], \]
where $M_0$ is the mass of a single tank, defined above as $187$~kt, and
$\Theta(x)$ is the Heaviside step function. We make an analogous definition for
the power of DUNE, again increasing by a factor of two after 6 years. As our
definition of cumulative run time $T$ would require 12 years to pass before 6
years of data had been collected by either of the experiments in the
combination of DUNE/2 + T2HK/2, we see the discontinuity in sensitivity due to
staging appear in two different places in our plots against run time: one for
an experiment alone, and one for DUNE/2 + T2HK/2.  This can be seen clearly in \eg\
\reffig{fig:MO_runtime}, where we mark the discontinuities with vertical dashed
lines. So as to better understand the impact of these upgrades, we
will also show the sensitivities against run time which would apply were they
absent. However, we stress that the full programme of upgrades is an integral
part of the collaborations' proposals and should be taken as part of their
baseline configurations.
 
Finally, in \refsec{sec:delta} we will deviate from these configurations (and
the labels in \reftab{tab:standard_run_times}) as we consider non-standard
exposures for the purpose of better exploring the complementarity of DUNE and
T2HK. This will be discussed in more detail in \refsec{sec:delta}.
 
\subsection{\label{sec:Statistical}Statistical method}

Our simulation uses GLoBES \cite{Huber:2004ka, Huber:2007ji} to compute the
event rates and statistical significances for the experiments discussed in the
previous section. We will now briefly recap the salient details of the
statistical model underlying the analysis.

Given the true bin-by-bin event rates $n_i$ for a specific experimental
configuration, we construct a $\chi^2$ function based on a log-likelihood
ratio,
\begin{equation}\label{chi-square} \chi^2(\vec{\theta},\xi_s,\xi_b)=2\sum_i
\left(\eta_i(\vec{\theta},\xi_s,\xi_b)-n_i+n_i\ln\frac{n_i}{\eta_i(\vec{\theta},\xi_s,\xi_b)}\right)+p(\xi_s,\sigma_s)+p(\xi_b,\sigma_b),
\end{equation} 
where $i$ runs over the number of bins, $\eta_i(\vec{\theta},\xi_s,\xi_b)$ is
the hypothesis event rate for bin $i$ and $E_i$ is the central bin energy. The
vector $\vec{\theta}$ has six components, corresponding to each of the three
mixing angles, one phase and two mass-squared splittings of the hypothesis. The
parameters $\xi_s$ and $\xi_b$ are introduced to account for the systematic
uncertainty of normalization for the signal (subscript $s$) and background
(subscript $b$) components of the event rate, and are allowed to vary in the
fit as nuisance parameters. For a given hypothesised set of parameters
$\vec{\theta}$, the event rate for bin $i$ is calculated as
\[ \eta_i(\vec{\theta},\xi_s,\xi_b)=(1+\xi_s)\times n_i+(1+\xi_b)\times b_i, \]
where $n_i$ and $b_i$ are the expected number of signal and background events
in bin $i$, respectively. The nuisance parameters are constrained by terms
$p(\xi,\sigma)= \xi^2/\sigma^2$, representing Gaussian priors on $\xi_s$ and
$\xi_b$ with corresponding uncertainties $\sigma_s$ and $\sigma_b$.
To test a given hypothesis against a data set, we profile out unwanted degrees
of freedom. This amounts to minimising the $\chi^2$ function \refeq{chi-square}
over these parameters whilst holding the relevant parameters fixed. We will
explain the statistical parameters of interest for each analysis in the
following sections, however, as an example we will be interested in how well
different hypothesised values of $\delta$ fit a given data set. In this case,
we would compute 
\begin{equation}\label{chi-square_prior}
\chi^2(\delta)=\min_{\{\vec{\theta}\neq \delta
,\xi_s,\xi_b\}}\left(\chi^2(\vec{\theta},\xi_s,\xi_b) + P(\vec{\theta})\right),
\end{equation}
where the notation $\vec{\theta}\neq\delta$ means all parameters other than
$\delta$. The function $P(\theta)$ is a prior, introduced to mimic the role of
data from existing experiments during fitting. In all fits that we perform,
unless explicitly stated otherwise, we use true values from the recent global
fit NuFit 2.2 (2016) \cite{Gonzalez-Garcia:2014bfa}.  
$P(\theta)$ comprises a sum of the 1D $\chi^2$ data provided by NuFit for each
parameter, except for $\delta$, and we switch between NO and IO priors
depending on the mass ordering of our hypothesis. 
This includes the correlations which are currently seen in the global data, and
our treatment goes beyond the common assumption of Gaussian priors, allowing
for both the degenerate solution and its relative poorness of fit to be more
accurately taken into account. The values of all parameters are permitted to
vary, including the different octants for $\theta_{23}$, the value of
$\delta_{CP}$ and the mass orderings, subject to the global constraints. Our
choice of true values depends on the mass ordering, and are given explicitly in
\reftab{tab:global_fit_parameters}, unless stated otherwise.
Note that the current best-fit values correlate the mass ordering
and the octant, with NO preferring the lower octant and IO, the higher octant.
This will affect our simulation, for example leading to poorer CPV sensitivity
for IO, and in \refsec{sec:results} we will show results for a
band of $\theta_{23}$ spanning both solutions to mitigate this asymmetry.

We point out that our treatment of the external data, which attempts to
accurately model the global constraints beyond the approximation of independent
Gaussians, leads to some differences between our results and those of previous
studies \cite{Acciarri:2015uup,HKDR,Fukasawa:2016yue}.  
The differences can be traced to two key features: first, we take into account
the significantly non-Gaussian behaviour of the global constraints at higher
significances. This is particularly relevant for the prior on $\Delta m^2_{21}$
and we will comment on this in more detail in \refsec{sec:MO} and
\refapp{app:MO_sensitivity}.
The second important feature of our priors is the strong correlation between
mass ordering and the octant of $\theta_{23}$. The current global data
disfavours the combination of IO and first octant (NO and second octant). This
fact is reflected in our priors; although a visible local minimum is always
present, it is never degenerate with the true minimum. In previous studies,
various treatments of this degeneracy have been employed, some which do not
allow the alternative minimum, and some which do not penalise it at all.  Our
method interpolates between these two extremes, and attempts to faithfully
describe the current global picture.
We will provide more detail on the specific differences between our results and
existing calculations of the sensitivity of DUNE, T2HK and their variant
designs on a case-by-case basis in the following sections.

\begin{table}[t]
\centering
\begin{tabular}{ |c|c|c| } 
 \hline
 Parameter & Normal ordering & Inverted ordering \\
  [0.5ex] 
 \hline\hline
 $\theta_{12}$ [$^\circ$] & $33.72^{+0.79}_{-0.76}$ & $33.72^{+0.79}_{-0.76}$ \\ 
 $\theta_{13}$ [$^\circ$] & $8.46^{+0.14}_{-0.15}$ & $8.48^{+0.15}_{-0.15}$ \\
 $\theta_{23}$ [$^\circ$] & $41.5^{+1.3}_{-1.1}$ & $49.9^{+1.1}_{-1.3}$ \\
 $\Delta m_{21}^2$ [$\times10^{-5}$ eV$^2$]& $7.49^{+0.19}_{-0.17}$  & $7.49^{+0.19}_{-0.17}$ \\
 $\Delta m_{31}^2$ [$\times10^{-3}$ eV$^2$]& $+2.526^{+0.039}_{-0.037}$ & $-2.518^{+0.038}_{-0.037}$ \\
 \hline
\end{tabular}

\caption{\label{tab:global_fit_parameters}The true values used in our fit,
unless otherwise stated explicitly, with their uncertainties (the $1\sigma$
range of the priors we have used in our fit). These are based on NuFit 2.2
(2016)~\cite{Gonzalez-Garcia:2014bfa}, and are similar to the parameters found
in other recent global fits (see \eg\ \cite{Forero:2014bxa, Capozzi:2016rtj}).\protect\footnotemark
}

\end{table}

\section{\label{sec:results}Sensitivity to mass ordering, CPV, non-maximal CPV, and octant}

In this section, we will present the results of our simulation studying the
sensitivity of the standard configurations of DUNE and T2HK. This means we use
the 2-horn optimised flux for DUNE with a staged beam upgrade after 6 years,
while for the T2HK detector we assume the installation of a second detector
module after 6 years.  More details of these configurations can be found in
\refsec{sec:details_DUNE} and \refsec{sec:details_T2HK}.
However, for comparison, we also include two unstaged options: where the
experiments continue without upgrading at the $6$ year mark. We
stress that these are not the baseline configurations of the experiments, and
that they are interesting for comparison purposes only. 
The run time and neutrino--antineutrino sharing for these configurations are
discussed in more detail in \refsec{sec:runtime}.
After considering these benchmark configurations and their complementarity, we
will return to the potential of alternative designs in
\refsec{sec:alternative_designs}.

\subsection{\label{sec:MO}Mass ordering sensitivity}

\footnotetext{An updated version of the NuFit global
fit (NuFit 3.0) was released after we had concluded this study. We have,
however, checked that no significant differences occur if we implement new
priors based on its results.}

The mass ordering is one of the central goals of the next generation of LBL
experiments; it is also one of the easiest to measure with this technology.
We quantify the ability to determine the mass ordering by computing the
following test statistic,
\begin{equation}
\Delta\chi^2_{\text{MO}}=\min_{\{\vec{\theta},\xi_s,\xi_b\}}\left[\chi^2(\text{sgn}\Delta
m^2_{31}=\text{true})-\chi^2(\text{sgn}\Delta m^2_{31}=\text{false})\right].
\end{equation} 
That is to say, the smallest value of the $\chi^2$ function for any parameter
set with the wrong ordering. All parameters are allowed to vary during
marginalisation whilst preserving the ordering. Although our composite
hypothesis violates the assumptions of Wilks' theorem
\cite{Qian:2012zn,Ciuffoli:2013rza}, and therefore invalidates the mapping
between $\sqrt{\Delta \chi^2}$ and $\sigma$-valued significance for
discrimination of the two hypotheses, we stick to convention in this section,
reporting the expected sensitivities for the median experiment in terms of
$\sqrt{\Delta\chi^2}$ and discussing it in terms of $\sigma$. For the reader
who is interested in the precise formulation of the statistical interpretation
of $\sqrt{\Delta \chi^2}$, see \eg\ \refref{Blennow:2013oma}.

\begin{figure}[t]
\centering
\includegraphics[width=0.8\textwidth, clip, trim = 1 3 5 5]{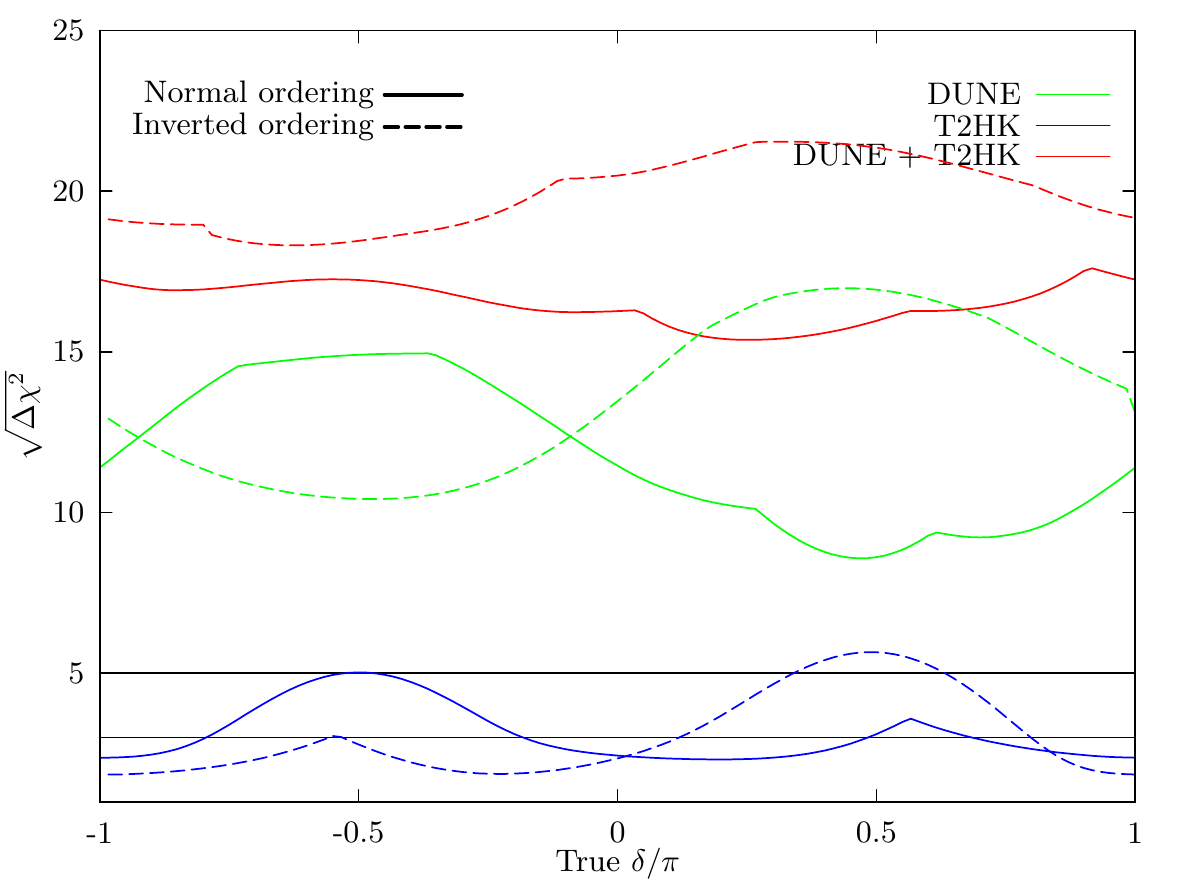}

\caption{\label{fig:both_MO}The sensitivity to the mass ordering for DUNE and
T2HK in isolation and combined for true normal ordering (solid) and inverted
ordering (dashed). This plot assumes the ``fixed run time''
configurations in \reftab{tab:standard_run_times} and the true
oscillation parameters given in \reftab{tab:global_fit_parameters}.}

\end{figure}

The sensitivity we find in \reffig{fig:both_MO} is very strong. DUNE, with its
large matter effects, can expect a greater than $8.5\sigma$ measurement of the
mass ordering after $10$ years for all values of $\delta$, with an average
sensitivity of around $12\sigma$ and a maximal sensitivity of
around $17\sigma$.  T2HK alone has limited access to this
measurement due to its shorter baseline, but can still expect a greater than
$3\sigma$ measurement for around $25\%$ of the possible values of $\delta$
after $10$ years of data-taking. The combination of DUNE and T2HK running for
10 years each can reach sensitivities of at least $15\sigma$,
with an average of around $18\sigma$.  
Care should be taken when interpreting such large significances; however, it is
clear that DUNE, and the combination of DUNE and T2HK, can expect a very strong
determination of the mass ordering.
We also note the strong complementarity here: for the values of $\delta$ where
DUNE performs the worst, the information from T2HK helps to raise the global
sensitivity by about $7\sigma$. Despite this interesting
interplay, the fact that this is such an easy measurement for experiments of
this type, means that we will not dwell on the question of optimising such a
measurement further. 

Our sensitivities in \reffig{fig:both_MO} deviate from previous published
values for DUNE, and we generally report a worse ability for DUNE to exclude
the ordering, with lower average sensitivity and visibly discontinuous
behaviour in the values of $\Delta \chi^2$.
This is due to the priors that we have imposed. Instead of a Gaussian
approximation to the global data, we implement the global 1D $\chi^2$
functions, as provided by NuFit \cite{Gonzalez-Garcia:2014bfa}. The true global
data has strongly non-Gaussian behaviour at high significance, and there exist
non-standard parameter sets which are not excluded at greater than $6\sigma$.
These parameter sets sometimes become the best-fitting wrong-ordering solution,
and must be excluded to rigorously establish the mass ordering. We discuss this
in more detail in \refapp{app:MO_sensitivity}. We point out, however, that our
priors do not always significantly affect the point of minimum sensitivity, and
DUNE still expects to see a greater than $5\sigma$ discovery for all true
values of $\delta$. 
However, the values of parameters at the minimum do depend on our assumptions.
For example, in \reffig{fig:both_MO} we have found for inverted ordering the
lowest MO sensitivity over $\delta$ is affected by the degeneracy due to our
prior, while for the normal ordering, the minimum is given by the conventional
parameter set.

\begin{figure}[t]
\centering
\includegraphics[width=0.99\textwidth,clip,trim=15 30 3 40]{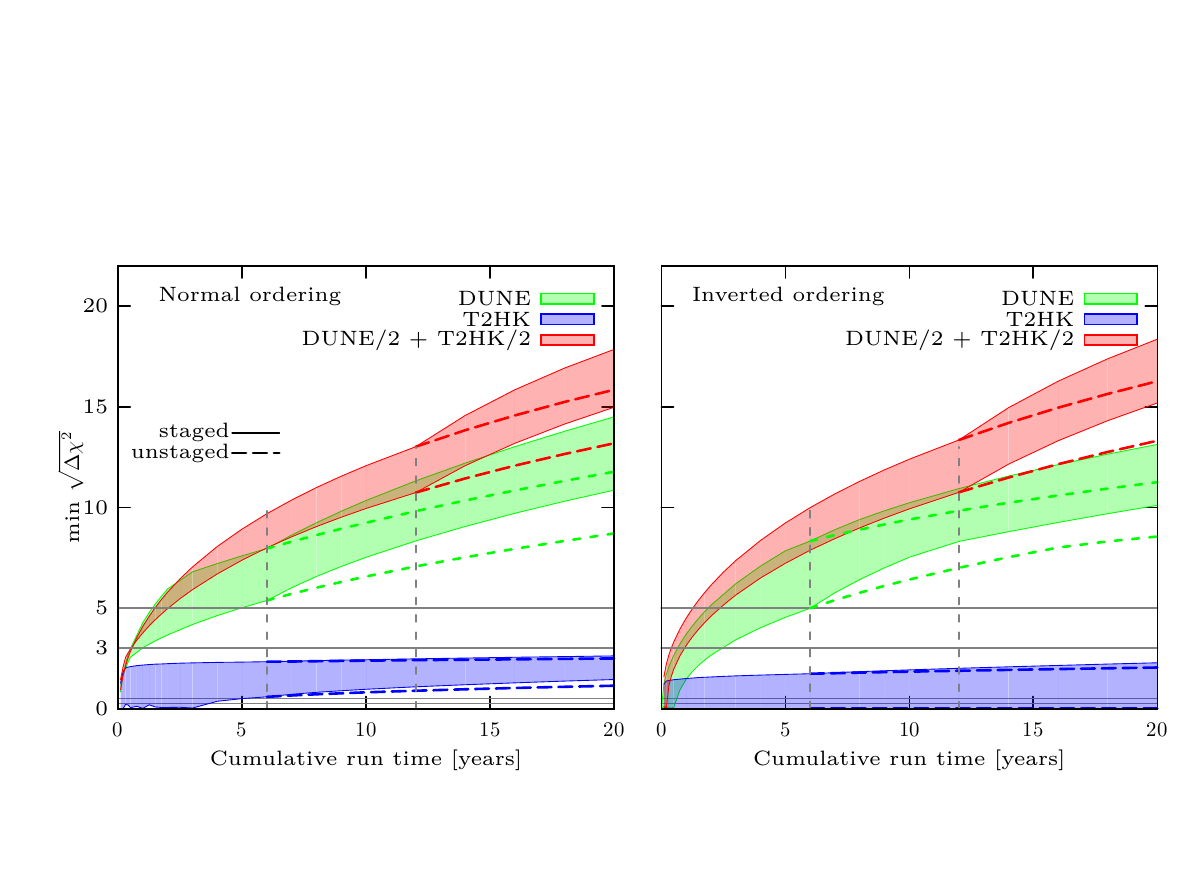}

\caption{\label{fig:MO_runtime}The least sensitivity for discovering mass
ordering, $\text{min}(\sqrt\Delta\chi^2)$, which can be
reached by DUNE, T2HK and their combination as a function of cumulative run
time. The width of the bands shows the sensitivity for
$40^\circ\le\theta_{23}\le50^\circ$. The left (right) panel assumes normal
(inverted) ordering. These plots assume the ``variable run time''
configurations in \reftab{tab:standard_run_times} and the true oscillation
parameters, apart from $\theta_{23}$, given in
\reftab{tab:global_fit_parameters}. The vertical lines mark the introduction of
a staged second detector for T2HK and/or a increase in the beam
power for DUNE. They lead to a notable discontinuity in sensitivity.}

\end{figure}

Another way to understand the complementarity of DUNE and T2HK is in terms of
minimal run time necessary to ensure a $\sqrt{\Delta\chi^2}>5$ measurement
regardless of the true value of $\delta$.  We plot this quantity in
\reffig{fig:MO_runtime}, for normal ordering (left) and inverted ordering
(right).  The shaded bands take into account the variation in sensitivity due
to the true value of $\theta_{23}$. DUNE alone takes between 2 and 6 years to
reach this sensitivity, while the combination of DUNE and T2HK always takes
less than 3 years (which if run in parallel is only 1.5 years).  T2HK running
alone cannot ensure a measurement of this significance over any plausible run
time.  We note the small discontinuity along the upper bound 
for normal (inverted) ordering after about 2 (5) years run time for DUNE.  
This marks the appearance of a degenerate solution due to
the non-Gaussianity of our priors as discussed before (and in more detail in
\refapp{app:MO_sensitivity}). We also show explicitly the difference in minimal
sensitivity for T2HK with (solid lines) and without (dashed lines) a second
staged detector module at Kamioka, as well as for DUNE with (solid lines) and
without (dashed lines) the upgraded accelerator complex. For T2HK, the increase
in performance is negligible, but DUNE as well as the combination of DUNE and
T2HK sees a notable performance increase.

\subsection{\label{sec:CPV}CP violation sensitivity}

To fulfil the central aim of the LBL programme, the experiments must be able to
rule out CP conservation over a large fraction of the true parameter space.
This would imply a non-zero Jarlskog invariant and rigorously establish CP
violation in the leptonic sector. Once again, we follow the conventional test
statistic and define the quantity
\begin{equation}
\Delta\chi^2_{\text{CP}}=\min_{\delta\in\{0,\pi\}}\Delta\chi^2(\delta),
\end{equation}
which amounts to studying the composite hypothesis of CP conservation
($\delta=0$ or $\delta=\pi$) \cite{Elevant:2015ska}. Although at
low-significance this test statistic is known to deviate from a $\chi^2$
distribution \cite{Blennow:2014sja}, we expect such effects to be small for the
experiments under consideration in this study and the interpretation of
$\sqrt{\Delta\chi^2}$ as $\sigma$-valued significances to be reasonable. 

\begin{figure}[t]
\includegraphics[width=0.49\textwidth]{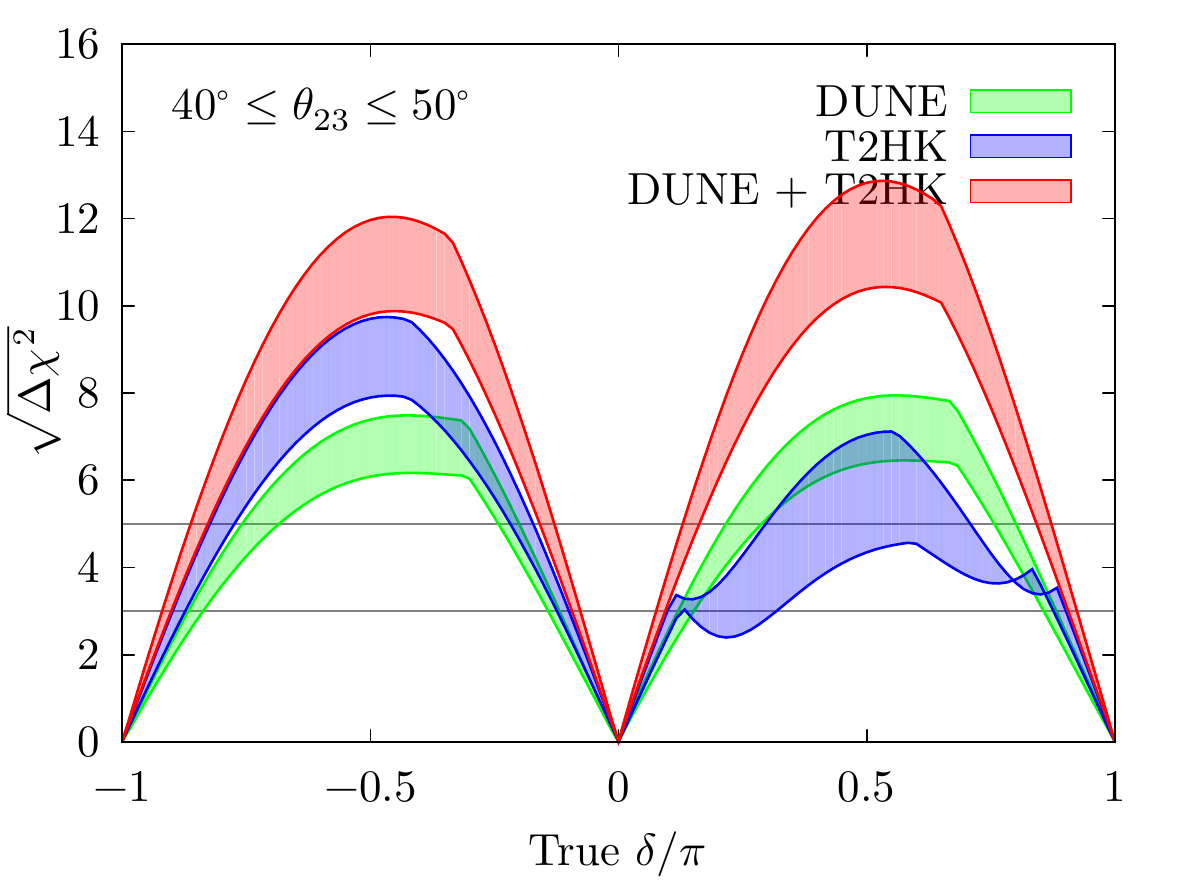}
\includegraphics[width=0.49\textwidth]{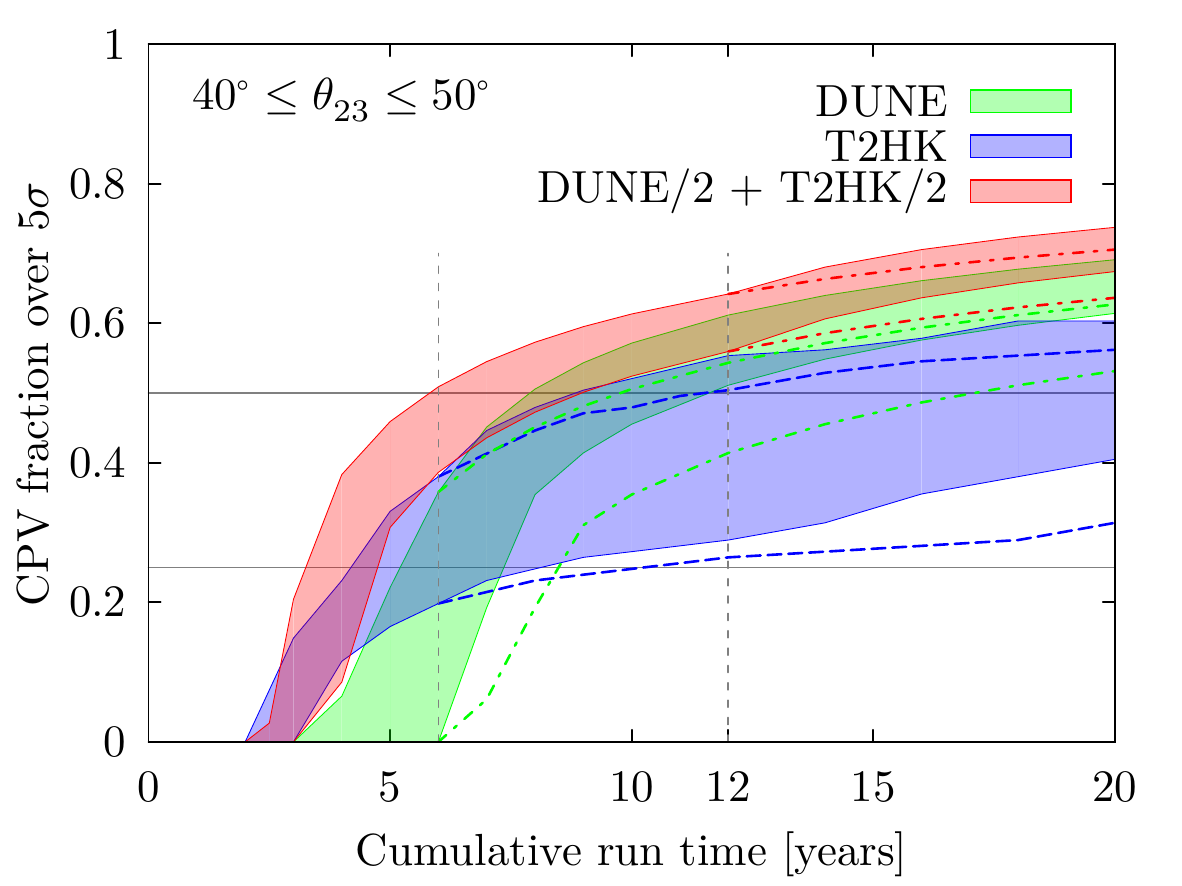}
\caption{\label{fig:both_CPV_th23}The sensitivity to CP violation for DUNE and
T2HK in isolation and combined as a function of delta (left) and the fraction
of $\delta$ parameter space for which greater than $5\sigma$ CPV discovery is
expected (right). We consider a range of true $\theta_{23}$ spanning both
octant solutions. The lower edge of the shaded regions corresponds to
$\theta_{23}>45^\circ$ due to a decrease in sensitivity arising from the
relative suppression of the CP sensitive terms in \refeq{eq:Asano_prob1}. 
The left (right) plot assumes the ``fixed run time'' (``variable run time'')
configurations in \reftab{tab:standard_run_times} and the true
oscillation parameters, apart from $\theta_{23}$, specified in
\reftab{tab:global_fit_parameters}.}
\end{figure}

For the discovery of CP violation, the true value of the mass ordering and
octant are relevant. We do not specify these values, and have studied the
sensitivity for all combinations of values.  We show in the left panel of
\reffig{fig:both_CPV_th23} the significance for exclusion of CP conservation
for the standard designs of the two facilities, in isolation and combination.
We find that both experiments have a high sensitivity to this measurement, with
at least a $3\sigma$ ($5\sigma$) discovery of CPV over $70$--$75\%$
($46$--$47\%$) of the parameter space for DUNE and $73$--$80\%$ ($26$--$51\%$)
for T2HK. 
For $0\le\delta\le\pi$, we see a notable difference in behaviour between DUNE
and T2HK: the sensitivity for T2HK is limited, and much more dependent on the
true value of $\theta_{23}$. This is due to the inability of T2HK to resolve
the mass ordering degeneracy, which leads to a degenerate approximately CP
conserving solution for these regions of parameter space\footnote{We note that
atmospheric neutrino oscillation data collected by HK may be able to help
resolve degeneracies and improve the experiment's sensitivity, but we do not
consider this option further.}. 
We point out that, as DUNE provides high MO sensitivity, the combination of
data from DUNE and T2HK does not suffer from this problem, and sees significant
improvements in sensitivity for these values of $\delta$.
Aside from this limitation, the general shape of these curves can be understood
by our discussion in \refsec{sec:precision}. Discovery potential for CPV is
closely related to the precision on $\delta$ at the CP conserving values, both
rely on distinguishing between \emph{e.g.} $\delta=0$ and other values. The
best sensitivity to CP conserving values of $\delta$ is at the first maximum,
where the majority of T2HK events are found and consequently it sees a better
sensitivity. 
Our plots have assumed NO, but the qualitative picture remains the same for IO:
in this case, the degeneracy occurs for the $-\pi\le\delta\le0$, but otherwise
the two regions of $\delta$ swap roles and the sensitivites are similar. We
note, however, that the current best-fit values of $\theta_{23}$ would lead to
additional suppression of CPV sensitivity for IO. The global data associates IO
with a value of $\theta_{23}$ in the higher octant, which predicts poorer
sensitivity to $\delta$.

As we mentioned in the last paragraph of \refsec{sec:Statistical}, our prior
correlates the allowed octant to the mass ordering, and this is responsible for
differences between our results and previously published work.
In Fig 6 of Ref.\ \cite{Fukasawa:2016yue}, there is almost no CPV sensitivity
for $0<\delta<\pi$ for T2HK, which has not been found in our results, while
their results for DUNE are similar to ours. This feature is explained as being
due to the lack of MO sensitivity at T2HK, allowing for degeneracies to limit
the sensitivity. In our simulation, however, T2HK alleviates this
problem by its strong determination of the octant and the correlation of the
global data. This lifts the degeneracy to higher significances, and allows a
higher sensitivity to be obtained before the limiting effect becomes relevant.

We find that DUNE performs slightly better in our simulation than is reported
in the left panel of Fig 3.13 in Ref.\ \cite{Acciarri:2015uup}. Around
$\delta=\pi/2$ ($-\pi/2$), their result shows the sensitivity is about $5.8$
($4.8$).\footnote{The range given in their work is for various beam designs.
The result for the design we consider is at the bottom of the range.} However,
our simulation finds a range of between $7.8$ to $9.0$ ($6$ to $8\sigma$) for
$\delta=-\pi/2$ ($=\pi/2$). There are two sources for this discrepancy.
Firstly, we are assuming a longer run time (10 years), for the purposes of
comparison between T2HK and DUNE. Secondly, our priors are based on newer data,
with updated central values and smaller $1\sigma$ intervals.
The CPV sensitivity for DUNE does not peak around $\delta=-\pi/2$ in the left
panel of Fig 3.13 in Ref.\ \cite{Acciarri:2015uup} like our results, due to the
relatively poor determination of the octant. DUNE does not have as strong
octant sensitivity as for the mass ordering, but our prior correlates the two,
helping to reduce the impact of this alternative minimum for values of $\delta$
around $\delta=-\pi/2$.
Finally, we find general agreement between our results and those of Fig.\ 119
in Ref.\ \cite{HKDR}. This is because the mass ordering is fixed during fitting
in Ref.\ \cite{HKDR}, which mitigates the impact of the mass ordering
degeneracy. This leads to superficial agreement between our two sets of results
when the degeneracy is not relevant, but discrepancies when it is. Our result
shows the sensitivity which is possible assuming only the current global data,
whereas assuming the MO is known would require new external data, perhaps from
another long-baseline experiment (or from a joint analysis with atmospheric
neutrino data). 

In the right panel of \reffig{fig:both_CPV_th23}, we show the fraction of
values of $\delta$ for which a $5\sigma$ exclusion of CP conservation can be
made as a function of run time. DUNE requires between $5$ and $7$ years of
data-taking to reach at least a $5\sigma$ measurement for  $25\%$ of the
possible values of $\delta$, while T2HK alone shows a stronger dependence on $\theta_{23}$ but expects to be able to make at
least a $5\sigma$ measurement for more than $25\%$ of the parameter space after
$8$ years.
The combination of DUNE and T2HK is shown as a function of cumulative run time,
the sum of the individual run times for each experiment, and as such
interpolates the two sensitivities. However, if run in parallel, the
combination of the two experiments performs stronger than either in isolation,
and expects a greater than $5\sigma$ measurement for more than $50\%$ of the
parameter space after between $1.5$ and $2.5$ years of parallel
data-taking.

\subsection{\label{sec:maximal_CP}Sensitivity to maximal CP violation}

Although the search for any non-zero CPV is the principle goal of the next LBL
experiments, understanding the value of $\delta$ is also highly relevant.
Current global fits \cite{Gonzalez-Garcia:2014bfa, Forero:2014bxa,
Capozzi:2016rtj} point towards maximal values of $\delta$, $\delta = \pm\pi/2$.
Of course, these should be treated with some scepticism: no single experiment
can claim evidence for this at an appreciable level.  However, determining if a
maximal CP violating phase exists will remain a high priority for the next
generation of long-baseline experiments. If established, it could be seen as an
``unnatural'' value advocated as evidence against anarchic PMNS matrices.
Indeed, it is also one of the most common predictions in flavour models with
generalised CP symmetries, and is often associated with close to
maximal values of $\theta_{23}$ in models with residual flavor symmetries. For
more discussion, see \eg\  \refref{King:2015aea, King:2017guk}.

\begin{figure}[t]
\centering
\includegraphics[width=0.49\textwidth]{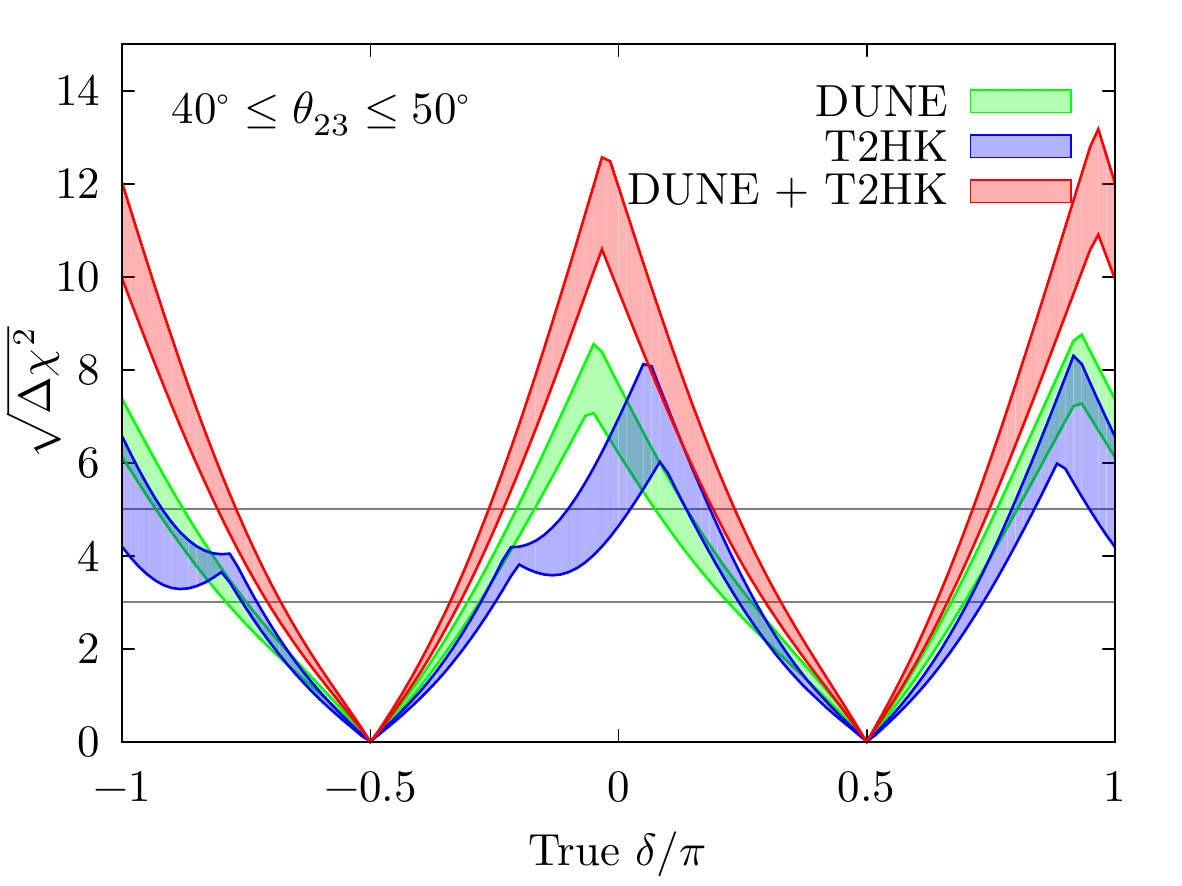}
\includegraphics[width=0.49\textwidth]{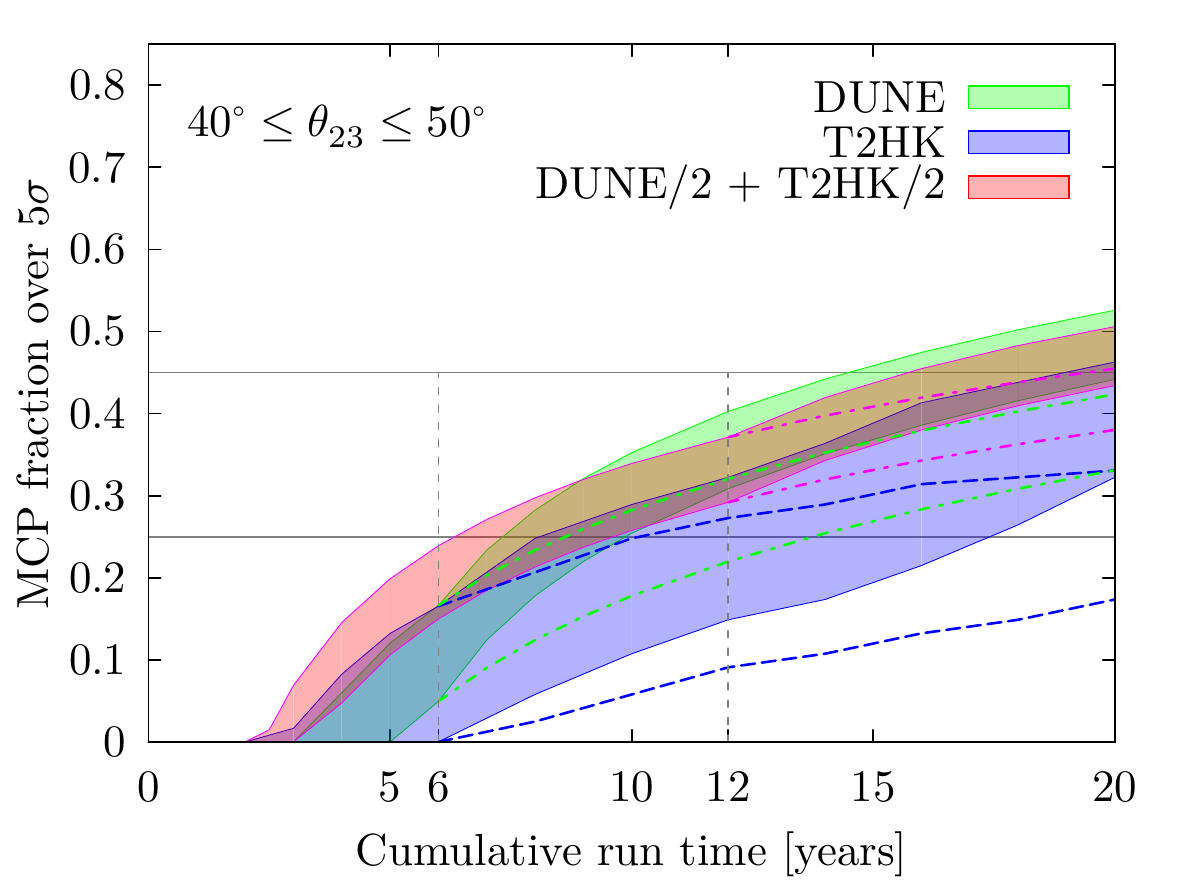}
\caption{\label{fig:both_MCP_th23} Left: the significance at which maximal CP
can be excluded for DUNE and T2HK in isolation and combined as a function of
true $\delta$. Right: the fraction of $\delta$-parameter space for which
maximal CP can be excluded as a function of run time. 
The left (right) plot assumes the ``fixed run time'' (``variable run time'')
configurations in \reftab{tab:standard_run_times} and the true
oscillation parameters, apart from $\theta_{23}$, specified in
\reftab{tab:global_fit_parameters}. }
\end{figure}

We have studied this question in \reffig{fig:both_MCP_th23} where we have
defined the quantity 
\begin{equation}
\Delta\chi^2_{\text{MCP}}=\min_{\delta\in\{-\frac{\pi}{2},\frac{\pi}{2}\}}\Delta\chi^2(\delta).
\end{equation}
This is analogous to $\Delta\chi^2_\text{CP}$ defined earlier, and gives us a
measure of the compatibility of the data with the hypothesis of maximal CP
violation.
On the left panel, we see the ability to exclude maximal CPV as a function of
the true value of $\delta$. There is a similar sensitivity for both facilities.
DUNE has the best performance for most cases, but T2HK still achieves the
highest significance exclusions for $-3\pi/4<\delta<-\pi/2$ and
$0<\delta<\pi/2$; 
although, its sensitivity is more affected by the value of $\theta_{23}$ and
the mass ordering.
In this way, the two experiments once again exhibit a complementarity, and the
combination of DUNE and T2HK inherits the best sensitivity of its two component
parts, expecting a $3\sigma$ exclusion of MCP for over $48$--$54\%$ of the
parameter space.

On the right panel of \reffig{fig:both_MCP_th23}, we show the fraction of true
values of $\delta$ for which a $5\sigma$ exclusion of maximal CP violation can
be achieved. By running in parallel for 10 years, DUNE and T2HK can expect a
coverage at this significance of around $42$--$50\%$ of the parameter space.
Once again we see T2HK's sensitivity is more dependent on $\theta_{23}$ and
generally lower than DUNE's.
 
\subsection{\label{sec:octant}Octant degeneracy and the precision on $\theta_{23}$}

Although we know that $\theta_{23}$ is around $45^\circ$, the current global
fit data allows for two distinct local minima, one below and one above
$45^\circ$. This ambiguity is known as the octant degeneracy and arises as the
disappearance channel of $\nu_\mu\rightarrow\nu_\mu$ is sensitive at
leading-order only to $\sin^22\theta_{23}$. However, the appearance channel
breaks this degeneracy at leading-order, and future long-baseline experiments
are expected to significantly improve our knowledge of $\theta_{23}$. In this
section, we study how well DUNE and T2HK will be able to measure $\theta_{23}$
as well as settling two central questions: is $\theta_{23}$ maximal, and which
is its correct octant? These questions are also of particular
theoretical significance as many models with flavour symmetries exist which predict
close to maximal values of $\theta_{23}$, and often the size of its deviation
from this point is in correlation to other parameters like $\delta$
\cite{King:2015aea, King:2017guk}. Therefore, determining the
octant (or maximality) of $\theta_{23}$ would be highly instructive in our
search to understand leptonic flavour.

\begin{figure}[t]
\centering
\includegraphics[width=0.49\textwidth]{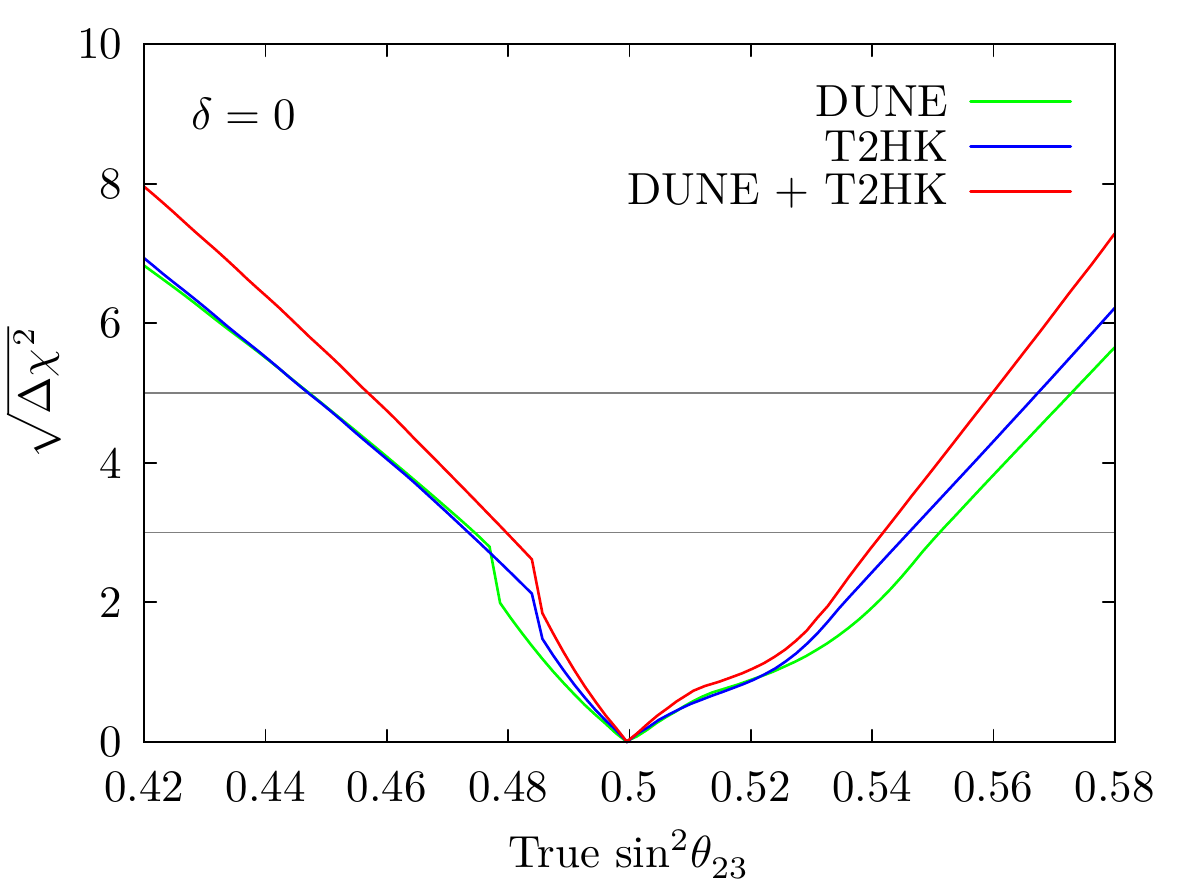}
\includegraphics[width=0.49\textwidth]{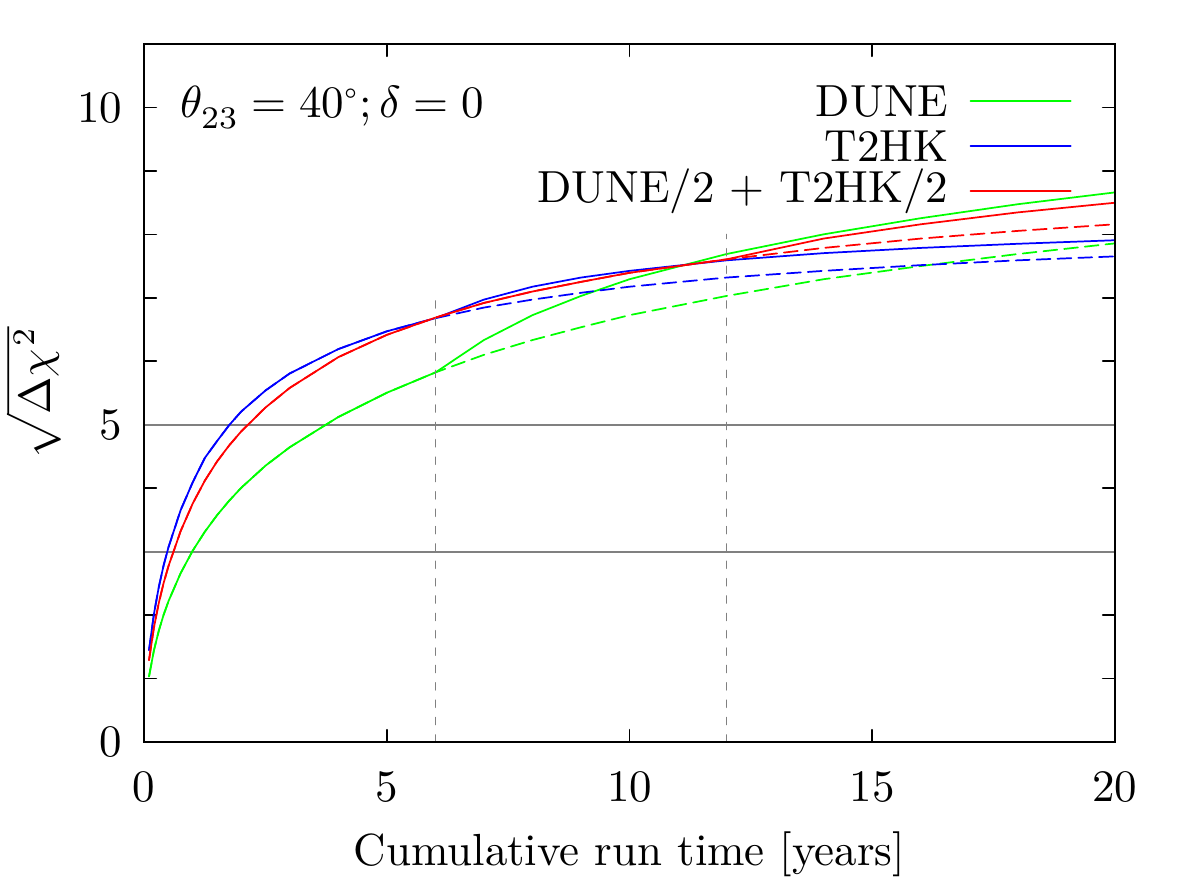}
\caption{\label{fig:OCT} The sensitivity to exclude the wrong octant for DUNE,
T2HK and their combination, as a function of $\sin^2\theta_{23}$ (left) and the
cumulative run time (right). These plots assume $\delta=0$ and normal mass ordering. 
The left (right) plot assumes the ``fixed run time'' (``variable run time'')
configurations in \reftab{tab:standard_run_times} and the true oscillation
parameters, apart from $\theta_{23}$, specified in
\reftab{tab:global_fit_parameters}. 
}
\end{figure}

The ability to exclude the wrong octant for DUNE, T2HK and their combination is
shown in \reffig{fig:OCT}. 
On the left, we show the sensitivity as a function of the true value of
$\theta_{23}$. In these plots we assume a fixed value of $\delta=0$.  The
impact of varying $\delta$ for these measurements is small, as the degeneracy
is broken at leading-order in the appearance channel, and the subdominant
effects of $\delta$ are less relevant.
The ability to exclude the wrong octant can reach up to 8$\sigma$ at the
extremes of the current $3\sigma$ range of $\theta_{23}$, and we see that
$3\sigma$ determinations of the upper (lower) octant can be expected for true
values of $\sin^2\theta_{23}$ less than 0.47--0.48 (greater than 0.54--0.55).
This corresponds to a $3\sigma$ determination of the octant for all values of
$\theta_{23}$ in the ranges $\theta_{23}\lesssim 43.3^\circ$--$43.8^\circ$ or
$\theta_{23}\gtrsim 47.3^\circ$--$48.4^\circ$.
On the right, we fix the true value of $\theta_{23}$ and show how the
sensitivity depends on cumulative run time. We see that the sensitivity quickly
plateaus, and the staging options make little difference. Overall, the
experiments expect to be able to establish the octant for this value of
$\theta_{23}$ after only 2 to 4 years. Although this plot assumes
$\theta_{23}=40^\circ$, changing the true value of $\theta_{23}$ leads to a
predictable change in sensitivity, as indicated in the left panel, but does not
qualitatively change the behaviour against run time.
We see that overall, T2HK performs better than DUNE for the determination of
the octant.
However, the difference in performance is marginal, and their combination after
10 years of data for each experiment, outperforms T2HK running alone for 20
years, but performs slightly worse than DUNE with 20 year of total run time.

In this simulation, we have not imposed a prior on $\theta_{23}$.
This process differs from Ref.\ \cite{Acciarri:2015uup}, in which they give a
gaussian prior for $\theta_{23}$. It also differs from the fitting method in
Ref.\ \cite{HKDR}, where they fit $\theta_{13}$, $\theta_{23}$ and the value of
$\Delta m^2_{31}$ without implementing any priors, but fix $\theta_{12}$,
$\Delta m^2_{21}$ and the mass ordering. 
In Ref.\ \cite{Fukasawa:2016yue}, the details of the fitting process are not
specified. 
Despite these differences, we see qualitatively similar behaviour between the
three sets of results.
We find the regions of $\theta_{23}$ where the octant cannot be determined at
$5\sigma$ to be $\theta_{23}\in [43^\circ,49.7^\circ]$,
$\theta_{23}\in[42^\circ,48.9^\circ]$, and $\theta_{23}\in[43^\circ,48.7^\circ]$ for DUNE, T2HK, and their combination,
respectively. 
In Fig.\ 3.18 of Ref.\ \cite{Acciarri:2015uup}, the equivalent region for DUNE
is $\theta_{23}\in[41^\circ,50^\circ]$, which is comparable to our work. 
In the middle panels of Fig.\ 5 in Ref.\ \cite{Fukasawa:2016yue}, the authors
estimate the region as $42.5^\circ<\theta_{23}<48.5^\circ$ for T2HK and the
combination of DUNE and T2HK, while for DUNE alone the range is slightly
smaller than in our simulation at $42^\circ<\theta_{23}<49^\circ$. 
Compare to our results, in Fig.\ 125 of Ref.\ \cite{HKDR}, we find the bigger range
at $5\sigma$ level is $0.44<\sin^2\theta_{23}<0.58$.

\begin{figure}[t]
\centering
\includegraphics[width=0.49\textwidth]{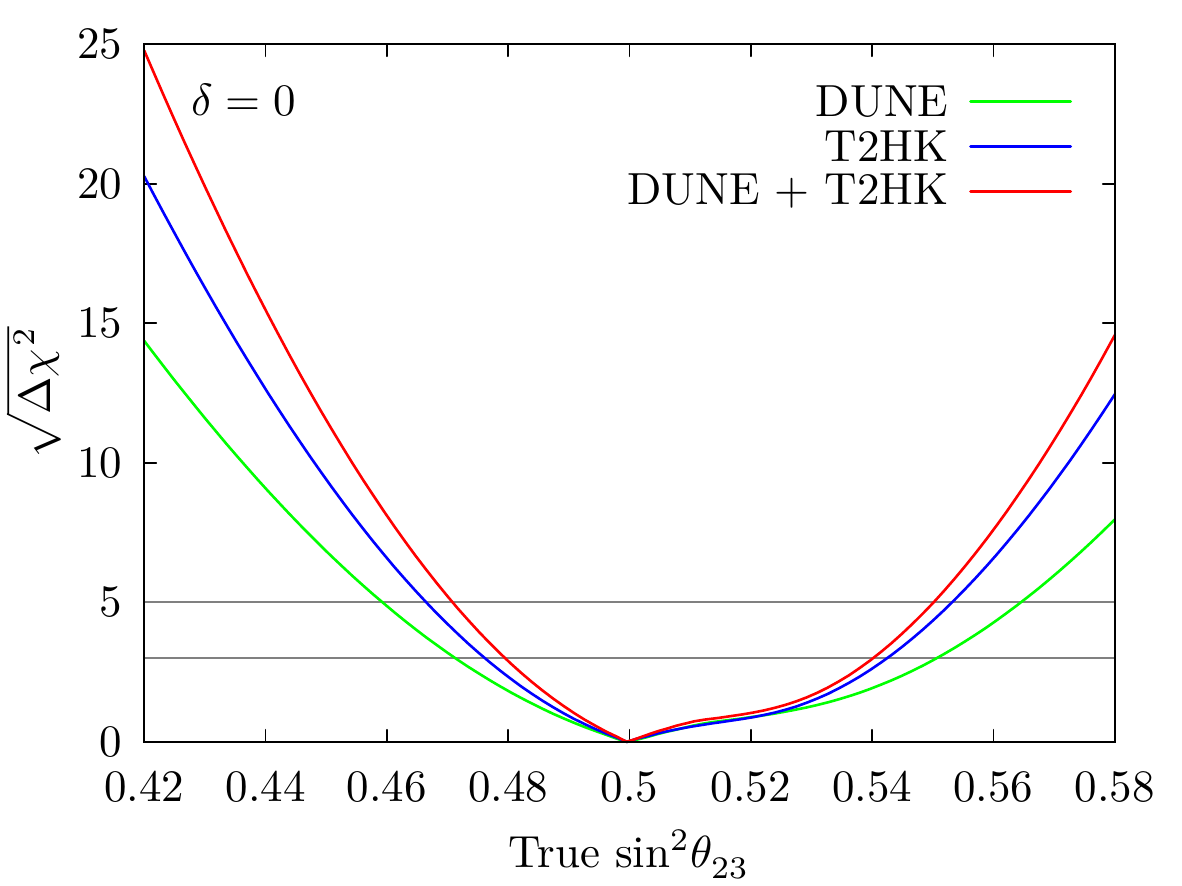}
\includegraphics[width=0.49\textwidth]{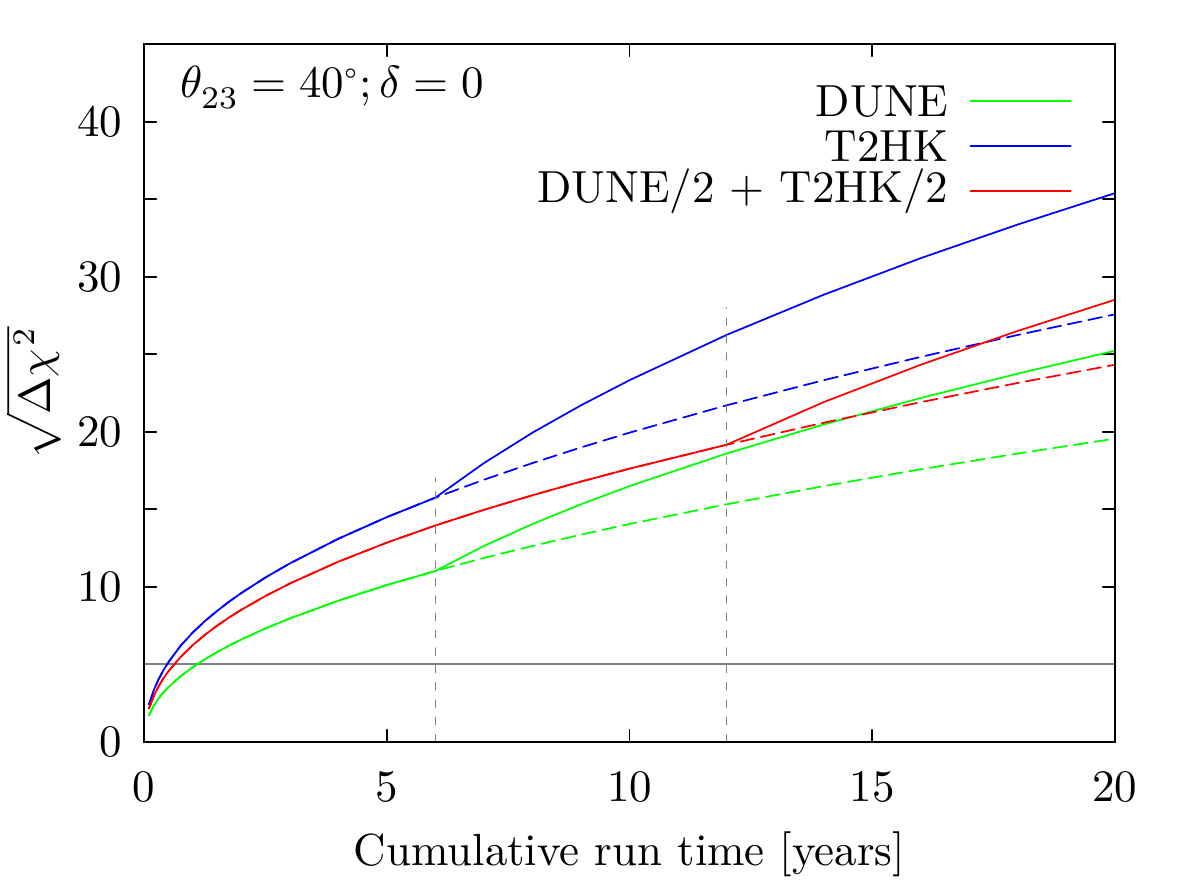}
\caption{\label{fig:MT23} The ability to exclude $\theta_{23}=45^\circ$ for 
DUNE, T2HK and their combination, against the true
value of $\sin^2\theta_{23}$ (left) and the cumulative run time (right). These plots assume 
$\delta=0$ and normal mass ordering. 
The left (right) plot assumes the ``fixed run time'' (``variable run time'')
configurations in \reftab{tab:standard_run_times} and the true
oscillation parameters, apart from $\theta_{23}$, specified in
\reftab{tab:global_fit_parameters}. 
}
\end{figure}

In \reffig{fig:MT23}, we show the analogous plots for the exclusion of maximal
$\theta_{23}$. We see that maximal $\theta_{23}$ can generally be excluded at
greater significance than the octant. 
T2HK can reach $5\sigma$ sensitivity for
$\sin^2\theta_{23}\lesssim0.47$ as well as for $\sin^2\theta_{23}\gtrsim 0.55$,
while DUNE can make an exclusion at the same statistical significance for
$\sin^2\theta_{23}\lesssim0.45$ and $\sin^2\theta_{23}\gtrsim0.56$.
Due to its poorer sensitivity, DUNE plays less of a role in the combination and
DUNE + T2HK follows the sensitivity of T2HK.
On the right, we show the sensitivity against cumulative run time.
Again, the combination of DUNE + T2HK performs similarly to T2HK when the
cumulative run time is divided by two, while DUNE performs slightly worse.  We
see that the staging of T2HK and DUNE plays a notable role, leading to
significantly higher sensitivities.

\begin{figure}[t]
\centering
\includegraphics[width=0.8\textwidth, clip, trim = 0 1 3 3]{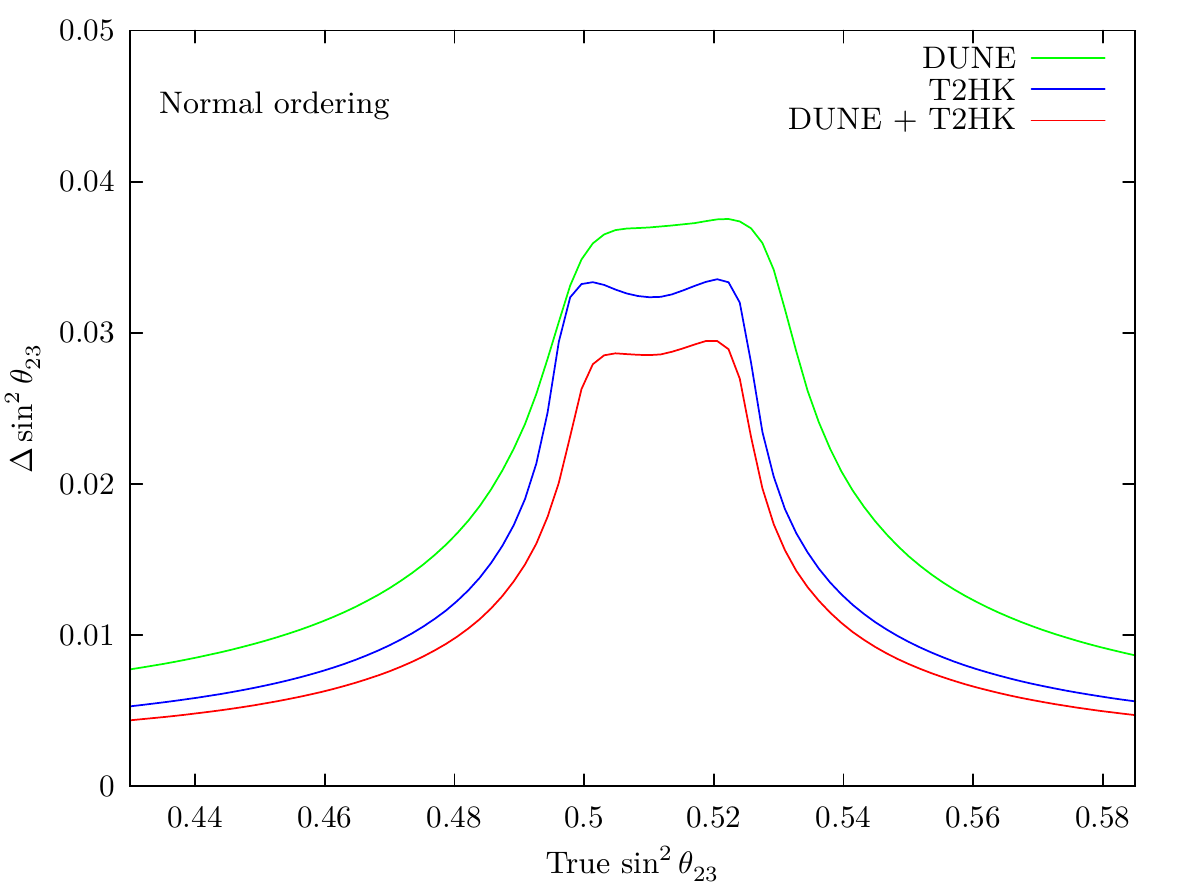}
\caption{\label{fig:T23} The expected 1$\sigma$ precision on
$\sin^2\theta_{23}$ as a function of true value of $\sin^2\theta_{23}$ from
$0.43$ to $0.585$ for DUNE, T2HK, and their combination, under the assumption
of normal ordering.
This plot assumes the ``fixed run time'' configurations in
\reftab{tab:standard_run_times} and the true oscillation parameters, apart from
$\theta_{23}$, specified in \reftab{tab:global_fit_parameters}. 
} 
\end{figure}

\begin{figure}[t]
\centering
\includegraphics[width=0.8\textwidth, clip, trim=1 10 35 20]{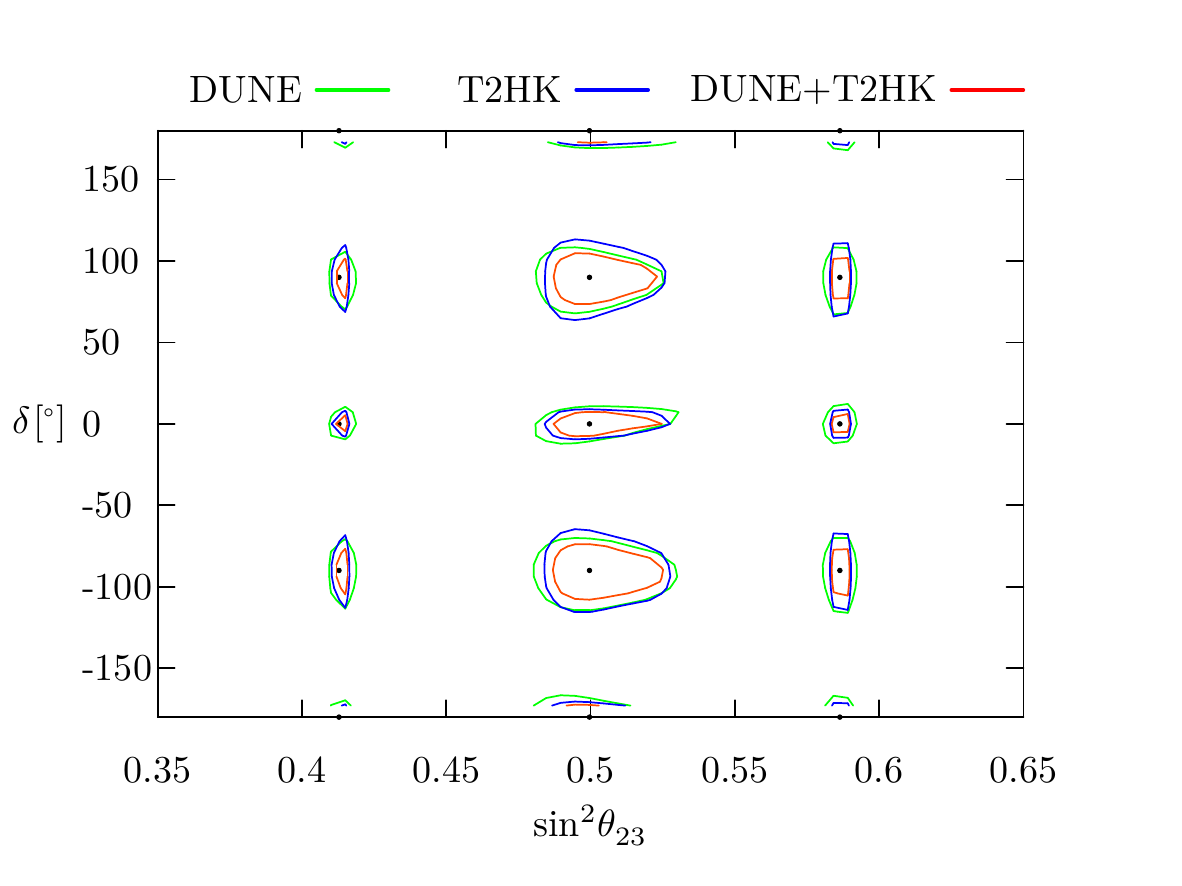}
\caption{\label{fig:T23_delta}The attainable 1$\sigma$ precision on
$\sin^2\theta_{23}$ and $\delta$ for DUNE, T2HK, and their combination. In each
case, the contours enclose the assumed true values for $\theta_{23}$ and
$\delta$, marked with a point.  
This plot assumes the ``fixed run time'' configurations in
\reftab{tab:standard_run_times} and the true oscillation
parameters, apart from $\theta_{23}$, specified in
\reftab{tab:global_fit_parameters}. 
} 
\end{figure}

We study the attainable precision on $\sin^2\theta_{23}$ in \reffig{fig:T23},
where we plot $\Delta(\sin^2\theta_{23})$ against the true value of
$\sin^2\theta_{23}$ for normal mass ordering. For all configurations, we see
the same behaviour: the uncertainty climbs up from about
$\sin^2\theta_{23}=0.48$ and falls down around $\sin^2\theta_{23}=0.54$,
peaking at $\sin^2\theta_{23}\sim0.51$. This is expected for a measurement
dominated by the disappearance channel, where the probability is proportional
to $\sin^2(2\theta_{23})$ and a leading-order analytic treatment would imply
the relation
\[   \Delta(\sin^2\theta_{23}) \propto \left|\tan(2\theta_{23})\right|, \]
which naively predicts a total loss of sensitivity at maximal mixing, analogous
to $\Delta\delta$ at $\delta =\pi/2$. This is mitigated by higher-order
effects, as well as the information from the appearance channel, which becomes
important around these values.
The drop in sensitivity seen in \reffig{fig:T23} is quite sharp, and for values
of $\sin^2\theta_{23}$ away from maximal mixing there is only modest variation
in precision.  For DUNE, $\Delta(\sin^2\theta_{23})$ is about $0.009$ at the
boundaries, and peaks up to the value $\sim0.038$. T2HK has better performance,
with $\Delta(\sin^2\theta_{23}) \sim0.005$ for $\sin^2\theta_{23}=0.43$ and
$0.585$.  As with DUNE, the worst performance for T2HK is near the peak at
$\sin^2\theta_{23}=0.5$ with $\Delta(\sin^2\theta_{23})\sim0.032$. For significant
deviations from $\theta_{23}=45^\circ$, the combination of DUNE and T2HK performs
very similarly to T2HK, as T2HK's high sensitivity drives that of the
combination. However, the improvement of including DUNE data is viewable around
the peak of $\Delta(\sin^2\theta_{23})$.
In these plots, we set $\delta=0$, although qualitatively similar behaviour
holds for other choices. There is, however, a correlation between the precision
on $\theta_{23}$ and $\delta$. We present an estimate of the joint precision on
$\theta_{23}$ and $\delta$ attainable at DUNE and T2HK in
\reffig{fig:T23_delta}.  In this plot, each ellipse shows the $1\sigma$ allowed
region for a set of true values inside its boundary taken from the sets
$\delta\in\{0^\circ, \pm 90^\circ, \pm 180^\circ\}$ and
$\theta_{23}\in\{40^\circ, 45^\circ, 50^\circ\}$.
T2HK generally performs slightly better for this measurement; although, at
times DUNE achieves a marginally better sensitivity to $\delta$, and the
combination of additional data from DUNE helps to reduce the T2HK contours. 
The best measurements will be obtained for large deviations from
$\theta_{23}$-maximality and values of $\delta$ close to the CP conserving
values, where DUNE (T2HK) can expect precisions on $\theta_{23}$ of $\Delta
\theta_{23} = 0.2^\circ$ ($\Delta \theta_{23} = 0.13^\circ$).  Conversely, the
worst precision comes from the values of $\theta_{23}$ near maximal mixing
where DUNE (T2HK) can expect larger uncertainties with $\Delta \theta_{23} =
2^\circ$ ($\Delta \theta_{23} =0.95^\circ$). 
Comparing our result in \reffig{fig:T23_delta} to Fig.\ 123 in \cite{HKDR}, we
find that our value for $\Delta\sin^2\theta_{23}$ is better than the official
result for T2HK, which we suspect is due to the differences in our treatment of
external data as mentioned previously.

\section{\label{sec:delta}Complementarity for precision measurements of $\delta$}

\begin{figure}[t]
\centering
\includegraphics[width=0.8\textwidth]{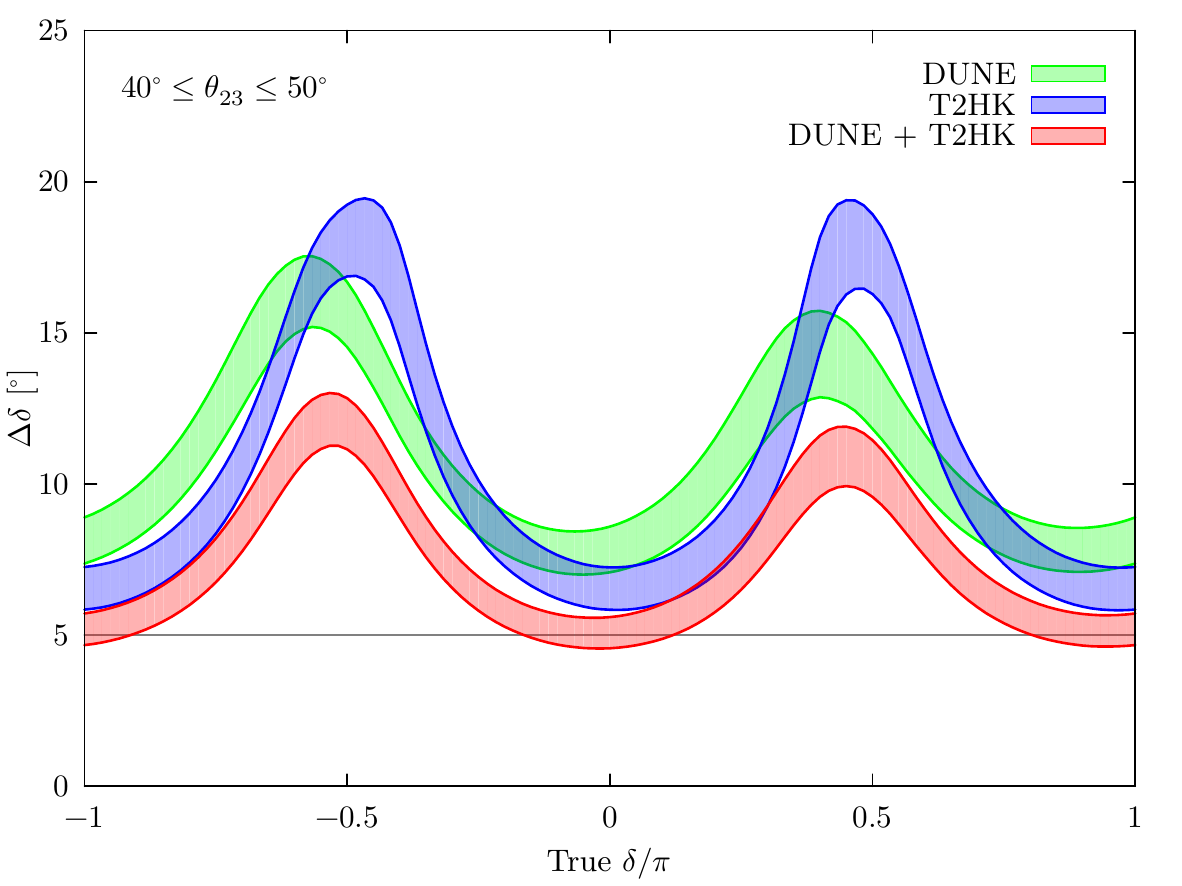}

\caption{\label{fig:both_Dd_th23}The $1\sigma$ precision on $\delta$ for DUNE
and T2HK in isolation and combination.  
This plot assumes the ``fixed run time'' configurations in
\reftab{tab:standard_run_times} and the true oscillation
parameters, apart from $\theta_{23}$, specified in
\reftab{tab:global_fit_parameters}.
}
\end{figure}

For the reasons outlined in \refsec{sec:precision}, we expect an interesting
interplay of sensitivities for a narrow-band and wide-band beam for the
determination of $\delta$. In this section, we study the complementarity of
DUNE and T2HK for precision measurements of $\delta$.
In \reffig{fig:both_Dd_th23}, we show the $1\sigma$ precision on $\delta$ which
is attainable by the standard configurations of DUNE and T2HK and their
combination.
We consider a range of true values of $\theta_{23}$ as this
significantly affects the ultimate precision. We see that for most of the
parameter space T2HK can attain a better precision, with values of $\delta$
between $6$ and $7^\circ$ for the CP conserving values of $\delta$ compared to
between $7.5$ and $9^\circ$ for DUNE. However, DUNE performs better
than T2HK for maximally CP violating values of $\delta$ up to $5^\circ$. 
This leads to an
effective complementarity between the two experiments, and their combined
sensitivity reduces $\Delta \delta$ as compared to the two experiments in
isolation by between $1$ and $6^\circ$ depending on the value of $\delta$.

We see therefore an improvement when combining the data from the two
experiments.  This was to be expected for a number of reasons. Firstly, there
is a simple statistical benefit of combination --- an increase in data reduces
the statistical uncertainty and allows for a more precise measurement. On top
of this, there is a synergistic benefit, where the two experiments mutually
improve the reconstruction of the parameter of interest. To try to understand
the synergy between DUNE and T2HK, we have run simulations where we mitigate
the statistical advantage through different normalization procedures so as to
expose the complementarity shown by the information available in each data set. 
As the experiments operate under such different assumptions, there is no
universal way to do this. 
There are many factors which influence an experiment's
sensitivity: for example, the total flux produced by the accelerator; the
effects of baseline distance on the flux; the detector's size, technology and
analysis efficiencies; not to mention the purely probabilistic effects of the
oscillation itself, which occurs over different baseline distances 
and at different energies.
In the next two sections, we consider different ways to normalise the
experiments which reveal different aspects of their sensitivities.

\subsection{Normalising by number of events}

We can remove the statistical advantage of combining two experiments by fixing
the number of events.  We will consider two ways of doing this, both based on
the total number of signal events $S$, composed of genuine appearance channel
events in the detectors. We define $S$ to be the sum of these events across
both neutrino and antineutrino mode appearance channels.

\begin{figure}[t] 

\centering \includegraphics[width=0.99\textwidth, clip, trim=8 20 5 60]{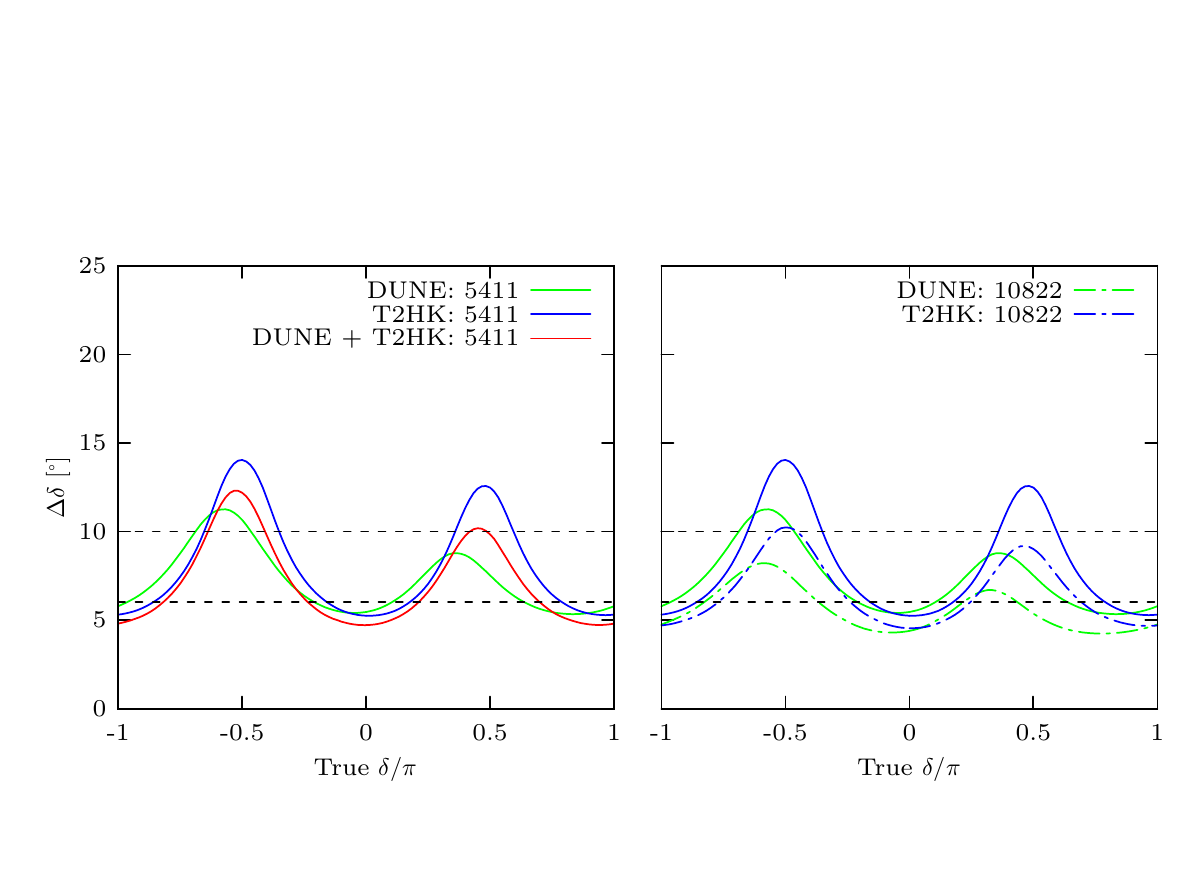} 
\caption{\label{fig:Dd_fix_S}Left: the precision attainable
by DUNE, T2HK and their combination with a fixed number (5411, the
average number expected by DUNE + T2HK) of appearance channel events. 
On the left, DUNE + T2HK denotes the ``fixed run time'' configuration in
\reftab{tab:standard_run_times}, which expects around 5411 events.
Right: the performance of DUNE and T2HK with double numbers of appearance
events (in brackets) compared to those with 5411 events. In both plots, all
unspecified parameters take the true values given in
\reftab{tab:global_fit_parameters}.
} 

\end{figure}

Our first normalization method fixes $S$. This is, of course, an unrealistic
goal in practice.  However, it answers an interesting hypothetical question:
would a given number of events be more informative if they came from DUNE or
T2HK?
We have run the simulation of T2HK and DUNE while fixing the number of events
in the appearance channel. This number varies with $\delta$, and so the
effective run time has been modified for each value of $\delta$ to keep the
observed events constant. In the left-hand panel of Fig.\ \ref{fig:Dd_fix_S},
we have fixed the number of appearance events to be 5411 for each
configuration, which is the average number of events expected for
the combination of DUNE and T2HK running for $20$ years cumulative run time.
We see that events at DUNE are more valuable than events at T2HK around
maximally CP violating values; however, around CP conserving values, the opposite
is true and T2HK has more valuable events.
We quantitatively assess this effect in the right-hand panel of
\reffig{fig:Dd_fix_S}. 
This plot compares the performance of DUNE and T2HK with a fixed 5411 events,
with the same experiments assuming double the number of events. The figure
shows that for DUNE to consistently outperform T2HK, it needs at least twice as
many events. The same is true to T2HK: it can only lead to better performance
for all values of $\delta$ once its has more than twice the exposure.
 
Our second normalization scheme is designed to include the effect of
the probability from the comparison with fixed event rates. The number of
appearance channel events, $S$, is to a good approximation proportional to the
oscillation probability,
\[   S \propto P(\nu_\mu \to \nu_e; \langle E\rangle), \]
where $\langle E\rangle$ denotes the average energy of the flux,
and we introduce a quantity $N$ denoting signal events with the effects due to the probability
removed,
\begin{equation}   N(\langle E\rangle) = S/P(\nu_\mu \to \nu_e; \langle E\rangle). \label{eq:N_def} \end{equation}
$N$ can be thought of as the constant of proportionality between
the number of signal events and the probability, and it is affected by many
factors, whose product is often referred to as the \emph{exposure} of the
experiment. These factors, such as run time, detector mass and power of the
accelerator, describe technical aspects of the experimental design and the
exposure is often taken as a proxy for run time in phenomenological studies of
neutrino oscillation experiments. However, there are other factors affecting
the coefficient $N$ such as the effects of cross-sections and detector
efficiencies, which also vary from experiment to experiment. Our definition of
$N$ accounts for all of the factors which affect the signal, apart from the
fundamental effect of the oscillation probability.
Equating $N$ assumes that all technical parameters are identical between the
two experiments, and allows us to study the effect of the oscillation
probability alone.
We find that fixing $N$\footnote{In practice, as we
are studying neutrino and antineutrino channels and our detector models have
binned energy spectra, we define an analogous quantity $N_i$ ($\overline{N}_i$)
for each energy $E_i$  ($\overline{E}_i$) in neutrino (antineutrino) mode. We
then define $N$ as the sum over $N_i + \overline{N}_i$.} leads
to little change from fixing $S$. DUNE still outperforms T2HK for values of
$\delta$ near maximal mixing, while T2HK performs best at CP conserving values.
Even isolating the effect of probability in this way, we arrive at the same
conclusion that events at DUNE are more informative about the value of $\delta$
than at T2HK around $\delta=\pm \pi/2$, while each event of T2HK has more
impact than when $\delta$ is CP conserving.

\begin{figure}[t] 

\centering \includegraphics[width=0.9\textwidth]{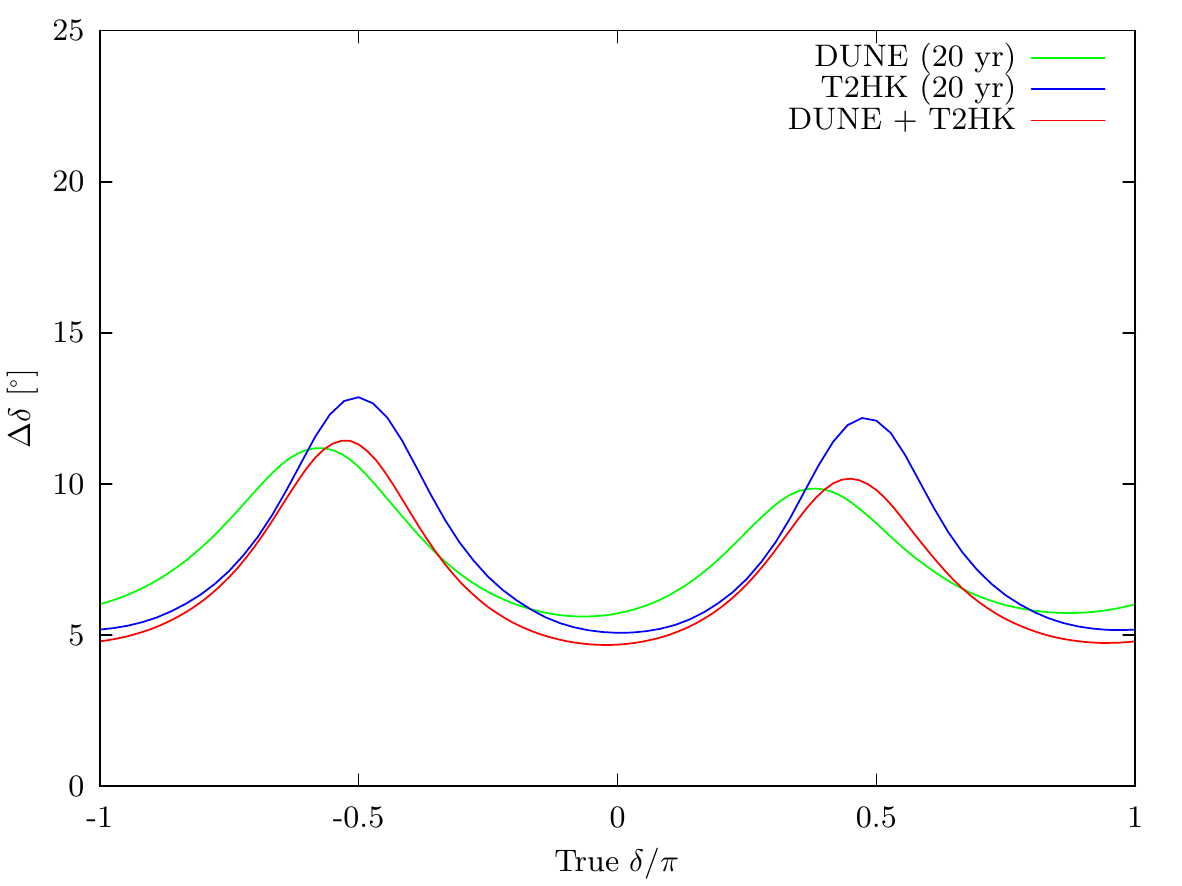}
\caption{\label{fig:Dd_runtime}The 1$\sigma$ precision on $\delta$ as a
function of the true value of $\delta$ for DUNE, T2HK and their combination
with the same cumulative run time of 20 years. 
The configuration of DUNE (20 yr) is defined by the ``variable
run time'' entry in \reftab{tab:standard_run_times}, with $T$ given in brackets
after the experiment's name, whereas DUNE + T2HK is the corresponding ``fixed
run time'' entry. Note that due to the staged upgrades of both
designs, DUNE (20 yr) and T2HK (20 yr) correspond to 6 years without the
planned upgrades followed by 14 years of upgraded running. 
This plot assumes normal mass ordering and all other unspecified true
parameters are given in \reftab{tab:global_fit_parameters}.}

\end{figure}

Comparing the expected precision on $\delta$ under our different normalization
conditions gives us an idea of the role played by the probability. We see that
generally, the conclusions are the same: when arranged to have equal
normalizations, T2HK does worse than DUNE for maximal CP violation, but
performs better at $\delta=0$ and $\pi$. This is true even if probability
is included in the normalization, so we infer the difference in performance
really does come from the spectrum. 
We conclude this section by noting that both normalization methods highlight
the same aspect of the two experiments: for equal events the two experiments
are very complementary, each providing the best measurement of $\delta$ for
around half of the parameter space. However, in its standard configuration,
DUNE expects fewer events than T2HK in the appearance channels. We will study
this in more detail in the next section.

\subsection{Normalising by run time}

Of course, one of the most pragmatic ways to normalise the experiments is by
run time. Would a decade of both experiments running in parallel be better than
two consecutive decades of DUNE (or T2HK)? To make this comparison, we assume
the same cumulative run time for the experiments running alone, and in
combination. In \reffig{fig:Dd_runtime} we show the results of our simulation.
The combination of DUNE and T2HK generally outperforms either
experiment running for twice as long. However, there are some small regions of
parameter space around maximal CP violating values of $\delta$ where $20$ years
of DUNE outperforms not only T2HK but also the combination of DUNE and T2HK. 
At these values of $\delta$, DUNE's wide-band beam performs best by
incorporating information from other energies. We also see this benefit in the
combination of DUNE and T2HK, which notably outperforms $20$ years of T2HK at
these values. 
This result tells us that the combination offers two advantages. First, running
the experiments in parallel allows us to collect two decades of data in half
the calendar time. This explains a significant part of the sensitivity
improvement; however, there is also a complementarity arising from the
different sensitivities of the two experiments. This is especially marked for
this measurement around the maximally CP violating values of $\delta$.

\begin{figure}[t] 
\centering
\includegraphics[width=0.9\textwidth,clip,trim=18 30 3 70]{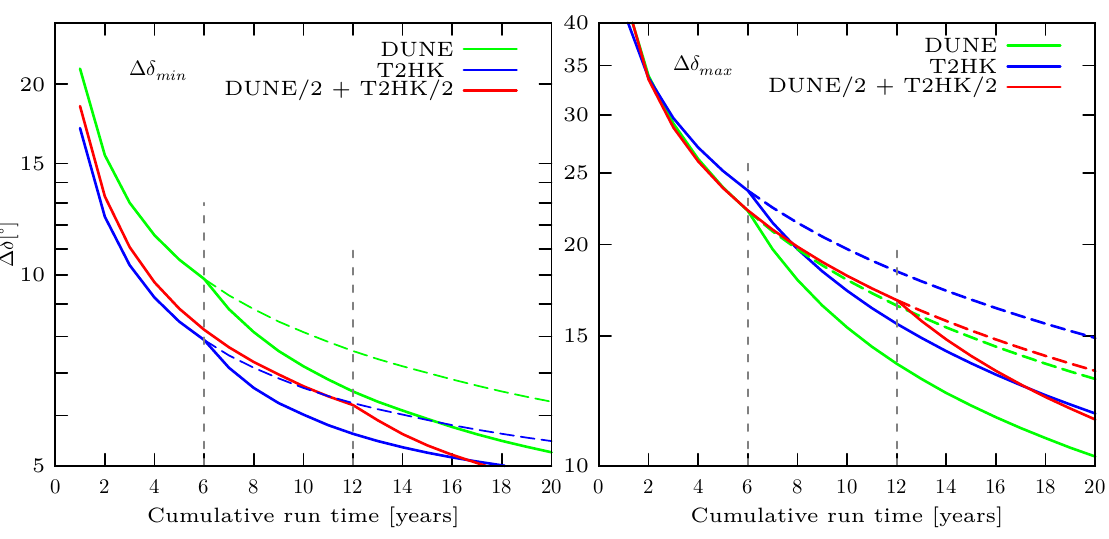}

\caption{\label{fig:Dd_exposure}$\Delta\delta_\text{min}$ (left)
and $\Delta\delta_\text{max}$ (right) at DUNE, T2HK and their combination as a
function of run time. These plots assume the ``variable run time''
configurations in \reftab{tab:standard_run_times} and the true oscillation
parameters appropriate for normal ordering as given in
\reftab{tab:global_fit_parameters}. We have checked that similar behaviour
obtains for inverted ordering.}

\end{figure}

The behaviour of $\Delta \delta$ for different experimental configurations as a
function of run time is shown in \reffig{fig:Dd_exposure}. We have
studied this for the maximum and the minimum values of $\Delta\delta$ (denoted
$\Delta\delta_\text{max}$ and $\Delta\delta_\text{min}$), which describe the
extremes of performance for the two experiments. We find that $\Delta
\delta_\text{max}$ is better at DUNE than T2HK for all run times, whereas the
situation is reversed for $\Delta \delta_\text{min}$. We note that for both
experiments, the staged upgrades lead to a strong improvement in the
sensitivity. If run in parallel, the combination of DUNE and T2HK expects
$\Delta\delta_\text{min}< 5^\circ$ and $\Delta \delta_\text{max}
\lesssim 11^\circ$ after 10 years.

To end this section, we compare the performance of the two experiments and
their combination through the minimal exposures required to obtain certain
physics goals.
In \reftab{tab:Dd_norm}, we show the value of $N$, see \refeq{eq:N_def}, the
number of signal events $S$ and the cumulative run time required to reach a
precision on $\delta$ of $10^\circ$ for $\Delta\delta_\text{max}$
and $\Delta\delta_\text{min}$. 
It is clear from our study in this section that to achieve a precision of
$10^\circ$ for $\Delta\delta_\text{max}$ will be a challenging measurement:
above $20$ years of data is necessary, requiring $12.5$ years of both
experiments running in parallel. For $\Delta\delta_\text{min}$ this is,
however, a feasible goal.  DUNE expects a similar measurement after a full
$5.8$ year data-taking period, while T2HK can achieve this goal in $3.3$ years.
The combination of DUNE and T2HK marginally improves on this, requiring only
$1.9$ years of parallel running.
 
\begin{table}[t]
\centering
\begin{tabular}{l|l|l|l|l|l|l|}
\cline{2-7}
~ & \multicolumn{3}{c|}{$\Delta\delta_{min}$} &\multicolumn{3}{c|}{$\Delta\delta_{max}$}  \\ \cline{2-7}
~ & DUNE  & T2HK  & Both  & DUNE  & T2HK & Both  \\ \cline{1-7}
\multicolumn{1}{|l|}{$\delta$} &  $354^\circ$ &  $0^\circ$ & $0^\circ$ & $255^\circ$ & $270^\circ$   & $264^\circ$ \\ \cline{1-7}
\multicolumn{1}{|l|}{$N$} &  26837 &   15868 &  21900 & 167497 & 332532  & 218995 \\ \cline{1-7}
\multicolumn{1}{|l|}{$S$}&  961 & 1034  & 739  & 6811  & 15653  & 8124 \\ \cline{1-7}
\multicolumn{1}{|l|}{Cumulative run time [years]} & 5.8  &3.3  & 3.8 & 21.1  & 27.1  & 25 \\ \cline{1-7}
\end{tabular}

\caption{\label{tab:Dd_norm}Exposures required for
$\Delta\delta_\text{max}$ and $\Delta \delta_\text{min}$ to reach $10^\circ$.
T2HK has the best precision on reasonable time scales due to its very high
event rate especially at $\delta=\pi$. DUNE marginally out performs T2HK for
maximally violating values of $\delta$. The year shown in this table, assumes
the ``variable run time'' configurations of \reftab{tab:standard_run_times}. The
combination ``Both'' assumes a scaling of the standard configuration of DUNE/2
+ T2HK/2.}

\end{table}
\subsection{Impact of systematic errors}

In the previous section, we have looked at the precision on $\delta$ under a
number of different assumptions. 
We have seen that T2HK has a larger number of events than DUNE, and for
the majority of the parameter space this leads to a
better expected precision on $\delta$. This means that the relationship
between statistical and systematic uncertainty will be quite different at the
different experiments and our assumptions about systematics, always a
contentious issue, may be significant.  In this section we try to understand
these effects and explore the impact on the expected precision on $\delta$
under differing systematics assumptions for the combination of DUNE and T2HK.

\begin{figure}[t] 

\centering \includegraphics[width=0.99\textwidth, clip, trim=10 20 5 60]{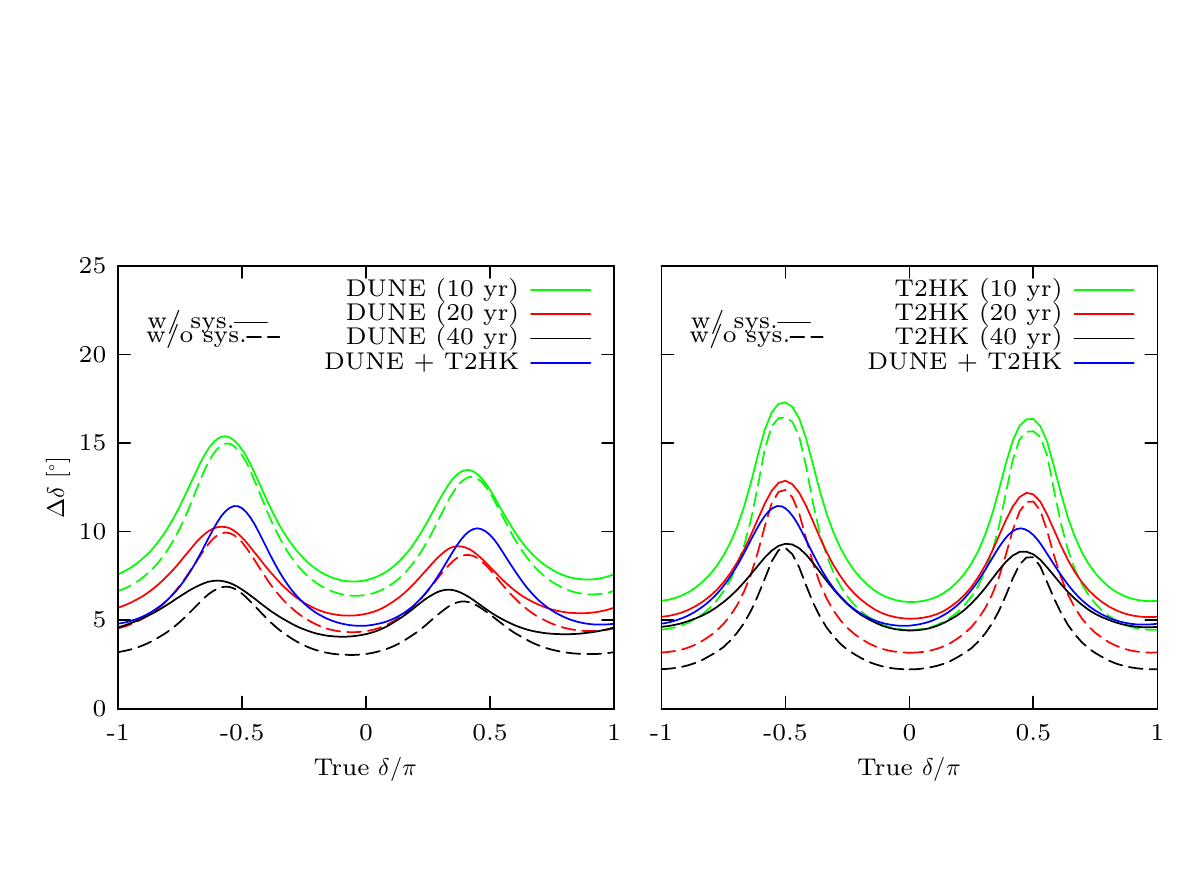} 
\caption{\label{fig:various_runtimes}Left (right): the
expected 1$\sigma$ precision on $\delta$ for DUNE (T2HK) with different run
times with and without systematics (solid and dashed, respectively) compared to
a reference design of our ``fixed run time'' configuration of DUNE + T2HK from
\reftab{tab:standard_run_times}. Note that in all cases, the experiments in
isolation have a staged upgrade after 6 years, and so see increasingly long
periods of upgraded running.  
}

\end{figure}

We can get a feel for the relevance of statistical versus systematic
uncertainty by seeing how the sensitivity scales with run time. In our model of
the systematics, we only consider effective signal and background normalisation
systematics for both DUNE and T2HK. In \reffig{fig:various_runtimes}, we show
the sensitivity to $\delta$ for different run times of the two experiments in
isolation, with and without systematic uncertainties. We see that there is
little impact from the systematic uncertainty at DUNE, and it continues to
further its sensitivity as we increase its run time.  This effect is quite
different for T2HK where systematics clearly have a more important role; for CP
conserving values, there is only modest improvement in
sensitivity after extensions of the experiment run time by a factor of 4.
This result neatly shows that DUNE is statistically limited while T2HK has more
reliance on its systematic assumptions (except for maximally CP violating
values of $\delta$). It is interesting to note that in both cases,
even after large increases in exposure, neither DUNE nor T2HK taken as a single
experiment can significantly improve on the sensitivity at CP conserving values
found by the combination of DUNE and T2HK running for only 10 years each.

\begin{figure}[t] 
\centering \includegraphics[width=0.89\textwidth]{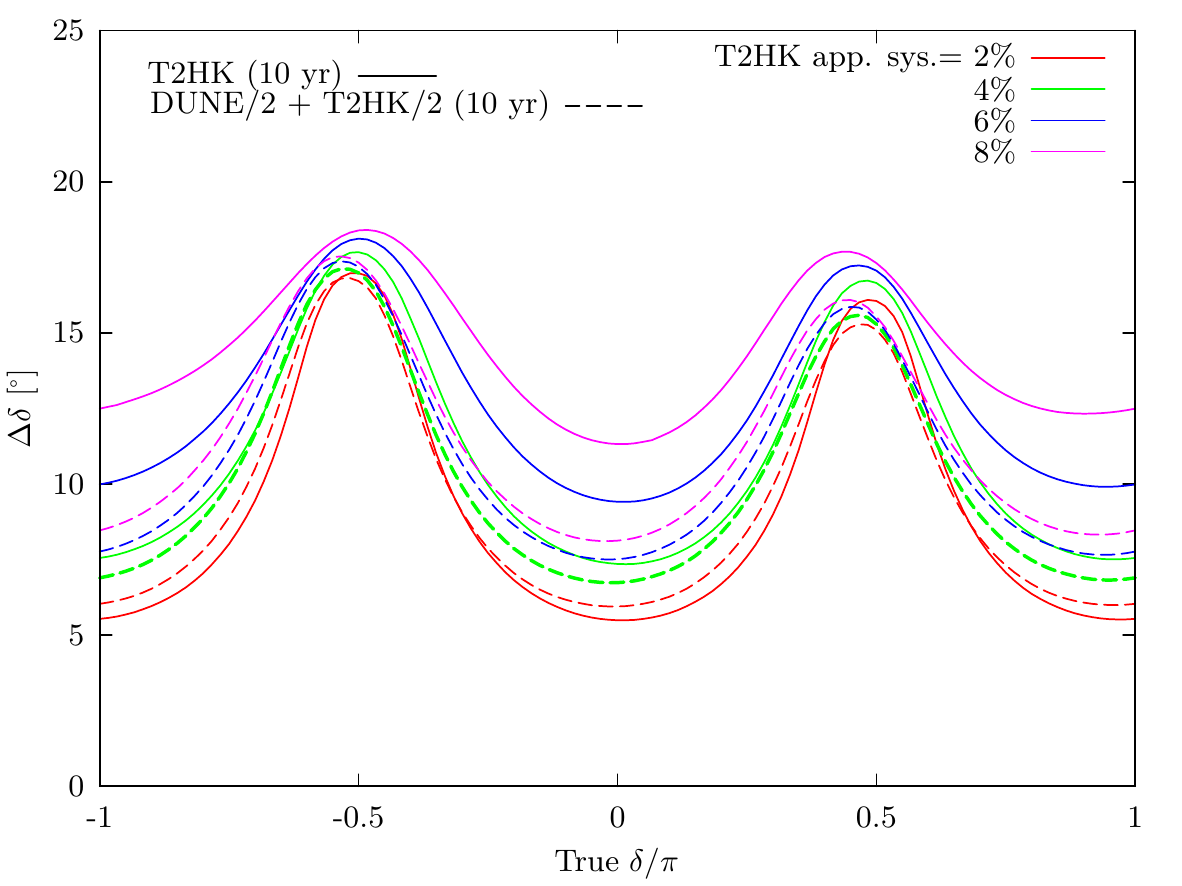}
\caption{\label{fig:systematics}$\Delta\delta$ for T2HK and the
combination of DUNE/2 + T2HK/2 each with 10 years cumulative run time for different
normalization systematic uncertainties on the appearance channel in T2HK (2\%,
4\%, 6\%, and 8\%). We hold the normalization systematics at 2\% for the
appearance channels of DUNE. 
The configurations in this plot are labelled ``variable run time'' in
\reftab{tab:standard_run_times} with the cumulative run time denoted in
brackets after their names.
This plot assumes normal ordering, but all other true parameters follow
\reftab{tab:global_fit_parameters}.} 
\end{figure}

Due to the limiting effect of systematic uncertainties suspected at T2HK, we
can expect that its performance is quite sensitive to our assumptions. To
understand how the combination of DUNE and T2HK can help reduce this
sensitivity, we have run simulations while varying the value of the
normalization systematics in T2HK. We study the case of 2\%, 4\%, 6\% and 8\%
normalization uncertainty at T2HK for the combination of DUNE and T2HK in
comparison to T2HK running for 10 years with the same systematic assumptions.
The results are shown in \reffig{fig:systematics}. We see that for $2\%$
systematic uncertainty, around $\delta=0$ and $\pi$, T2HK dominates the
precision on $\delta$ and is limited strongly by the systematics, meaning that
doubling the run time leads to scant improvement. As the systematic uncertainty
on T2HK increases, we see more of an advantage of including DUNE. Although at
$4\%$ systematics the lines are almost identical, for $6\%$ systematics the
improvement in precision at $\delta=0$ is around $2^\circ$ (an improvement of
around $10\%$). 
We conclude that T2HK is systematically limited around CP conserving values of
$\delta$, and including DUNE data can help to mitigate the effect of larger
uncertainties. At maximally CP violating value of $\delta$, we see little
impact of our systematic assumptions.
 
\section{\label{sec:alternative_designs}Impact of potential alternative designs}

As part of their continual optimisation work, both the DUNE and T2HK
collaborations have considered modifications of their reference designs, aiming
to further the physics reach of their experiments. As mentioned in
\refsec{sec:details_DUNE}, DUNE has considered an optimised beam based on a
3-horn design, and a novel beam concept, nuPIL. 
For T2HK, the redesign efforts are focused on the location of the second tank.
Originally foreseen as being installed at Kamioka 6 years after the experiment
started to take data, the possibility of installing the detector
in southern Korea has been mooted \cite{Hagiwara:2005pe, Hagiwara:2006vn,
Kajita:2006bt, Ishitsuka:2005qi}.
In this section, we discuss the impact of these redesigns on the physics reach
of the experiments, both alone and in combination, via the results of our
phenomenological discussion and simulations. We focus on the mass ordering, CPV
discovery, MCP and precision measurements of $\delta$. We point out that we do
not discuss measurements of $\theta_{23}$ further, as we have found that there
is little difference between the alternative designs under consideration.

\subsection{\label{sec:alt_run_times}Experimental run times and $\nu$ : $\overline{\nu}$ ratios}

In all plots that follow, we assume that DUNE and its variants will run with
equal time allocated to neutrino and antineutrino mode, while T2HK and T2HKK
will always follow the 1:3 ratio of their standard configuration. We also
assume that there is no staged implementation of any of the variants of T2HKK,
and that both detector modules start collecting data at the same time. For DUNE
and the lines labelled T2HK, we assume our standard configurations which
implement a staged upgrade at 6 years. Note that this means that when comparing
T2HKK with DUNE or the single-tank T2HK, T2HKK benefits from an increase in
exposure.

The run time configurations for these alternative designs follow those of the
``variable run time'' options in \reftab{tab:standard_run_times}, albeit with
variant fluxes for each experiment. All variants of DUNE, T2HK and T2HKK when
run on their own are assumed to have a cumulative run time of 10 years. When a
variant of DUNE is run in combination with a variant of T2HK, we assume that
the cumulative run time is divided equally between the two experiments in the
same way as DUNE/2 + T2HK/2 in \reftab{tab:standard_run_times}. This means that
when not plotted against $T$, the combination of DUNE and T2HK will have
$T=20$, corresponding to 10 years running time for each of the two experiments.

\subsection{Mass ordering}

\begin{figure}[t]
\centering{
\includegraphics[width=0.95\textwidth]{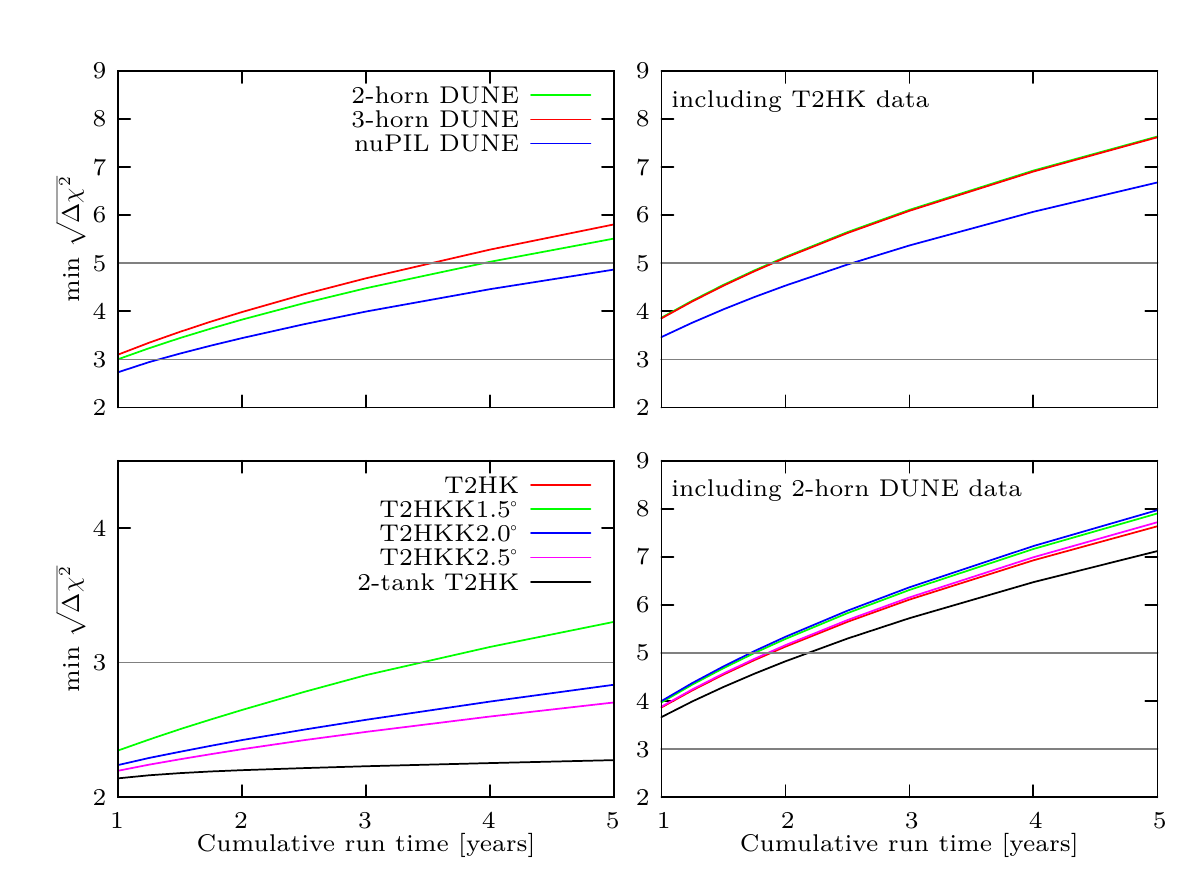}}
\caption{\label{fig:DUNE_redesign_MOtoyear_DUNE} Top (bottom) row: The minimum
statistical significance of mass ordering discrimination for DUNE (T2HK) with
various beam designs. 
On both rows, the left-hand panels show the performance of the alternative
designs in isolation, while the right-hand panels show the impact of an
alternative design on the combination of DUNE and T2HK by incorporating the
standard T2HK and DUNE designs on the top and bottom rows, respectively. The
configurations assumed here are described in \refsec{sec:alt_run_times} and the
true oscillation parameters are given in \reftab{tab:global_fit_parameters}.
Full details of the assumed exposures can be found at the start of \refsec{sec:alt_run_times}, and that in the top-right panel, the blue and green lines overlap. }
\end{figure}

As shown for the standard configurations in \refsec{sec:MO}, identifying the
mass ordering is almost guaranteed for experiments on this scale. However, we
see a large difference in performance between DUNE and T2HK due to the
difference in baseline distance. The alternative beams of the DUNE
collaboration do little to change this picture. The results of our simulation
are shown in \reffig{fig:DUNE_redesign_MOtoyear_DUNE}, in which we show the
minimum sensitivity to the mass ordering as a function of cumulative run time. 
The left column of panels shows the performance of the alternative designs for
DUNE (top) and T2HK (bottom). 
We see that for DUNE, the 3-horn and 2-horn designs do better at the minimum
sensitivity by about $1\sigma$ compared to the nuPIL design.
We see that the 3-horn design can reach greater than $5\sigma$ significance
after around $3.3$ years run time, while the 2-horn design achieves the same
significance after around 4 years, and  nuPIL requires above $5$ years.
For T2HK and its alternative designs the picture is quite different.  The T2HK
design cannot achieve sensitivity above $2\sigma$ for these run times. However,
placing a second tank in Korea will allow T2HKK to see larger matter effects
over the $1000$--$1200$ km baseline: the sure-fire way to sensitivity to the
mass ordering.  Moreover, the possibility of placing the second detector at a
different off-axis angle, could produce a wider beam, or a narrow beam whose
peak is shifted away from the first maximum.  This interplay of factors could
qualitatively alter our picture of mass ordering sensitivity at HK(K).
We see a greater variation in performance as the fluxes are varied, but as we
saw before, lower overall sensitivities. Due to the larger matter effects
associated with the Korean detector, we might expect increased sensitivity to
the mass ordering over the standard T2HK design; however, we do not see an
enhancement of this kind. 
We understand this effect as due in part to the limited data collected by T2HKK
at the longer baseline. Fewer events associated with neutrinos travelling the
longer baseline are detected as the beam suffers significant suppression due to
dispersion over the longer distance\footnote{The flux is dispersed by an
inverse square law as baseline increases; subsequently, a Korean detector sees
around $11\%$ of the flux seen at Kamioka.} as can be seen in
\reftab{tab:total_number}. With WC technology, we know that the advantage comes
from scale, and such a limitation on event numbers means that longer baselines
will not be competitive unless operated for a longer period of time. 
Moreover, the matter effect is relatively suppressed compared to the effect at
DUNE due to the lower energies of the J-PARC beam. And it has been shown in
\refref{Ishitsuka:2005qi} that it is not sufficient to allow for a separation
of the two degenerate solutions in all cases at fixed energies. However, the
most important contribution of a Korean second detector is the very different
spectral information it provides from a detector at Kamioka. This helps to
provide sensitivity to the ordering, and we see that the T2HKK$1.5^\circ$
option expects to push the sensitivity above $3\sigma$ after around $3$ years.
Although we do not show the full MO sensitivity against $\delta$ in
\reffig{fig:DUNE_redesign_MOtoyear_DUNE}, we can draw a limited comparison
between our work and Fig.\ 18 in Ref.\ \cite{T2HKK}. Our results find slightly
lower sensitivities: for T2HKK$1.5^\circ$, the difference is about $1\sigma$,
while for off-axis angles of $2.5^\circ$ and $2.0^\circ$ the difference is
smaller than $1\sigma$.

The sensitivity is seen to increase as the Korean detector is moved to smaller
off-axis angles. This can be explained by the different flux profiles of the
T2HKK options. As the detector is moved towards the beam axis, the events
sample the oscillation probability increasingly close to the first maximum.
This is where the mass ordering is most visible in the presence of matter effects
and we see an accordingly stronger discovery potential.

On the right column of \reffig{fig:DUNE_redesign_MOtoyear_DUNE}, we show how
the alternative designs impact the combination of the two experiments.
Including T2HK data reduces the difference in performance between the three
DUNE beam designs, which all expect a minimum sensitivity of $5\sigma$ after
about $2$ years. For T2HK, the inclusion of DUNE data, pushes the overall
sensitivity above $5\sigma$ for the first time, with an extra Korean detector,
DUNE + T2HKK expects a greater than $5\sigma$ measurement for all values of
$\delta$ with around $2$ years run time.

\subsection{CPV and MCP sensitivity}

\begin{figure}[t] 

\centering

\includegraphics[width=0.99\textwidth]{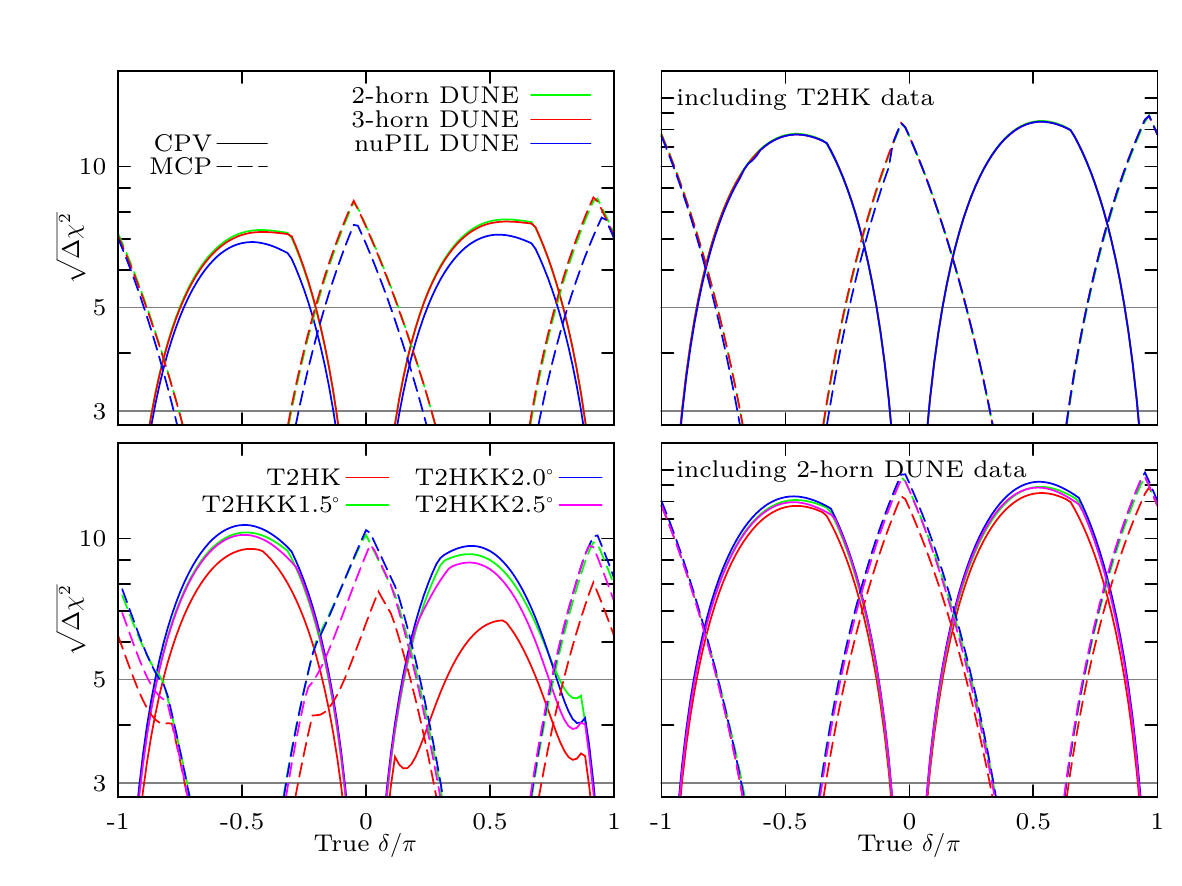}

\caption{\label{fig:Redesign_CPV_MCP} The sensitivity to CPV (solid) and MCP
(dashed) as a function of $\delta$ for various designs of DUNE (top row) and
T2HK (bottom row).  
The exposures assumed here are described in
\refsec{sec:alt_run_times} and the true oscillation parameters are
given in \reftab{tab:global_fit_parameters}.} 

\end{figure}

\begin{figure}[t]
\centering
\includegraphics[width=0.99\textwidth]{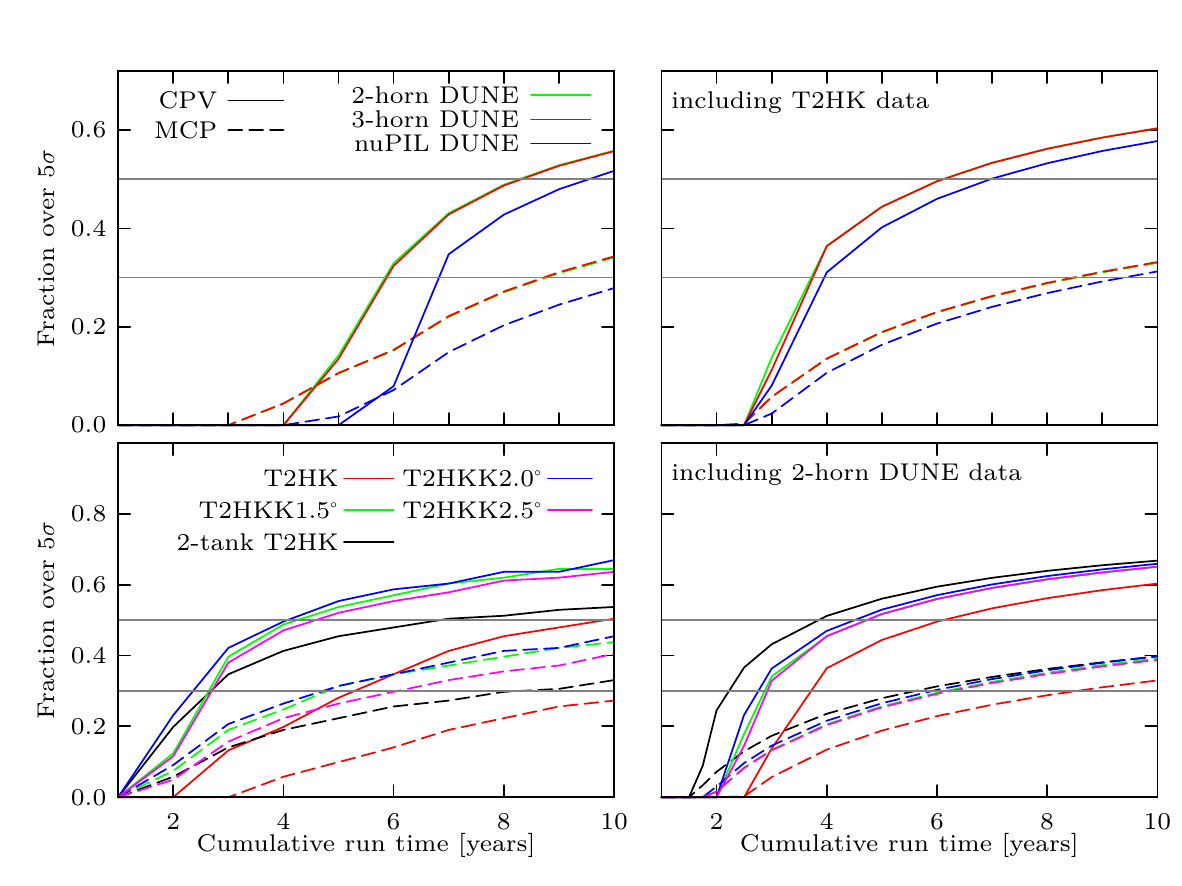}
\caption{\label{fig:Redesign_CPV_MCP_toyear}The fraction of true $\delta$ values 
for which we expect a CPV sensitivity (solid) and MCP sensitivity (dashed) over 5 $\sigma$, against
cumulative run time. 
The exposures assumed here are described in
\refsec{sec:alt_run_times} and the true oscillation parameters are
given in \reftab{tab:global_fit_parameters}.%
} 
\end{figure}

The sensitivity to CPV is understood to depend upon the energy of the events
observed, meaning that modifying the flux spectrum, for example with a narrower
beam from nuPIL or a beam located at the second maximum for T2HKK, could lead
to significant changes in the physics reach of the design. 
In the top-left panel of \reffig{fig:Redesign_CPV_MCP} we compare the
performance of the standard and alternative DUNE designs. CPV and MCP
sensitivities are shown for the three beam options as a function of $\delta$ in
solid and dashed lines, respectively. 
We find that the 2-horn and 3-horn designs perform similarly for CPV
and MCP measurements, and nuPIL performs slightly worse, by about $1\sigma$.
The top-right panel shows how these sensitivities are changed as information
from the standard configuration of T2HK is included. We see that due to T2HK's
strong sensitivity to the parameter $\delta$, the impact of alternative designs
for DUNE is greatly reduced. Maximal sensitivities to CPV of above $11\sigma$
are found for the maximal values of $\delta\in\{\frac{\pi}{2},
\frac{3\pi}{2}\}$.

For T2HKK we compare three off-axis angles for the Korean detector to the
standard configuration in the bottom row of \reffig{fig:Redesign_CPV_MCP}. On
the left panel, we show the performance of these alternative designs in
isolation. We see that the experiments perform comparably, but the best
performance comes from the T2HKK2.0$^\circ$ flux. As can be seen in
\reffig{fig:t2hk_fluxes_LE}, this flux is the best aligned with the second
maximum, suggesting that it is the access to events which sample this part of
the oscillation spectrum which lead to the increase in sensitivity.  The
increase in sensitivity for $-\pi\le\delta\le0$ is modest between T2HK and
T2HKK. We understand this again due to the suppression in event rates for a
Korean detector: although possessing valuable information, they are seen in
relatively small numbers, and their impact is limited.  However, there is a
notable difference for $0\le\delta\le\pi$, as the Korean detector helps to lift
the degeneracy which limits the performance of T2HK. 
In the bottom-right panel of \reffig{fig:Redesign_CPV_MCP}, we see the
sensitivity to CPV and MCP for combinations of DUNE and T2HKK.  In these
simulations, the degeneracy is lifted by the inclusion of DUNE data, and there
is little difference between the alternative designs for T2HKK
aside from an overall improvement in the sensitivities by between
$1$ and $2\sigma$. 

In \reffig{fig:Redesign_CPV_MCP_toyear}, we have computed the fraction of
values of $\delta$ for which CP conservation or maximal CP violation can be
excluded at greater than $5\sigma$ confidence. The top-left panel shows the
performance of the alternative DUNE beam designs in isolation. 
The 3-horn and 2-horn designs have almost identical sensitivities for all run
times, with a CPV fraction greater than that of nuPIL by between $10$--$30\%$
and an MCP fraction higher by around $10\%$. If we consider $30\%$ to be a
benchmark CPV fraction, the 3-horn and 2-horn designs expect to reach this
sensitivity after around $5$ years, while nuPIL takes around $7$ years.
Excluding MCP is a harder measurement for all beam designs, and exposures of
greater than $10$ years would be required to achieve a $30\%$ coverage of
$\delta$ parameter space at $5\sigma$. The top-right panel shows how the
alternative DUNE designs are affected by the inclusion of T2HK data.
Thanks to the good CPV and MCP sensitivity of T2HK, we see the improvement for
the combination, especially for nuPIL by up to $10\%$. 
We also find a relative suppression of the difference between variants ---
ultimately, DUNE offers less to this configuration and its precise design is
less important. These combinations expect to reach a CPV fraction of $30\%$
($50\%$) after about $4$ ($6$) years. For the exclusion of MCP, a $30\%$
fraction will be approximately reached after 9 years run time. 

The bottom row of \reffig{fig:Redesign_CPV_MCP_toyear}, shows analogous plots
for T2HK and T2HKK. On the left, these alternative designs are considered in
isolation, and we have also included a 2-tank T2HK line for comparison which
assumes two tanks collecting data at Kamioka from the start of the experiment.
There is very little difference between the T2HKK designs, although they all
show an increase in CPV and MCP fraction over the T2HK design.  
T2HKK expects a CPV fraction of over $50\%$ after less than $4$ years, while
T2HK requires around $10$ years for the same sensitivity (and 2-tank T2HK
around $7$ years). MCP fractions of greater than $30\%$ are possible after $5$
and $11$ years for T2HKK and T2HK, respectively.
Compared with the results shown in the upper panels in Fig.\ 20 in Ref.\
\cite{T2HKK}, we find the same ranking of designs. However, we also find
sensitivities around $2\sigma$ higher near $\delta=\pm\pi/2$. 
We suspect this quantitative difference is due to our priors, as in Ref.\
\cite{T2HKK}, it is pointed out that priors for $\delta_{CP}$, $\theta_{23}$
and $\Delta m^2_{31}$ are not implemented. However, we use priors on all
variables apart from $\delta$, and our simulation has slightly less leeway to
accommodate degenerate solutions, and a correspondingly improved ability to
exclude CP conserving parameter sets.
It is interesting to point out that, for both DUNE and T2HK, differences in
design have a greater impact on the highest sensitivity to CPV and MCP, as seen
in \reffig{fig:Redesign_CPV_MCP}, than on the long-term average performance
encapsulated in the CPV/MCP fraction at $5\sigma$. This can be seen in
\reffig{fig:Redesign_CPV_MCP} as the width of the sensitivity curves remaining
unchanged, while the peak is raised or lowered. The sensitivity of the peak
corresponds to different rising behaviour in
\reffig{fig:Redesign_CPV_MCP_toyear}, but the curves can be seen to quickly
plateau for T2HK. For DUNE, this effect is less marked, and suggests increasing
run time would still lead to increases in sensitivity.

On the right panel, we show the performance for the combination of DUNE data
with the T2HK variants. As in the bottom-left panel, we see that the T2HKK
designs perform similarly, with T2HKK2.0 performing marginally better.  The
inclusion of DUNE data here makes little change to the sensitivities. In fact,
as we define cumulative run time as the sum of the individual DUNE and T2HKK
run times, we see an apparent decrease in performance. Scaled appropriately for
parallel data collection, we find that DUNE + T2HKK expect a $5\sigma$ CPV
fraction of greater than $50\%$ after around $2$ years compared to $4$ years
for T2HKK alone. We note that there is a notable change in the
performance of the T2HK design with two tanks at Kamioka operated for the
duration of the experiment. Without DUNE data, this configuration performs more
poorly than the T2HKK designs; however, with the inclusion of DUNE data, it
becomes the best option. This can be understood as DUNE resolving the
degeneracy and T2HK maximising its CPV measurement by a large increase of
data at shorter baselines.

To conclude this section, we note that almost all of the experiments, when
running in isolation, can expect the exclusion of one of CP conservation or
maximal CP violation for all values of $\delta$ at $4\sigma$ and $5\sigma$ for
DUNE and T2HK variants, respectively.  This can be seen clearly in
\reffig{fig:Redesign_CPV_MCP}, where the intersections between CPV and MCP
lines are above the $3$ or $5\sigma$ horizontal lines. This is true for all
alternative designs, while the combination of DUNE and T2HK ensures that one of
these facts would be established with a significance greater than $6\sigma$.
The exception is for T2HK alone which, due to the degeneracy, falls short in
some some regions of parameter space.

\subsection{Precision on $\delta$}

\begin{figure}[t] 

\centering
\includegraphics[width=0.99\textwidth]{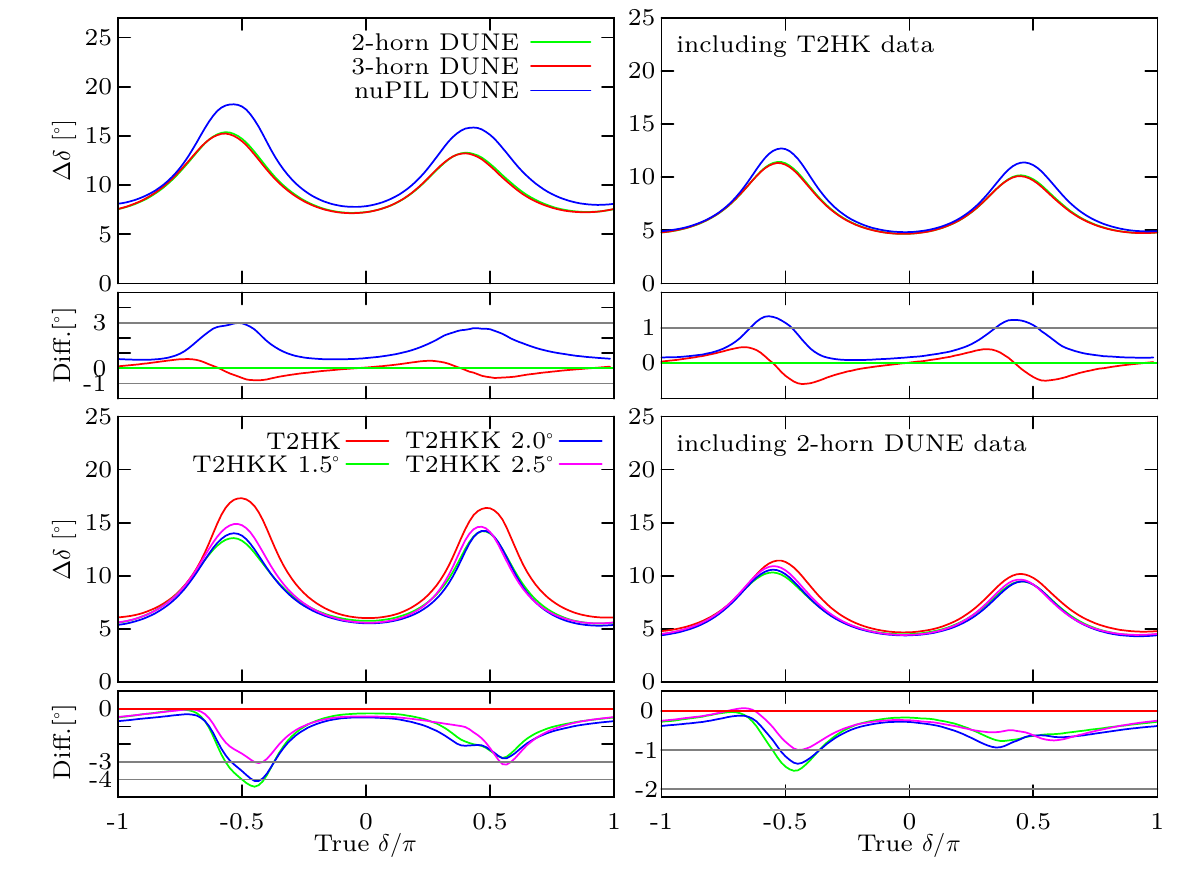}

\caption{\label{fig:Dd_redesign}The 1$\sigma$ precision on
$\delta$ for variants of DUNE (top row) and T2HK (bottom row). In the left
column, these designs are considered in isolation while on the right, we
combine variant designs of one experiment with the standard configuration of
the other. Our configurations are described in \refsec{sec:alt_run_times}.
These plots assume normal mass ordering and the remaining true
parameters are specified in \reftab{tab:global_fit_parameters}.} 

\end{figure}

We show the difference in $\Delta\delta$ for the alternative designs in the
left column of \reffig{fig:Dd_redesign}. 
We find that for DUNE, the 3-horn design works similarly to 2-horn design;
although, the 3-horn design performs slightly better in the 2nd and 4th
quadrant and for maximal CP violation,  while the 2-horn design expects smaller
$\Delta\delta$ in all other cases. 
These designs expect a precision on $\delta$ somewhere between $8$ and
$18^\circ$ after their full data taking period.  The performance of the nuPIL
design depends significantly on the true value of $\delta$. For values near
maximal CP violation $\delta = \pm\frac{\pi}{2}$, nuPIL performs worse than the
standard design. This can be understood due to the narrowing of the beam, which
when focused on first maximum, has insufficient events from other energies to
mitigate the poor sensitivity around maximal CP violating phases.
On the top-right panel of \reffig{fig:Dd_redesign}, we show the impact that the
DUNE redesigns have on the combination of DUNE and the standard configuration
of T2HK. As shown in \refsec{sec:delta}, data from T2HK improves the resolution
on $\delta$ for DUNE, and we see a correspondingly small impact of alternative
beam designs for DUNE. Notably, we do however see the worsening of performance
around maximal CP violating values of $\delta$ for the combination of nuPIL and
T2HK.

The expected sensitivity of $\Delta \delta$ for the alternative designs for
T2HKK are shown on the bottom-left panel of \reffig{fig:Dd_redesign}. Here we
see that all designs with a far detector allow for a significant improvement in
the precision on $\delta$, generally seeing the best performance coming from
the $1.5^\circ$ or $2.0^\circ$ off-axis angle fluxes. We see a slight loss of
performance for larger off-axis angles, which may be associated with the peak
of the flux falling beyond the second maximum into a region of hard to
identify, fast oscillations. 
Our result for $\Delta\delta$ is very close to that shown in the upper panels
of Fig.\ 23 in Ref.\ \cite{T2HKK}, and we agree on the ranking among
alternative designs. This is notable, given the differences induced by our
priors in other variables of interest, but is explained by the fact that our
priors differ in their global structure more than in their local structure. It
is this local structure which dictates $\Delta \delta$, as at low significance
the Gaussian approximation works well and multiple minima are irrelevant. 
On the right panel of \reffig{fig:Dd_redesign}, the combination is shown with
different T2HKK fluxes and the standard DUNE configuration. Once again, we see
that T2HKK dominates the combination, and therefore the shapes of these curves
closely follow those on the left panel.

\begin{figure}[t]
\centering
\includegraphics[width=0.7\textwidth]{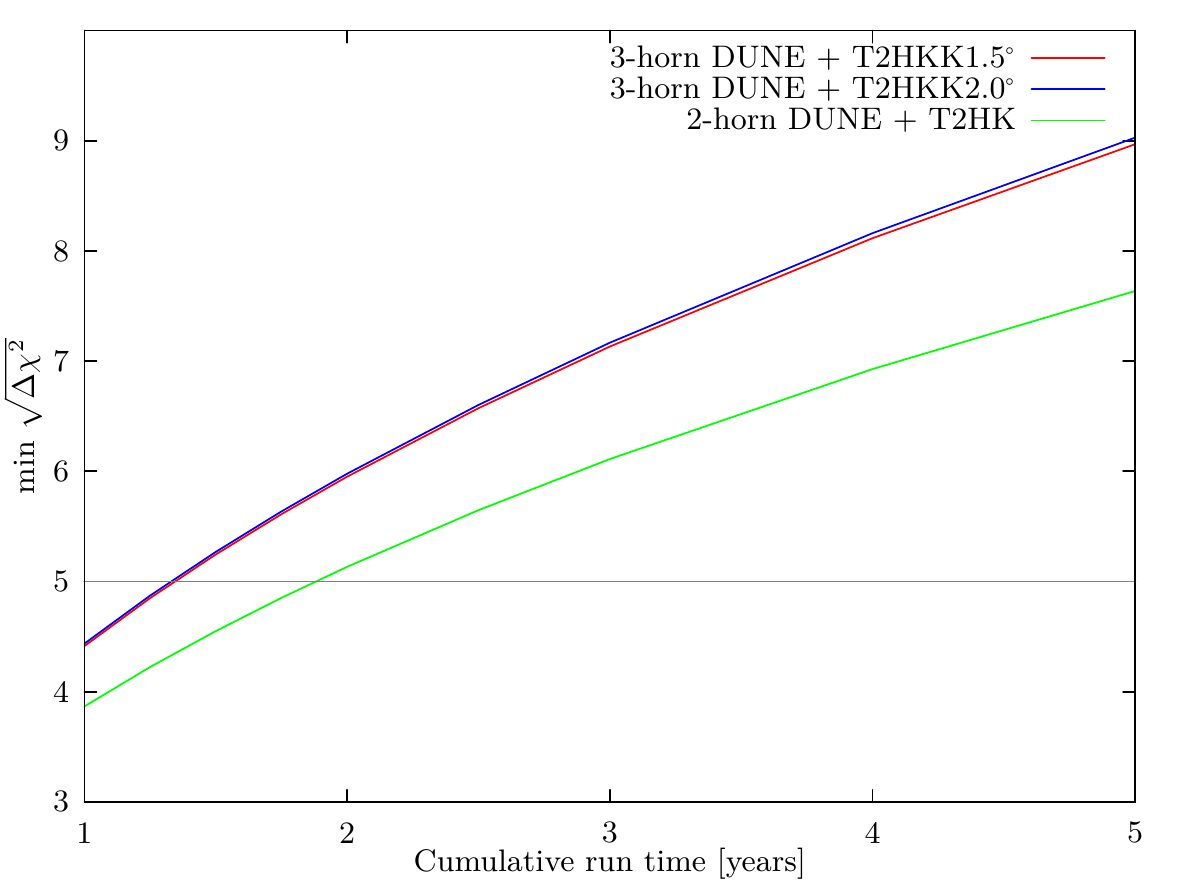}

\caption{\label{fig:optimised_MO}The minimum mass ordering sensitivity for
the combination of DUNE with the 3 horn flux and T2HKK$1.5^\circ$ (red)
and T2HKK$2.0^\circ$ (blue) compared with the standard configurations of DUNE
with 2-horn flux and T2HK with a single tank at Kamioka (green).
The configurations assumed here are described in
\refsec{sec:alt_run_times} and the true oscillation parameters are
given in \reftab{tab:global_fit_parameters}.
}

\end{figure}

\subsection{Optimal configuration}

In the preceding sections, we have studied how the alternative designs of T2HKK
and DUNE could impact the physics reach for key measurements, considering both
the experiments in isolation and in certain combinations. 
We have seen that for DUNE, the 2-horn and 3-horn designs perform similarly,
with the greatest difference occurring for the measurements of the mass ordering
and $\Delta\delta$. Both designs still expect very high significance
measurements of the mass ordering. However, as we see in
\reffig{fig:both_Dd_th23}, the 3-horn design can achieve marginally better
values of $\Delta\delta$ when $\delta$ in the 2nd and 4th quadrants, which is
where T2HK performs worse than DUNE. We therefore take the 3-horn design to be
the optimal choice for DUNE, with the 2-horn a close second.
T2HKK in contrast performs best with a flux positioned between $1.5$ and
$2.0^\circ$ degrees off axis. Here it maximizes its sensitivity to CP
violation, its ability to exclude maximal CP violation and to make precision
measurements of $\delta$ around CP conserving values. Whereas so far we have
only considered alternative designs for one experiment in combination with the
standard design of the other, in this section we report the physics reach of
the optimal combination of DUNE 3-horn and T2HKK1.5 (and T2HKK2.0).

\begin{figure}[t]
\centering
\includegraphics[width=0.49\textwidth]{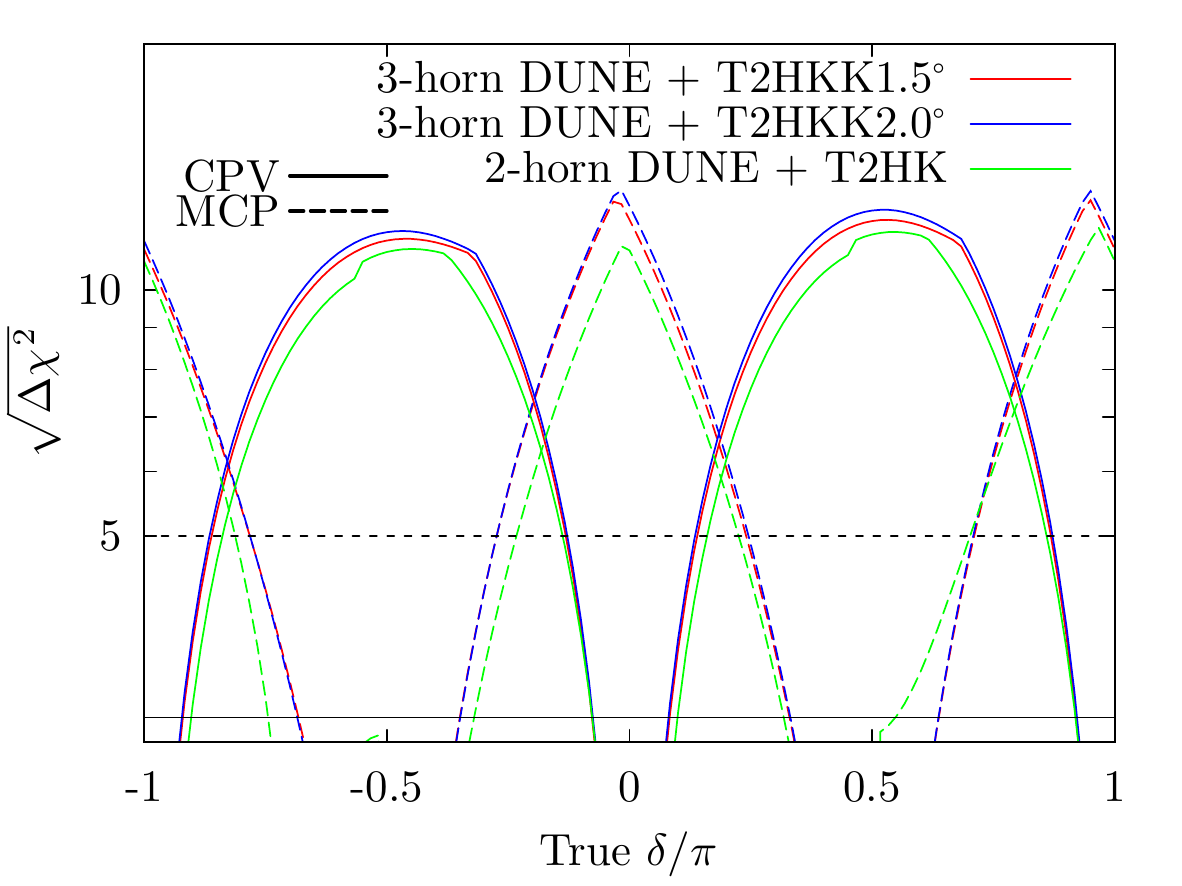}
\includegraphics[width=0.49\textwidth]{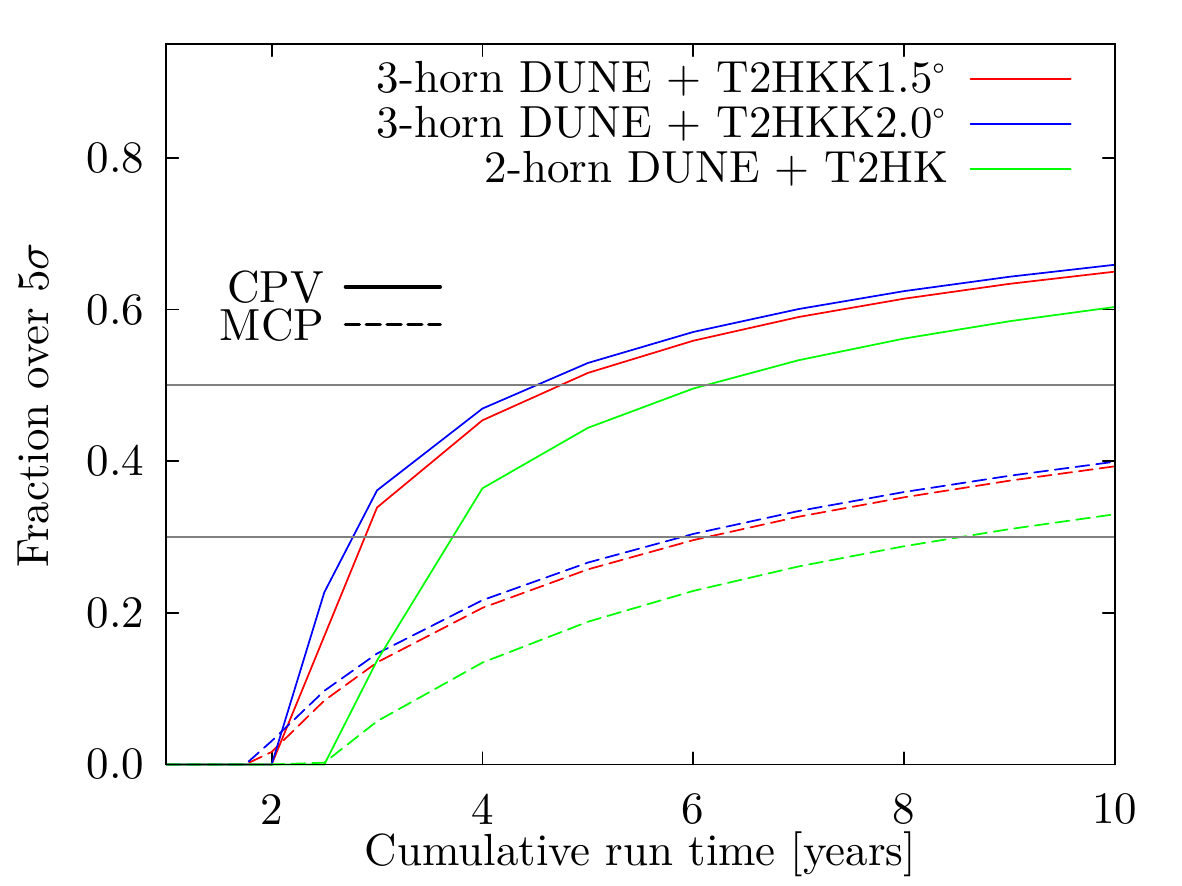}

\caption{\label{fig:optimised_CPV_MCP} Left: the CPV and MCP sensitivity for
the combination of DUNE with the 3-horn flux and T2HKK$1.5^\circ$($2.0^\circ)$. For
reference, we also show the combination of the two standard designs: DUNE with 2
horn beam and T2HK (green). Right: the fraction of $\delta$ parameter space for
CPV (MCP) sensitivity over $5\sigma$ for the same configurations as on the left 
panel.
The configurations assumed here are described in
\refsec{sec:alt_run_times} and the true oscillation parameters are
given in \reftab{tab:global_fit_parameters}.
}

\end{figure}

\begin{figure}[t] \centering
\includegraphics[width=0.7\textwidth]{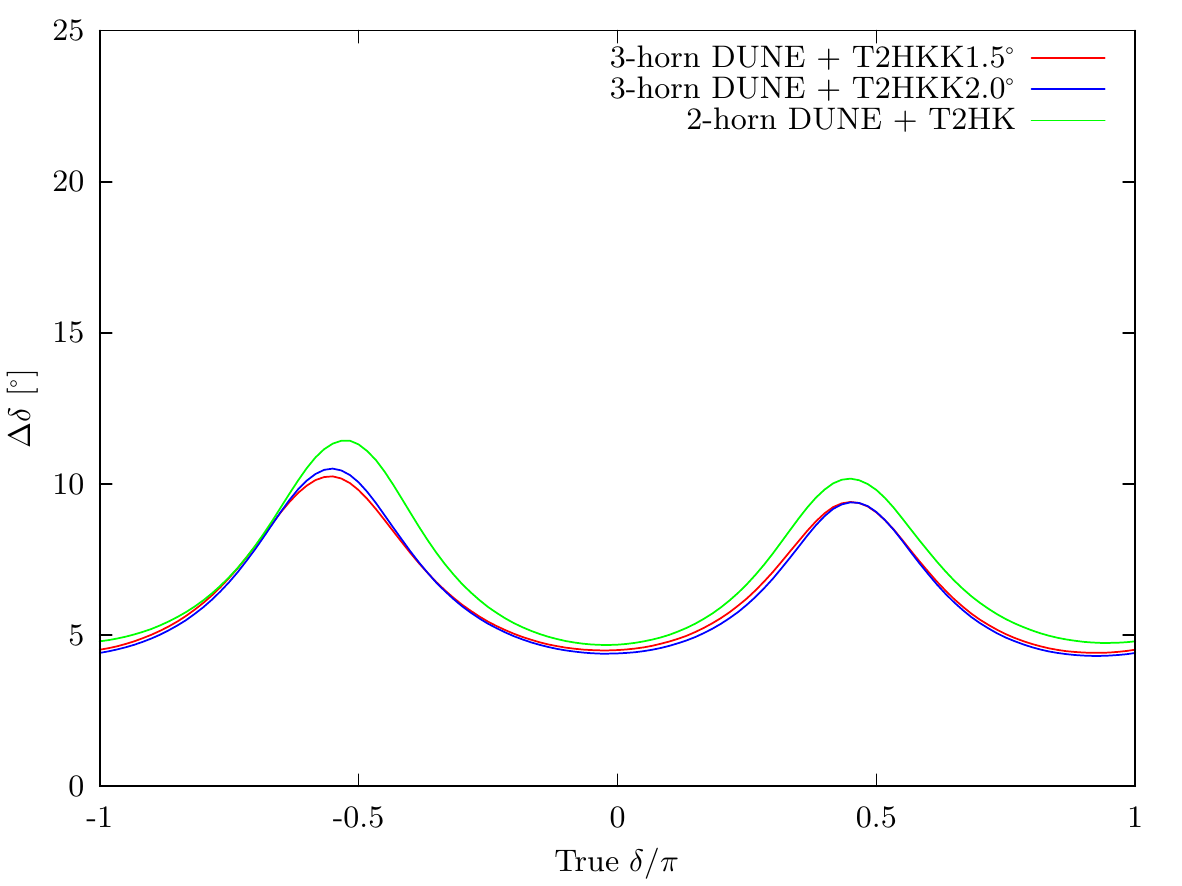}
\caption{\label{fig:optimised_Dd} The $1\sigma$ error on $\delta$ for the
combination of DUNE with the 3-horn flux and
T2HKK$1.5^\circ$($2.0^\circ$) shown in red (blue). For reference, we also show
the combination of the two standard designs: DUNE with 2-horn beam and T2HK
with one tank (green). 
The configurations assumed here are described in
\refsec{sec:alt_run_times} and the true oscillation parameters are
given in \reftab{tab:global_fit_parameters}.
}

\end{figure}

In \reffig{fig:optimised_MO}, we show the minimum sensitivity expected for the
mass ordering for this optimal configuration of DUNE + T2HKK. A $4\sigma$
measurement is expected after less than a year, which increases to $5\sigma$
after 1.5 years. In \reffig{fig:optimised_CPV_MCP}, we show the
significance at which we can expect to exclude CP conservation (solid) and
maximal CP violation (dashed). These are expected to reach a maximal
significance of $11\sigma$ and $12\sigma$, respectively. The advantage of the
combination is clearer when the performance is viewed in terms of the minimal
run time required for the exclusions to be made at $5\sigma$. The combination of
DUNE + T2HKK expects to have greater than $5\sigma$ exclusion of CP
conservation for more than $25\%$ ($50\%$) of the parameter space after $2.5$
($5$) years of cumulative run time. 
For the exclusion of maximal CP violation, longer run times are
required: about $6$ years ensures the exclusion for more than $25\%$ of values
of $\delta$.
For the precision on $\delta$, shown in \reffig{fig:optimised_Dd}, we see that
the optimal combination of DUNE + T2HKK could expect a measurement around a CP
conserving value with an uncertainty of only $4.5^\circ$. This worsens for
maximally CP violating values of $\delta$ to around $10^\circ$.

\section{\label{sec:conclusions}Conclusions}

DUNE and T2HK will lead the way in key measurements of the neutrino oscillation
parameters. These long-baseline experiments will make high statistics
determinations of the mass ordering, the first precision measurements of
$\delta$, and have an excellent chance to establish the presence of fundamental
CP violation in the leptonic sector.
In this article, we have studied the expected performance of these two
experiments, including possible alternative designs which have been recently
suggested. We see that, thanks to their different designs, both the energy
profiles of the beam and the different baseline distances chosen, DUNE and T2HK
have different sensitivities to the mass ordering and the value of $\delta$, leading to a
natural complementarity.

DUNE, with its long baseline and significant matter effects, excels at
measuring the mass ordering. It can expect a greater than $5\sigma$
determination after between $2$ and $6$ years depending on the true value of
$\theta_{23}$.  T2HK is limited in its sensitivity for this measurement, but
the combination of data collected at T2HK with the DUNE data reduces the impact
of the worst-case scenario, significantly reducing the required run times. DUNE
+ T2HK can expect the same measurement in less than $3$ years regardless of
$\theta_{23}$. 
The roles are reversed for measurements of the CP phase $\delta$.  Although
T2HK has high sensitivity for this measurement, it suffers from a degeneracy
linked to the mass ordering which may limit its performance for some values of
$\delta$.  In isolation, T2HK expects to be able to exclude CP conservation at
greater than $5\sigma$ for more than $50\%$ of the parameter space after around
$5$ years in the best case scenario. DUNE alone would require at least $11$
years of data for the same measurement, but the combination of the two
experiments, assumed to collect data in parallel, would take at most $5$ years.
This is a particularly clear example of synergy between the two designs, as the
degeneracy limiting T2HK's sensitivity can be lifted by the inclusion of DUNE
data.
A similar but less pronounced synergy is present for the measurement of maximal
CP violation, where the MO degeneracy again affects T2HK's sensitivity.
However, the combination of DUNE + T2HK mitigates this limitation and can
exclude MCP for between $42$--$50\%$ of the parameter space after 10 years
parallel data taking. 
For the measurement of the octant sensitivity, we find that  to exclude the
upper octant solution at 5$\sigma$ with a true value of $\theta_{23}=40^\circ$,
T2HK needs about 2 years, while DUNE requires a slightly longer run time.
This pattern is repeated for the exclusion of maximal mixing, where for the
true value $\theta_{23}=40^\circ$, $5\sigma$ exclusion at DUNE takes around 2
years, while T2HK can make this exclusion in only 1 year. 
For these measurements, the performance of the combination of DUNE and T2HK
generally follows the sensitivity of T2HK, although some small benefit is found
from the inclusion of extra data.
We have also studied the precision on $\sin^2\theta_{23}$, where there is a
strong dependence on the true value of $\theta_{23}$, with the worst precision
close to maximal mixing, as expected for a measurement driven by the
disappearance channel. At the peak, $\Delta(\sin^2\theta_{23})$ for DUNE is
about $0.041$ while T2HK can improve this, peaking around $0.032$.
Extending our study to the 1$\sigma$ joint precision on $\delta$ and
$\sin^2\theta_{23}$, we see the measurement to these two parameters are largely
independent, due to the disappearance channel driving the fits to $\theta_{23}$
and the appearance channel dictates $\delta$. The precision gets worse at
$\theta_{23}=45^\circ$, as seen before, and improves as we move from this
maximal value. For $\theta_{23}=40^\circ$ or $50^\circ$, the precision on
$\theta_{23}$ is around $0.2^\circ$ ($0.13^\circ$) for DUNE (T2HK). However,
near maximal mixing the value increases to $\Delta\theta_{23}=2^\circ$
($0.95^\circ$) for DUNE (T2HK).
 
We have stressed in particular the sensitivity to $\delta$, studying the
behaviour of the $1\sigma$ uncertainty on $\delta$, $\Delta \delta$, in some
detail. We find that for equal event rates, the two experiments
perform comparably, with each having the best sensitivity for around half of
the parameter space. For fixed run times of 10 years, however, T2HK has on
average the best sensitivities and expects $\Delta \delta$ to lie between $6$
and $18^\circ$. 
We have shown that T2HK is not intrinsically more sensitive to $\delta$, but
increases its sensitivity through large statistics. DUNE on the other hand, is
limited by lower event rates, suggesting that it may be able to
improve its sensitivity with further data collection. However, to provide
uniformly improved precision on $\delta$, T2HK would require between $2$ and
$3$ times as many events as DUNE.
Beyond the question of statistics, we have discussed the complementarity of the
two experiments for precision measurements of $\delta$. DUNE's wide-band beam
helps to compensate for a loss of sensitivity at the first oscillation maximum,
which hampers T2HK's performance. We find that DUNE performs best for maximally
CP violating values of $\delta$ and T2HK, in contrast, prefers CP conserving
values. When combined, these experiments complement each other, and the global
sensitivity to $\delta$ is well covered by the two technologies: we expect DUNE
+ T2HK to reach $4.5^\circ\lesssim\Delta\delta\lesssim 11^\circ$ for all values
of $\delta$ after 10 years of running in parallel. 

We have also considered potential alternative designs for T2HK and DUNE. T2HK
may locate its second detector module in southern Korea, while DUNE has been
associated with two beam designs beyond its 2-horn design: a 3-horn optimised
design and the nuPIL design. Although the nuPIL design is no longer being
actively pursued by the collaboration, we have shown that this novel technology
leads to interesting phenomenology which highlights the flux dependence of an
experiment's sensitivities to key measurements.
We have investigated the ability of these designs to determine the mass
ordering, to exclude CP conservation and maximal CP violation, and to measure
$\delta$. These alternatives are promising extensions of the current physics
programme, and lead to modest improvements in all measurements studied in this
work. We have identified the combination of DUNE (3-horn) and T2HKK with a flux
between $1.5^\circ$ and $2.0^\circ$ off-axis as the optimal choice; although,
the difference between the performance of the 2-horn and 3-horn designs is not
very significant. Assuming parallel running, the optimal combination expects to
discover the mass ordering at $5\sigma$ after only $0.7$ years, to be able to
exclude CP conservation at $5\sigma$ for more than $50\%$ of the parameter
space after $2.5$ years, and to measure $\delta$ around CP conserving
(maximally violating) values with an uncertainty of around $4.5^\circ$
$(10^\circ)$ after its full data-taking period.

We conclude that DUNE and T2HK have a natural complementarity, thanks to key
differences in their designs. Although design modifications, such as nuPIL for
DUNE or the location of T2HK's second detector in Korea, have quite distinct
features which could upset the existing synergy, we find that the combination
of the two experiments is quite robust. 
Sensitivity to the mass ordering will come primarily from DUNE, sensitivity to
CPV sees a larger contribution from T2HK (although due to the mass ordering
degeneracy the sensitivity is notably improved by DUNE data, or perhaps
data from atmospherics), but precision on $\delta$ is a bit more nuanced
with wider-band information being preferred for maximally CP violating values
of $\delta$, and high statistics first maximum measurements preferred for CP
conserving values. Overall, the global physics program greatly benefits from
breadth and variation in design. 


\acknowledgments

We would like to thank Thomas Schwetz-Mangold for pointing out an
inconsistency in the analysis of CPV sensitivity in an earlier version of this
article. We also thank Alan Bross, Ao Liu, Michel Sorel, Mark Thomson, and
Elizabeth Worcester for valuable comments and experimental input on DUNE and
nuPIL. Thanks also to the Hyper-Kamiokande proto-collaboration for information
incorporated into our simulations.

The authors acknowledge partial support for this work from ELUSIVES ITN
(H2020-MSCA-ITN-2015, GA-2015-674896- ELUSIVES), and InvisiblesPlus RISE
(H2020-MSCARISE-2015, GA-2015-690575-InvisiblesPlus).
In addition, PB, SP and TC would like to acknowledge support from the European
Research Council under ERC Grant ``NuMass'' (FP7-IDEAS-ERC ERC-CG 617143), and
SP gratefully acknowledges partial support from the Wolfson Foundation and the
Royal Society.

\appendix
\section{\label{app:t2hk_details}Further details of T2HK simulation}

Our model of the T2HK detector significantly deviates from previous work. In
this appendix, we give some further details of its implementation which where
glossed over in the main text and a comparison with the collaboration's
simulation.

\subsection{Energy bins}

Our model of the T2HK detector(s) features $12$ energy bins. Bin 1 collects all
events below 0.35 GeV. The next 5 bins are 0.1 GeV wide, collecting events from
0.35-0.85 GeV. The next two bins are 0.2 GeV wide, followed by a single bin of
0.25 GeV width. There are then 3 increasingly broad bins, from 1.5 to 3.5, 3.5
to 6 and an overspill bin from 6 to 10 GeV.  

\subsection{Channel systematic uncertainties}

Our model of the systematic uncertainty at T2HK uses two general normalisation systematics for the signal and background of each channel. The precise systematic errors used in our simulation are given, channel by channel, in \reftab{tab:hk-systematics}.

\begin{table}[t]
\centering%

\begin{tabular}{|l|c|c|c|c|}
\hline
           & $\nu_\mu\to\nu_e$ & $\nu_\mu\to\nu_\mu$ & $\anu_\mu\to\anu_e$ & $\anu_\mu\to\anu_\mu$ \\
\hline
Signal     & 2.4\%             & 2.7\%               & 2.925\%               & 2.7\% \\
Background & 2.4\%             & 2.7\%               & 2.925\%               & 2.7\% \\
\hline
\end{tabular}
\caption{Systematic errors used for T2HK simulation.}\label{tab:hk-systematics}
\end{table}

\subsection{Comparison with published event rates}

\begin{figure}[t]

\centering

\includegraphics[width=0.9\textwidth]{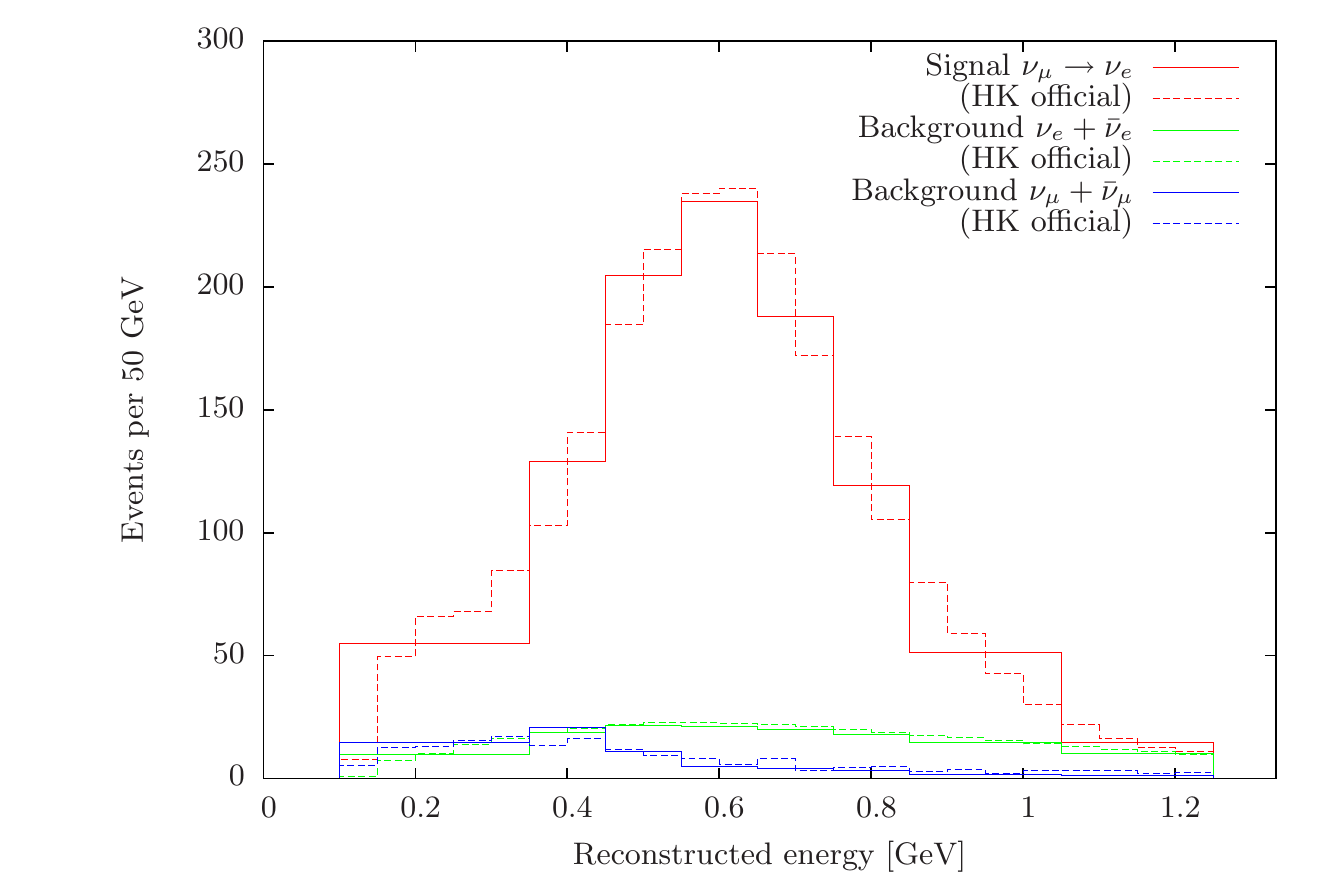}

\caption{\label{fig:T2HK_app_spec}T2HK appearance spectrum from our simulation compared 
to official event rates \cite{HKDR}. Note that the finer binning of the rates published by T2HK are shown for reference, 
but this finer granularity is not used in their oscillation fits; the binning we have used 
in our own fits has been chosen to match that of the official T2HK studies}

\end{figure}

In \reffig{fig:T2HK_app_spec} we compare the event rates from our simulation to
the official rates published by T2HK. The official simulation does not use
GLoBES, and our reproduction is a non-trivial check to show that the signal and
background modelling in our simulation is faithful. Additional checks have also been
made to ensure that our simulations are able to reproduce the final sensitivities 
of official simulations, once we have modified our simulation to match the priors
and chosen fitted parameters of the official simulations.

\section{\label{sec:total_number}Total number of events for all configurations}
In \reftab{tab:total_number}, we show the expected total rates for events and
backgrounds for all configurations, discussed in this work. We adopt the true
values according to NuFit 2.2, shown in Tab. \ref{tab:global_fit_parameters},
but assume $\delta=0$. Two mass orderings are considered. For all cases,
cumulative run time is set 10 years. For DUNE, we take events from 0.5 GeV to 8
GeV, while for the other configurations we take from 0.1 GeV to 1.2 GeV.

\begin{table}[t]
%
\begin{tabular}{|l|c|c|c|c|c|c|c|c|}
\hline
&\multicolumn{2}{c|}{$\nu_\mu\to\nu_e$} & \multicolumn{2}{c|}{$\nu_\mu\to\nu_\mu$}& \multicolumn{2}{c|}{$\anu_\mu\to\anu_e$} & \multicolumn{2}{c|}{$\anu_\mu\to\anu_\mu$}  \\\cline{2-9}
& NO & IO& NO & IO & NO & IO& NO & IO\\
\hline
2-horn DUNE (total)&2353 & 1589 & 13269& 13189 & 667 &1210 & 
13180& 13095 \\\hline
2-horn DUNE (BG)& 486& 502 & 200 &203  & 253&252 & 
111&112 \\\hline
3-horn DUNE (total)& 2317&1561 & 13773&13774 & 587&1087 & 
5125&5081  \\\hline
3-horn DUNE (BG)& 488&504 & 199&203 & 228&227 & 
90&92 \\\hline
nuPIL DUNE (total)& 1209&721 & 5756&5801 & 230&580 & 
2079&2077 \\\hline
nuPIL DUNE (BG)& 111&116 & 84&85 & 43&42 & 
38&38 \\\hline
staged T2HK (total)&2294&2514 & 9221&9157 & 2093&2715 & 
10997&10855 \\\hline
staged T2HK (BG)& 522&525 & 619&619 & 695&694 & 
805&805 \\\hline
1tank T2HK (total)&1638&1795 & 6587&6540 & 1495&1939 & 
7855&7754 \\\hline
1tank T2HK (BG)& 373&375 & 442&442 & 496&495 & 
575&575 \\\hline
T2HKK$1.5^\circ$ (total)&207&196 & 3151&3066 & 288&275 & 
4453&4362 \\\hline
T2HKK$1.5^\circ$ (BG)& 96&96 & 117&117 & 148&148 & 
176&176 \\\hline
T2HKK$2.0^\circ$ (total)&163&154 & 1913&1854 & 198&194 & 
2331&2256 \\\hline
T2HKK$2.0^\circ$ (BG)& 51&51 & 53&53 & 71&71 & 
63&63 \\\hline
T2HKK$2.5^\circ$ (total)&121&116 & 1269&1283 & 135&146 & 
1322&1328 \\\hline
T2HKK$2.5^\circ$ (BG)& 29&29 & 36&36 & 37&37 & 
41&41 \\\hline

\end{tabular}
\caption{The total rate of events and backgrounds for all configurations with 
cumulative run time of 10 years, assuming $\delta=0$ for normal ordering (NO) 
and inverse ordering (IO). The true values are adopted according to the best 
of NuFit 2.2, shown in Tab. \ref{tab:global_fit_parameters}.
For all configurations of DUNE, we take events from 0.5 GeV to 8 GeV, while
for the others we take from 0.1 GeV to 1.2 GeV.}\label{tab:total_number}
\end{table}

\section{\label{app:MO_sensitivity}Mass ordering sensitivity at high significance}

\begin{figure}[t]

\centering
\includegraphics[width=0.49\textwidth, clip, trim=0 3 5 5]{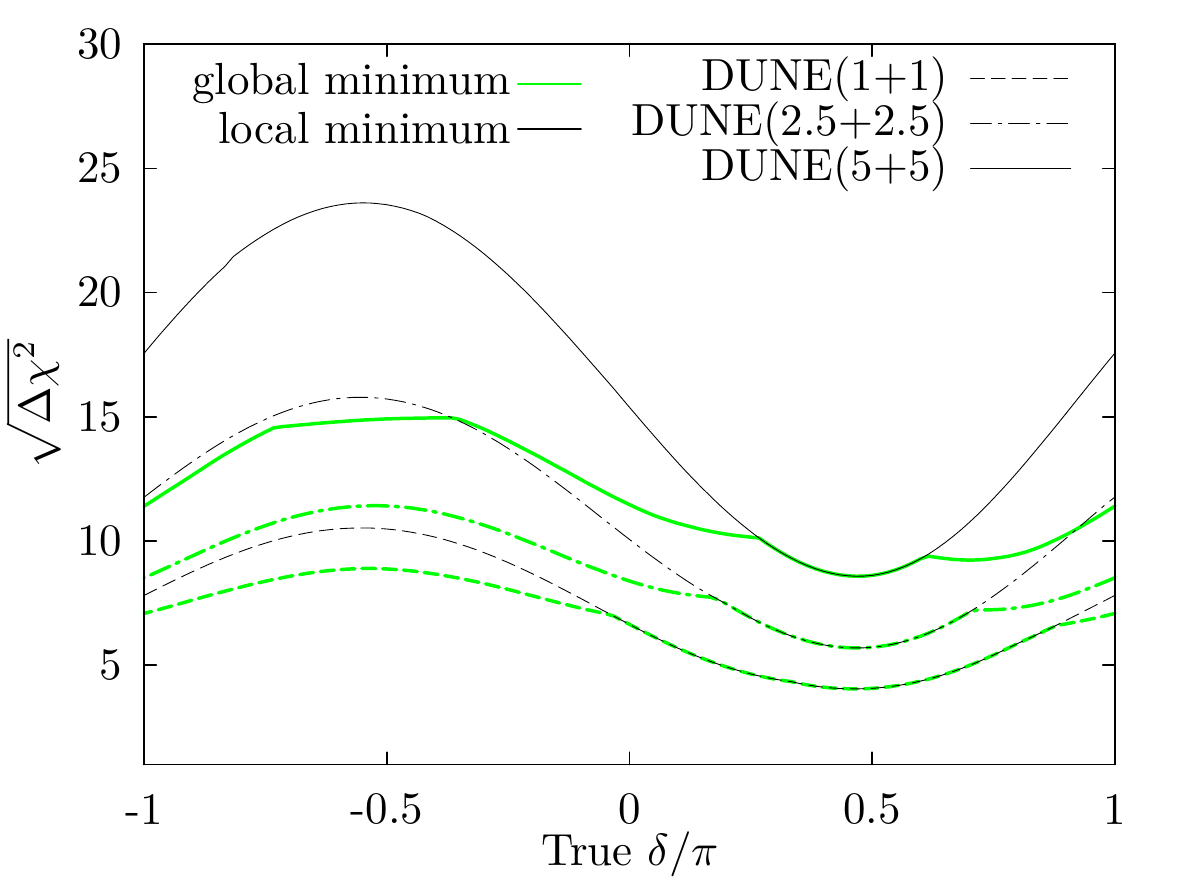}
\includegraphics[width=0.49\textwidth]{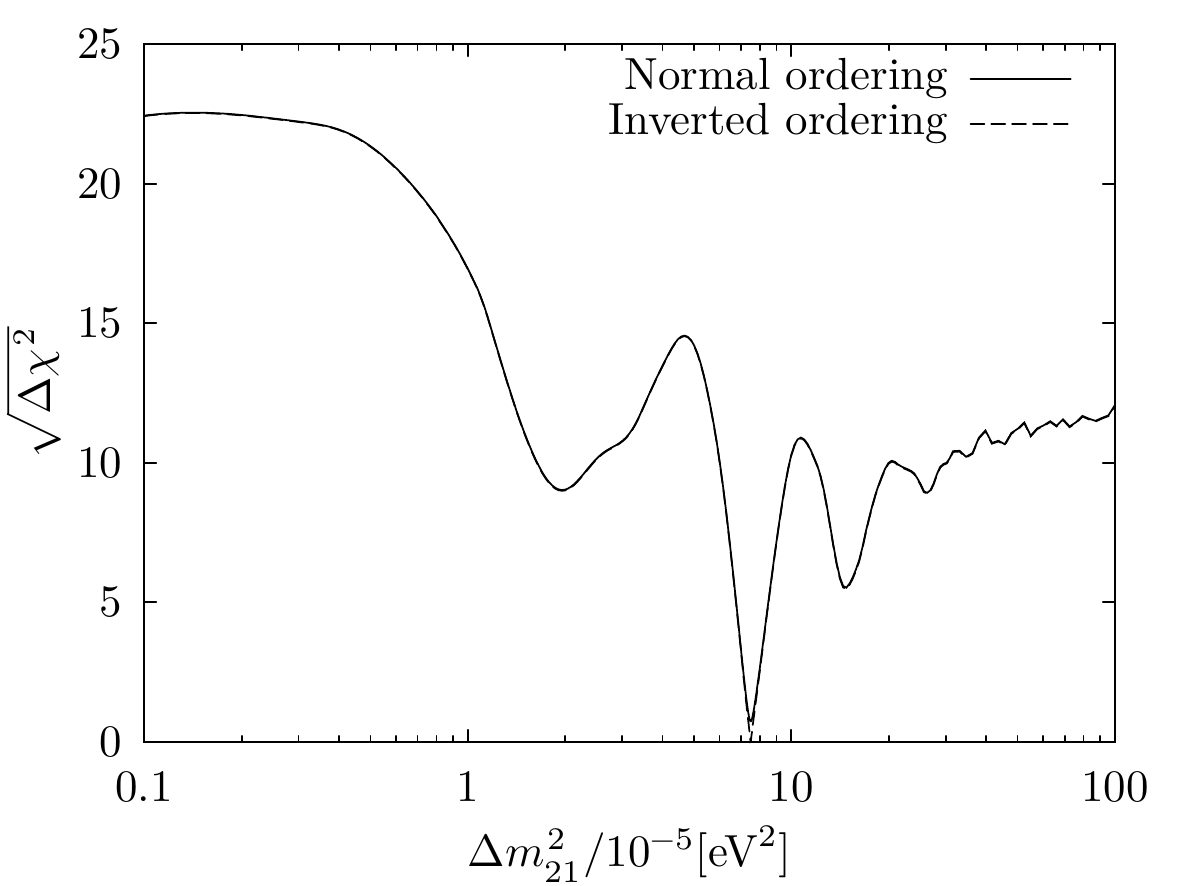}

\caption{\label{fig:nufit_prior}Left: Comparison of our reported sensitivities
based on the global minimum and more common published versions which show only
a local minimum. This is due to the presence of additional wrong-ordering
minima at high significance in the global data. Right: The prior on $\Delta
m^2_{21}$ provided by the NuFit global fit \cite{Gonzalez-Garcia:2014bfa}.
Additional local minima with significances around $5\sigma$
($\Delta m^2_{21}\approx 1.6\times 10^{-4}$ eV$^2$) and $10\sigma$ ($\Delta
m^2_{21}\approx 2.6\times 10^{-4}$ eV$^2$) lead to the unusual behaviour in our
reported mass ordering sensitivities.}

\end{figure}

The sensitivity to mass ordering is conventionally reported as the difference
between the value of a $\chi^2$ statistic for the true parameter set and the
close degenerate set with the atmospheric mass splitting changed by the
following mapping,
\[  \Delta m^2_{31} \to -\Delta m^2_{31} + \Delta m^2_{21}. \]
This local minimum becomes a worse and worse fit as data is collected, and
reaches a $\Delta \chi^2$ value of above $8\sigma$ within a few years of
running DUNE. This method computes the decreasing quality of a poor fit to the
data; however, there are lots of parameter sets which are poor fits to the
current data, and many cannot be excluded with a significance greater than
$8\sigma$.
Statistically speaking, to establish the mass ordering we must exclude all
possible parameter sets with that ordering regardless of the other parameter
values. In some circumstances, this may mean the local minimum identified above
is not the true global wrong-ordering minimum.
We find this problem is relevant for DUNE as soon as the local minimum
approaches a $5\sigma$ exclusion. This is because the global prior for the
solar mass-squared splitting, $\Delta m^2_{21}$ has a second minimum at around
this significance. 
The long-baseline experiments considered in this paper, offer no sensitivity to
this parameter themselves, and rely on the priors to help constrain it. We have
plotted the prior that we have used in our simulations in
\reffig{fig:nufit_prior}, where the second minimum can be seen just above the
global minimum. For DUNE to exclude the wrong mass ordering at above $5\sigma$,
we must ensure it considers all values of $\Delta m^2_{21}$ allowed by the
global data at this significance. We find that DUNE can often exclude this
minimum only at lower significance than the more obvious local minimum
corresponding to the expected degeneracy. This causes the lower significances,
and discontinuous behaviour, that we have reported in \refsec{sec:MO}. On
average, this reduces the expected significance of the mass ordering
measurement by around $5\sigma$.

Of course, predicting any sensitivities at high significance requires good
control over all other aspects of the statistical modelling, and we do not
pretend that our method correctly models all uncertainties up to very small
fluctuations. However, we point out this particular subtlety as a concrete
example of how the oft quoted sensitivity is not quite what it seems: it is the
confidence at which we can expect to exclude a particular local minimum, not to
the best-fitting set of parameters with the wrong ordering. The difference in
these quantities starts to become relevant at for DUNE at very modest
exposures. In the left panel of \reffig{fig:nufit_prior}, we show the
difference in $\Delta\chi^2$ values for the local minimum and the full set of
wrong ordering parameter sets (green), which starts to be visible after only
$2$ years run time. 
We hope that this example helps to highlight some of the complexities of making
precise statements with high confidence sensitivities.
 

\bibliographystyle{apsrev4-1}
\bibliography{lib}{}

\end{document}